\documentclass[fleqn,usenatbib]{mnras}

\usepackage{newtxtext,newtxmath}
\usepackage[T1]{fontenc}

\DeclareRobustCommand{\VAN}[3]{#2}
\let\VANthebibliography\thebibliography
\def\thebibliography{\DeclareRobustCommand{\VAN}[3]{##3}\VANthebibliography}

\usepackage{graphicx}	% Including figure files
\usepackage{amsmath}	% Advanced maths commands
\usepackage{comment}
\usepackage{tablefootnote}

\newcommand{\modelone}{this work}

\defcitealias{PS20}{PS20}

\title[Hybrid cooling model]{\textsc{Hybrid-chimes}: A model for radiative cooling and the abundances of ions and molecules in simulations of galaxy formation}

\author[Sylvia Ploeckinger et al.]{
Sylvia Ploeckinger,$^{1}$\thanks{E-mail: sylvia.ploeckinger@univie.ac.at}
Alexander J. Richings,$^{2,3}$
Joop Schaye,$^{4}$
James W. Trayford,$^{5}$
Matthieu Schaller,$^{4,6}$\newauthor
and Evgenii Chaikin${^4}$
\\
$^{1}$Department of Astrophysics, University of Vienna, T\"urkenschanzstrasse 17, 1180 Vienna, Austria\\
$^{2}$Centre for Data Science, Artificial Intelligence and Modelling, University of Hull, Cottingham Road, Hull, HU6 7RX, UK\\
$^{3}$E. A. Milne Centre for Astrophysics, University of Hull, Cottingham Road, Hull, HU6 7RX, UK \\
$^{4}$Leiden Observatory, Leiden University, PO Box 9513, 2300 RA Leiden, the Netherlands \\
$^{5}$Institute of Cosmology and Gravitation, University of Portsmouth, Dennis Sciama Building, Burnaby Road, Portsmouth PO1 3FX, UK\\
$^{6}$Lorentz Institute for Theoretical Physics, Leiden University, PO Box 9506, NL-2300 RA Leiden, The Netherlands
}

\date{Accepted XXX. Received YYY; in original form ZZZ}

\pubyear{2025}

\begin{document}
\label{firstpage}
\pagerange{\pageref{firstpage}--\pageref{lastpage}}
\maketitle

\begin{abstract}
Radiative processes play a pivotal role in shaping the thermal and chemical states of gas across diverse astrophysical environments, from the interstellar medium (ISM) to the intergalactic medium. We present a hybrid cooling model for cosmological simulations that incorporates a comprehensive treatment of radiative processes, including parameterizations of the interstellar radiation field, cosmic ray rates, and dust physics. The model uses the \textsc{chimes} chemical network and combines on-the-fly non-equilibrium calculations with quasi-equilibrium cooling rates. The quasi-equilibrium rates account for the time-dependent free electron fractions of elements tracked in non-equilibrium, balancing computational efficiency with physical accuracy.
We evaluate the performance under various conditions, including the thermal evolution of primordial gas at the cosmic mean density, the properties of the warm and cold neutral media in Milky Way-like galaxies, and the atomic-to-molecular hydrogen transition. 
We demonstrate that thermal equilibrium predictions for the neutral phases of the ISM underestimate the median gas pressures in simulations of isolated galaxies by up to 0.5~dex. Finally we find that the atomic-to-molecular hydrogen transition is shifted to lower densities by up to 1~dex if oxygen is not included in the chemical network. Our work provides a robust framework for studying the multi-phase ISM and its role in galaxy formation and evolution.
\end{abstract}


%%%%%%%%%%%%%%%%%%%%%%%%%%%%%%%%%%%%%%%%%%%%%%%%%%

%%%%%%%%%%%%%%%%% BODY OF PAPER %%%%%%%%%%%%%%%%%%

\section{Introduction}

Radiative processes are fundamental to the evolution of baryonic matter in the Universe, governing the thermal and chemical states of gas across a wide range of environments. These processes are not only critical for the formation of stars and galaxies but also serve as essential tools for interpreting spectroscopic observations. Emission and absorption lines from ions and molecules provide insight into the composition, physical properties (e.g., density, temperature), and dynamics (e.g., turbulence, inflows, outflows) of gas within galaxies (the interstellar medium, ISM, see review by \citealp{Saintonge2022}), around galaxies (the circumgalactic medium, CGM, see reviews by \citealp{Tumlinson2017,FG2023CGM}), and in the vast intergalactic medium (the IGM, see review by \citealp{McQuinn2016}). Understanding the interplay between radiative cooling, chemical reactions, and the ambient radiation field is therefore crucial for connecting theoretical models to observations and for advancing our understanding of galaxy formation and evolution.

Determining the species fractions and radiative cooling rates of gas relies on a series of intertwined chemical reactions and interactions with the radiation field (see e.g. appendix B in \citealp{chimes2014optthin} for a reaction network that includes 157 species). A full implementation of these processes in simulations would require solving large chemical networks coupled with radiative transfer calculations for each resolution element at every timestep (see e.g.~\citealp{Katz2022}), which is typically too computationally expensive for simulations of small volumes and not feasible for large cosmological volumes. 

Several simplifying assumptions are often made to model radiative cooling and heating processes. The assumption of collisional ionization equilibrium drastically decreases the complexity by ignoring photo-ionization and photo-heating and results in ion species and cooling rates that can be pre-tabulated for each element and only depend on the gas temperature \citep[e.g.,][]{Cox1969, BoehringerHensler1989, SutherlandDopita1993}. Including a redshift-dependent radiation field (e.g. a UV/X-ray background from distant galaxies and quasars: e.g., \citealp{HM2001, HM2012, FG20}) and assuming optically thin gas requires tables with more dimensions but does not significantly add to the runtime of a simulation because both rates and fractions can be pre-tabulated (e.g. with \textsc{cloudy} \citealp{Cloudy1998PASP, Cloudy2023}), as done by \citealp{Wiersma2009} for the \textsc{owls} simulations \citep{Schaye2010OWLS} and only need to be interpolated at runtime. While this approach is computationally efficient and widely used in state-of-the-art cosmological simulations (e.g., \textsc{eagle}: \citealp{EagleSchaye2015}, \textsc{horizonAGN}: \citealp{HorizonAGN2016}, \textsc{romulus}: \citealp{Romulus2017}), optionally augmented with a simple prescription for the self-shielding of neutral hydrogen \citep{Rahmati2013}, it cannot capture non-equilibrium effects that arise in rapidly evolving environments. 

Non-equilibrium chemical models, which follow the time-dependent evolution of ions and molecules, offer a more accurate representation of the ISM, CGM, and IGM but their computational cost scales super-linearly with the number of included species. Reduced chemical networks that include a small subset of species are often used to calculate fractions and rates of selected chemical species in non-equilibrium. For example, the \textsc{Krome} library \citep{Grassi2014, Krome2016} allows the user to select relevant species and builds optimized routines that solve the associated system of ordinary differential equations (ODEs) to follow the non-equilibrium species abundances and the thermal state of the gas over time. A reduced network of nine primordial and seven metal species, solved with \textsc{Krome}, has been used in simulations of both isolated galaxies \citep{Sillero2021} and cosmological zoom-in simulations of a dwarf galaxy (for $z\ge6$; \citealp{Lupi2020M}). The \textsc{grackle} chemistry and cooling library \citep{Grackle2017} follows the non-equilibrium primordial chemistry and cooling for hydrogen, deuterium and helium species. The non-equilibrium cooling rates are supplemented by pre-tabulated metal cooling rates that assume chemical equilibrium\footnote{We refer to steady-state chemistry and ionization equilibrium as  `chemical equilibrium' or `equilibrium chemistry' for brevity.}. \textsc{grackle} has been applied in cosmological simulations, such as the \textsc{Agora} project \citep{Agora2014} and \textsc{Simba} \citep{Dave2019}. Another example of a reduced chemical network is \textsc{Hyacinth} \citep{Hyacinth2024}, which follows selected hydrogen, helium, carbon, and oxygen species. A more minimal approach, where only a subset of primordial species is followed in non-equilibrium, is used e.g. in \textsc{tng} \citep{TNGPillepich2018} and the \textsc{lyra} simulations \citep{LyraI2021}.

In \citet{PS20} (hereafter: \citetalias{PS20}) we presented species fractions as well as cooling and heating rates that were calculated with a slightly modified\footnote{We reduced the photo-dissociation rate of H$_2$ by cosmic rays, which has since been implemented in more recent versions of \textsc{cloudy}, see \citet{Shaw2020} for details.} version of \textsc{cloudy} v17.01 \citep{Ferland2017}. Assuming an interstellar radiation field, a cosmic ray rate and a shielding column density that depend only on local gas densities (temperature, density, metallicity) and redshift (for the homogeneous background radiation field), we provided pre-tabulated cooling and heating rates that can be incorporated in simulations without a full radiative transfer model. These tables have been used by e.g. the \textsc{lyra} \citep{LyraI2021} and \textsc{flamingo} \citep{Flamingo2023} projects. In their appendix A2, \citet{PS20} included a suggestion for coupling the pre-tabulated cooling rates, that assume chemical equilibrium, to a non-equilibrium chemistry network. To avoid inconsistencies, it is necessary to use the same chemical network, atomic data, and assumptions (e.g. for shielding and radiation fields) for both the equilibrium and non-equilibrium rates.

\textsc{Chimes}\footnote{\href{https://richings.bitbucket.io/chimes/home.html}{richings.bitbucket.io/chimes/home.html}} is a chemical network \citep{chimes2014optthin, chimes2014shielded} which includes 157 chemical species in total, but with the option to actively track only a subset of the elements H, He, C, N, O, Ne, Si, Mg, S, Ca, and Fe. The full network has been used in simulations of isolated low-mass galaxies \citep{Richings2016, Richings2016GMC, Richings2022} and of fast molecular outflows in quasars \citep{Richings2018ChimesPEupdate}. While \textsc{chimes} solves for the non-equilibrium species abundances, it can also be run in equilibrium mode, which has been used to post-process simulations \citep{Keating2020}.

In this work, we present a hybrid non-equilibrium cooling model based on \textsc{chimes}: \textsc{hybrid-chimes}. While \textsc{hybrid-chimes} may be used with parameters supported by \textsc{chimes}, here we build on the framework of \citetalias{PS20} and use the Jeans column density \citep{Jeans1902, Schaye2001ApJ559} to scale the normalization of the interstellar radiation field (ISRF) and the cosmic ray rate, and to include self-shielding by gas and dust.  We evaluate the performance of our model under various conditions, including the thermal evolution of primordial gas at the cosmic mean density, the warm and cold neutral medium in Milky Way-like galaxies, and the transition from atomic to molecular hydrogen. 

This paper is organized as follows. In Section~\ref{sec:method}, we describe the methodology, starting with a description of the model input parameters (Section~\ref{sec:inputparameter}) and key updates to the \citetalias{PS20} model. In Section~\ref{sec:equilibriumabundances}, we demonstrate the calculation of equilibrium abundances in our model and highlight the systematic differences between using full and reduced chemical networks that we identified (Section~\ref{sec:equilibriumabundances:red}). We present the hybrid non-equilibrium model, \textsc{hybrid-chimes}, in Section~\ref{sec:hybrid} and revisit the included cooling and heating processes in Section~\ref{sec:processes}. In Section~\ref{sec:results}, we present results from a series of tests and simulations. First, we compare the thermal evolution of primordial gas at the cosmic mean baryon density with observations (Section~\ref{sec:thermevol}). In Section~\ref{sec:neutral}, we present the expected thermal equilibrium pressures in the ISM from the balance of cooling and heating rates and their dependence on variations in the assumed parameter values. Finally, in Section~\ref{sec:galaxysimulations}, we use the hybrid model in simulations of isolated galaxies and discuss the mismatch between the actual pressures of the multi-phase ISM in the simulations and the thermal equilibrium pressures from Section~\ref{sec:neutral}. We discuss the implications of our findings in Section~\ref{sec:summary}.

As in \citetalias{PS20}, the grid spacing in temperature, density, and metallicity, as well as most of the tabulated properties, are in $\log$ base 10. Throughout the paper $\log$ refers to $\log_{\mathrm{10}}$ and we use $10^{-50}$ as a floor value for property values that would be zero. All bins are generally equally spaced in $\log$ but the metallicity dimension has an additional bin for primordial abundances ($Z=0$, or: $\log Z/\mathrm{Z}_{\odot} = -50$), and we use the solar metallicity ($\mathrm{Z}_{\odot} = 0.0134$) and solar abundance ratios from \citet{Asplund2009}. 

\section{Method}\label{sec:method}

We use the chemical network software package \textsc{chimes} \citep{chimes2014optthin, chimes2014shielded}, to calculate both equilibrium and non-equilibrium species abundances of selected elements. The resulting cooling and heating rates, as well as the species abundances, depend on the shielded radiation field, the cosmic ray rate, and the dust content. The hybrid model presented in this work, \textsc{hybrid-chimes}, has been developed for simulations that do not include full radiative transfer, and for which neither the local radiation field nor the local shielding column density is known. 
Our model therefore relies on approximations that depend solely on local gas properties that can be calculated in a standard hydrodynamic scheme, such as the density, temperature, and pressure of each gas resolution element. 

For simulations of individual galaxies, a constant intrinsic radiation field from stars within the galaxy (the interstellar radiation field, ISRF), may be a valid assumption. However, in cosmological simulations that simultaneously follow the evolution of different types of galaxies, from dwarf galaxies with low star formation rates to starburst galaxies, a radiation field that captures the effects of the varying levels of star formation is needed.   

In this section we discuss the model input parameters (Section~\ref{sec:inputparameter}), the resulting species abundances in chemical equilibrium (Section~\ref{sec:equilibriumabundances}) and present the hybrid cooling model, which combines equilibrium and non-equilibrium cooling rates (Section~\ref{sec:hybrid}). Finally, we summarize the included heating and cooling processes (Section~\ref{sec:processes}).

\subsection{Model input parameters}\label{sec:inputparameter}

In neutral gas, the interplay between radiation and gas chemistry is complex. Neutral gas is typically not optically thin and dust and molecule formation further complicate the picture. For a detailed modelling of all processes within the ISM, a combination of radiative transfer (RT) and a non-equilibrium chemistry network is required to follow the emitted radiation through the gas, determine its absorption, and calculate the species abundances as well as cooling and heating rates for gas along the photon path. 

In this work, we discuss an approximate treatment to include critical processes in neutral gas for simulations that do not include full RT + chemistry. Following \citetalias{PS20},
we first define the reference column density, $N_{\mathrm{ref}}$ (Section~\ref{sec:Nref}), which is a function of pressure and serves as a typical coherence length scale of self-gravitating gas. The  strength of the interstellar radiation field (ISRF, Section~\ref{sec:ISRF}), the cosmic ray rate (CR, Section~\ref{sec:CR}), and the gas shielding column density, $N_{\mathrm{sh}}$ (Section~\ref{sec:Nsh}) are assumed to be a function of $N_{\mathrm{ref}}$. The exponents of the power law scaling relations as well as the normalizations are input parameters and summarized in Table~\ref{tab:fiducialparameter}.

\subsubsection{Reference column density}\label{sec:Nref}

The reference column density, $N_{\mathrm{ref}}$, is based on the coherence scale of self-gravitating gas: the local Jeans column density, $N_{\mathrm{J}}$ \citep{Schaye2001ApJ559, Schaye2001ApJ562}, which has been shown to reproduce the shielding lengths in cosmological, radiative transfer simulations for the transition from ionized to neutral gas \citep{Rahmati2013}. $N_{\mathrm{J}}$ depends on the gas pressure which was assumed to be purely thermal in \citetalias{PS20} and was defined as 

\begin{eqnarray}\label{eq:NJthermal}
    N_{\mathrm{J,thermal}} &=& \sqrt{\frac{\gamma X_{\mathrm{H}}^2}{G m_{\mathrm{H}}^2}P_{T}} = 
     \sqrt{\frac{\gamma k_{\mathrm{B}} X_{\mathrm{H}} T n_{\mathrm{H}}}{\mu G m_{\mathrm{H}}^2 }} \nonumber\\ 
    &\approx& 8.6 \times 10^{20}\,\mathrm{cm}^{-2} \left ( \frac{n_{\mathrm{H}}}{1\,\mathrm{cm}^{-3}} \right)^{1/2} \left ( \frac{T}{1000\,\mathrm{K}} \right)^{1/2} \times \nonumber \\ &&\left ( \frac{\mu_{\mathrm{const}}}{1.24} \right)^{-1/2}  \left ( \frac{X_{\mathrm{H,const}}}{0.75} \right)^{1/2}  \left ( \frac{\gamma_{\mathrm{const}}}{5/3} \right)^{1/2} ,
\end{eqnarray}

\noindent
with the thermal pressure, $P_{\mathrm{T}} = n_{\mathrm{H}}\,k_{\mathrm{B}}\,T/(X_{\mathrm{H}}\,\mu)$, the hydrogen number density, $n_{\mathrm{H}}$, the ratio of specific heats, $\gamma$, the Boltzmann constant, $k_{\mathrm{B}}$, the gas temperature, $T$, the hydrogen mass fraction, $X_{\mathrm{H}}$, the mean particle mass, $\mu$, the gravitational constant, $G$, and the hydrogen particle mass, $m_{\mathrm{H}}$. For simplicity, we let the reference column density depend only on gas density and temperature and use constant values for $\mu = \mu_{\mathrm{const}} = 1.24$, $X_{\mathrm{H}} = X_{\mathrm{H,const}} = 0.75$, and $\gamma = \gamma_{\mathrm{const}} = 5/3$.

We include the option to add a turbulent, non-thermal pressure component by replacing $P_{\mathrm{T}}$ with  $P_{\mathrm{NT}} = n_{\mathrm{H}} \, m_{\mathrm{H}} \, v_{\mathrm{turb}}^2/ X_{\mathrm{H}}$ and introduce the turbulent Jeans column density, defined as  

\begin{eqnarray}\label{eq:NJturb}
    N_{\mathrm{J,turb}} &=& \sqrt{\frac{\gamma X_{\mathrm{H}}^2}{G m_{\mathrm{H}}^2}P_{\mathrm{NT}}} = v_{\mathrm{turb}} \sqrt{\frac{\gamma X_{\mathrm{H}} n_{\mathrm{H}}}{m_{\mathrm{H}}G}} \nonumber \\
    &\approx& 2.0 \times 10^{21}\,\mathrm{cm}^{-2} \left ( \frac{v_{\mathrm{turb}}}{6\,\mathrm{km\,s}^{-1}} \right )\left ( \frac{n_{\mathrm{H}}}{1\,\mathrm{cm}^{-3}} \right )^{1/2} \times \nonumber \\
    &&\left ( \frac{X_{\mathrm{H,const}}}{0.75} \right)^{1/2} \left ( \frac{\gamma_{\mathrm{const}}}{5/3} \right)^{1/2} ,
\end{eqnarray}

\noindent
with the 1D turbulent velocity dispersion, $v_{\mathrm{turb}}$. The unresolved turbulence, described by $v_{\mathrm{turb}}$, acts dynamically by increasing the column density of a coherent self-gravitating structure\footnote{In addition, turbulence causes the Lyman-Werner lines to shift, which suppresses H$_2$ self-shielding. For this process, the turbulence is parametrized with the Doppler broadening parameter $b_{\mathrm{turb}} = \sqrt{2} v_{\mathrm{turb}}$ (see Section~\ref{sec:heatingH2}).}.

The fiducial value of $v_{\mathrm{turb}} = 6\,\mathrm{km\,s}^{-1}$ is typical for the warm neutral medium (WNM) in the Galaxy, as measured from observed \ion{H}{I} linewidths \citep{KalberlaHaud2018}\footnote{\citet{KalberlaHaud2018} report a mean line width for the WNM component in their data of $\mathrm{FWHM}_{\mathrm{WNM}} = 23.3\,\mathrm{km\,s}^{-1}$ which corresponds to a total 1D velocity dispersion of $v_{\mathrm{total}} = \mathrm{FWHM}\,/\, 2.35 \approx 10 \,\mathrm{km\,s}^{-1}$ (relation between $\mathrm{FWHM}$ and $v_{\mathrm{total}}$ discussed in e.g. \citealp{HaudKalberla2007}). The turbulent 1D velocity dispersion of the WNM is $v_{\mathrm{turb,WNM}} = \sqrt{v_{\mathrm{total}}^2 - 2 k_{\mathrm{B}} T_{\mathrm{WNM}} / (\mu \,m_{\mathrm{H}})} $, which leads to $v_{\mathrm{turb,WNM}}\approx 6\,\mathrm{km\,s}^{-1}$ for $T_{\mathrm{WNM}}\approx 5000\,\mathrm{K}$.}. For the cold neutral medium (CNM), the fiducial value of $v_{\mathrm{turb}} = 6\,\mathrm{km\,s}^{-1}$ represents unresolved turbulence on a length-scale of $\lesssim 100 \,\mathrm{pc}$ following the velocity dispersion - size relation of \ion{H}{I} clouds by \citet{Larson1979}, with the normalization ($1.2\,\mathrm{km\,s}^{-1}$ at a length-scale of $1~\mathrm{pc}$) from \citet{Wolfire2003}. Simulations with spatial resolutions much better than $\approx 100\,\mathrm{pc}$ may use a smaller values for $v_{\mathrm{turb}}$ in cold gas.

Together, we define the Jeans column density as 

\begin{equation}\label{eq:NJ}
    N_{\mathrm{J}} = \mathrm{max}(N_{\mathrm{J,thermal}}, N_{\mathrm{J,turb}})\;. 
\end{equation}

\noindent
For the reference column density, we first define $N_{\mathrm{ref}}'$ which largely equals the Jeans column density, but is limited\footnote{In \citetalias{PS20}, $N_{\mathrm{ref}}$ is furthermore limited by a maximum length scale at low densities, but here this is replaced by a redshift-dependent limit on the maximum shielding length $l_{\mathrm{max}}$ (Section~\ref{sec:Nsh}) and low-density cutoffs for the ISRF (Section~\ref{sec:ISRF}) and CR rates (Section~\ref{sec:CR}).} to a maximum column density, $N_{\mathrm{max}}$,

\begin{equation}\label{eq:Nrefprime}
    N_{\mathrm{ref}}'= \mathrm{min} \left ( N_{\mathrm{J}}, \,N_{\mathrm{max}} \right )\;.
\end{equation}

\noindent
The final prescription for $N_{\mathrm{ref}}$ includes an asymptotic transition to a minimum column density, $N_{\mathrm{min}}$, at high temperatures,

\begin{equation}\label{eq:Nref}
     \log N_{\mathrm{ref}} = \log N_{\mathrm{ref}}' - \frac{ \log N_{\mathrm{ref}}' -  \log N_{\mathrm{min}}}{1 + \left ( \sqrt{T_{\mathrm{min}} T_{\mathrm{max}}} /T \right )^2}\;,
\end{equation}

\noindent
where the transition is characterized by a minimum and maximum temperature, $T_{\mathrm{min}}$ and $T_{\mathrm{max}}$. In the fiducial model, $N_{\mathrm{max}} = 10^{24}\,\mathrm{cm}^{-2}$, $T_{\mathrm{min}}= 10^4\,\mathrm{K}$, $T_{\mathrm{max}}= 10^5\,\mathrm{K}$, and $N_{\mathrm{min}} = 3.1\times10^{15}\,\mathrm{cm}^{-2}$. This transition avoids the extrapolation of scalings related to the ISM (i.e. the radiation field, cosmic ray rates, and shielding column densities) to gas in the circum- or intergalactic medium. Compared to \citet{PS20}, the transition starts at higher temperatures ($T_{\mathrm{min}}= 10^3\,\mathrm{K}$ in \citetalias{PS20}) to ensure that the intended scalings include the warm neutral medium.

\begin{table*}
    \caption{Overview of the fiducial parameter values in this work (column 3) and in \citetalias{PS20} for their fiducidal model (UVB\_dust1\_CR1\_G1\_shield1). Parameters in boldface are varied in Section~\ref{sec:results}. }
    \centering
    \begin{tabular}{llccl}
        \hline
        Description & Parameter & \modelone & PS20   &   \\
        \hline
        \multicolumn{5}{c}{Reference Column Density} \\
        \hline
        {\bf 1D turbulent velocity dispersion} & $v_{\mathrm{turb}}$ & $6\,\mathrm{km\,s}^{-1}$& $0\,\mathrm{km\,s}^{-1}$ & \\

        Hydrogen mass fraction & $X_{\mathrm{H,const}}$ & 0.75 &0.7563& \\
        Ratio of specific heats & $\gamma_{\mathrm{const}}$ &5/3 &5/3& \\
        Mean particle mass & $\mu_{\mathrm{const}}$ & 1.24&1.2328& \\
        Minimum column density & $N_{\mathrm{min}}$  & $3.1\times10^{15}\,\mathrm{cm}^{-2}$& $3.1\times10^{15}\,\mathrm{cm}^{-2}$& \\
        Maximum column density & $N_{\mathrm{max}}$  & $10^{24}\,\mathrm{cm}^{-2}$& $10^{24}\,\mathrm{cm}^{-2}$ & \\
        Minimum temperature for transition from shielded to optically thin gas&$T_{\mathrm{min}}$  & $10^4\,\mathrm{K}$ & $10^3\,\mathrm{K}$&\\
        Maximum temperature for transition from shielded to optically thin gas&$T_{\mathrm{max}}$  & $10^5\,\mathrm{K}$ & $10^5\,\mathrm{K}$&\\    
        \hline
        \multicolumn{5}{c}{Interstellar radiation field} \\
        \hline
        {\bf Normalization of the ISRF}  & $R_{\mathrm{ISRF}}$ & 1 & 0.54& \\
        Power law exponent for low column densities & $\alpha_{\mathrm{ISRF}}$ &1.4& 1.4& \\
        Power law exponent for high column densities & $\beta_{\mathrm{ISRF}}$ &0 & 1.4& \\
        Transition column density for double power law& $N_{\mathrm{t,ISRF}}$ & $10^{22}\,\mathrm{cm}^{-2}$ & $10^{22}\,\mathrm{cm}^{-2}$ & \\
        Overdensity of low-density cutoff &$\Delta_{\mathrm{cut,ISRF}}$ & 100 & 0 & \\
        Physical density of low-density cutoff & $n_{\mathrm{cut,ISRF,phys}}$ & $10^{-2}\,\mathrm{cm}^{-3}$ & $10^{-2}\,\mathrm{cm}^{-3}$ ($z>7.5$) & \\
        \hline
        \multicolumn{5}{c}{Cosmic ray rate} \\
        \hline  
        {\bf Normalization of the CR rate} & $R_{\mathrm{CR}}$ & 1 & 0.41& \\
        Power law exponent for low column densities & $\alpha_{\mathrm{CR}}$ &1.4& 1.4& \\
        Power law exponent for high column densities & $\beta_{\mathrm{CR}}$ &0.0 &1.4& \\
        Transition column density for double power law& $N_{\mathrm{t,CR}}$ & $10^{21}\,\mathrm{cm}^{-2}$ & $10^{21}\,\mathrm{cm}^{-2}$ & \\   
        Overdensity of low-density cutoff &$\Delta_{\mathrm{cut,CR}}$ & 100 &0& \\
        Physical density of low-density cutoff & $n_{\mathrm{cut,CR,phys}}$ & $10^{-2}\,\mathrm{cm}^{-3}$ & 0 & \\
        \hline
        \multicolumn{5}{c}{Shielding column density} \\
        \hline        
        {\bf Normalization of the shielding column density} & $R_{\mathrm{sh}}$   & 0.5 & 0.5&\\
        {\bf Maximum shielding length (low \boldmath$z$) }&$l_{\mathrm{lowz}}$ & $50 \,\mathrm{kpc}$ & $50 \,\mathrm{kpc}$ &\\
        {\bf Maximum shielding length (high \boldmath$z$) }&$l_{\mathrm{highz}}$ & $10\,\mathrm{kpc}$ & $50 \,\mathrm{kpc}$ &\\ 
        Redshift of transition from $l_{\mathrm{highz}}$ and $l_{\mathrm{lowz}}$ & $z_{\mathrm{t,lsh}}$ & 7 & - &  \\
        \hline
        \multicolumn{5}{c}{Dust} \\
        \hline        
        {\bf Maximum dust boost for H$_2$ formation} & $b_{\mathrm{dust,max}}$     & 1 & 1 &\\
        Depletion strength& $F_{\star}$ & 1 & 1&\\ 
        Dust-to-gas ratio for $N_{\mathrm{ref}}\ge N_{\mathrm{t,dust}}$ & $\mathcal{DTG}$   & $5.3\times10^{-3}$ & $5.6\times10^{-3}$&\\
        Effective grain size for $N_{\mathrm{ref}}\ge N_{\mathrm{t,dust}}$ & $d_{\mathrm{eff}}$ & $4\times10^{-22}\,\mathrm{mag\,cm^2}$ & $4.72\times10^{-22}\,\mathrm{mag\,cm}^2$ &\\
        Power law exponent for low column densities & $\alpha_{\mathrm{dust}}$ &1.4&1.4& \\
        Transition column density for double power law& $N_{\mathrm{t,dust}}$ & $10^{20}\,\mathrm{cm}^{-2}$ & $3.65\times10^{20}\,\mathrm{cm}^{-2}$& \\ 
        Minimum gas density for dust boost &$n_{\mathrm{b,min}}$     & 1 & - &\\
        Maximum gas density for dust boost &$n_{\mathrm{b,max}}$     & 10 &- &\\    
        \hline
    \end{tabular}
    \label{tab:fiducialparameter}
\end{table*}

\subsubsection{Radiation fields}\label{sec:ISRF}

\paragraph*{Metagalactic radiation field.}
The Universe is permeated with radiation from distant galaxies and quasars, which together create a UV/X-ray background (UVB). We use the \citet{FG20} background, modified as described in \citetalias{PS20}, which depends on redshift, $z$, as the fiducial UVB. 

In Appendix~\ref{sec:app:highz} we show the dependence of the thermal equilibrium properties of primordial gas at $z=9$ on the assumed UVB. Assuming a radiation field with drastically increased H$_2$ photo-dissociating Lyman-Werner radiation at $z>7$ (see Fig.~\ref{fig:LymanWerner}), following \citet{Incatasciato2023}, results in lower H$_2$ cooling rates and therefore higher equilibrium temperatures (Fig.~\ref{fig:Teqhighz}). This may impact the properties of `REionization-Limited \ion{H}{I} Clouds' (RELHICs, \citealp{BenitezLlambay2017}). The tabulated rates from both UVB variations are made public (see data availability statement).

\paragraph*{Interstellar radiation field.}
In the interstellar medium, the radiation field can be described by a diffuse interstellar radiation field (ISRF), produced by stars within each galaxy, in addition to the UVB from distant objects. As reference, we use the ISRF at the position of the Sun, presented in \citet{Black1987} as a combination of the solar vicinity radiation field \citep{Mathis1983} and a Galactic soft X-ray background \citep{Bregman1986}. The UV radiation from stars in this work is based on a full sky survey survey by the ESRO TD-1 satellite \citep{Gondhalekar1980} and integrated to calculate the total intensity of direct starlight \citep{Mezger1982}. Additional components of the solar vicinity radiation field include diluted black-body radiation fields with $T = 7500\,\mathrm{K}$, 4000~K, and 3000~K, to model the contribution of F-stars, K-stars, and M giants, respectively, as well as radiation re-emitted by dust (see \citealp{Mathis1983} for details).

Following \citetalias{PS20}, we assume that the spectral shape of the incident ISRF is constant but that its normalization scales with the typical star formation surface density rate for the local gas surface density, and therefore $\propto N_{\mathrm{ref}}^{\alpha_{\mathrm{ISRF}}}$ for which the exponent $\alpha_{\mathrm{ISRF}} = 1.4$ follows the Kennicutt-Schmidt relation \citep{Kennicutt1998SchmidtLaw}. Expanding the prescription in \citetalias{PS20}, we include the option to switch to a second exponent $\beta_{\mathrm{ISRF}}$ for high gas surface densities ($N_{\mathrm{ref}} > N_{\mathrm{t,ISRF}}$). This allows for a scaling that deviates from the assumption of self-gravitating structures through the Jeans column density for pressure-confined (molecular) clouds. In the fiducial model, we use a power law slope of $\beta_{\mathrm{ISRF}}=0$, which represents a constant (i.e. saturated) ISRF. 

The strength of the ISRF, expressed here as $I_{\mathrm{ISRF}}'$, defined as the number density, $n_{\mathrm{\gamma}}'$, of hydrogen ionizing photons ($>13.6\,\mathrm{eV}$), relative to the \citet{Black1987} ISRF with $n_{\mathrm{\gamma,B87}} = 3.7\times 10^{-4}\,\mathrm{photons\,cm}^{-3}$, is

\begin{equation}\label{eq:nISRF}
  I_{\mathrm{ISRF}}' \equiv \frac{n_{\mathrm{\gamma}}'}{n_{\mathrm{\gamma,B87}}} = 19 \,R_{\mathrm{ISRF}}
    \begin{cases}
      \left ( \frac{N_{\mathrm{ref}}}{N_{\mathrm{t,ISRF}}} \right )^{\alpha_{\mathrm{ISRF}}} & \text{if}\; N_{\mathrm{ref}} < N_{\mathrm{t,ISRF}}\\
      \left ( \frac{N_{\mathrm{ref}}}{N_{\mathrm{t,ISRF}}} \right )^{\beta_{\mathrm{ISRF}}} & \text{if}\; N_{\mathrm{ref}} \ge N_{\mathrm{t,ISRF}} \; ,
    \end{cases}
\end{equation}

\noindent
with the normalization, $19\,R_{\mathrm{ISRF}}$, at the transition column density, $N_{\mathrm{t,ISRF}}$, between the two power laws with slopes $\alpha_{\mathrm{ISRF}}$ (for $N_{\mathrm{ref}} < N_{\mathrm{t,ISRF}}$) and $\beta_{\mathrm{ISRF}}$ (for $N_{\mathrm{ref}} \ge N_{\mathrm{t,ISRF}}$). 

In the fiducial model ($R_{\mathrm{ISRF}} = 1$), the ISRF strength increases super-linearly with gas column density ($\alpha_{\mathrm{ISRF}} = 1.4$) for $N_{\mathrm{ref}}< N_{\mathrm{t,ISRF}} = 10^{22}\,\mathrm{cm}^{-2}$ and matches the original \citet{Black1987} ISRF for solar neighborhood conditions, i.e. $I_{\mathrm{ISRF}}' = 1$ for the reference column density equal to the average total hydrogen column density within a cylindrical radius of 1~kpc at the position of the Sun ($N_{\mathrm{ref}} = N_{\odot} = 1.22\times10^{21}\,\mathrm{cm}^{-2}$, \citealp{McKee2015}). At higher column densities, ($N_{\mathrm{ref}}\ge  10^{22}\,\mathrm{cm}^{-2}$), the ISRF normalization saturates ($\beta_{\mathrm{ISRF}} = 0$).

The physical densities of the local ISM can correspond to small over-densities at high redshift. For example, a physical density of $n_{\mathrm{H}} = 10^{-1}\,\mathrm{cm}^{-3}$, typical for the warm neutral medium, corresponds to an over-density of $\Delta < 100$ for $z\gtrsim 9$. In order to avoid including the ISRF in non-ISM gas, we add density cutoffs, both in physical density ($n_{\mathrm{cut,ISRF,phys}}$) and in the redshift-dependent over-density, $n_{\mathrm{cut,ISRF,over}}$, defined as

\begin{equation}
    n_{\mathrm{cut,ISRF,over}} = \Delta_{\mathrm{cut,ISRF}} \,\rho_{\mathrm{crit}}\,\Omega_{\mathrm{b}} \frac{X_{\mathrm{H,const}}}{ m_{\mathrm{H}}} \;,
\end{equation}

\noindent
with the over-density parameter for the cutoff, $\Delta_{\mathrm{cut,ISRF}}$, the critical density $\rho_{\mathrm{crit}}(z)$, and the present-day cosmic baryon fraction $\Omega_{\mathrm{b}}$. In the fiducial model $n_{\mathrm{cut,ISRF,phys}} = 10^{-2}\,\mathrm{cm}^{-3}$ and $\Delta_{\mathrm{cut,ISRF}} = 100$. The radiation field is suppressed for $n_{\mathrm{H}}<n_{\mathrm{cut,ISRF}}$ with 

\begin{align}
 n_{\mathrm{cut,ISRF}}&=  \mathrm{max} ( n_{\mathrm{cut,ISRF,phys}}, n_{\mathrm{cut,ISRF,over}})  \nonumber\\
 &=  \mathrm{max} ( n_{\mathrm{cut,ISRF,phys}}, \Delta_{\mathrm{cut,ISRF}} \,\rho_{\mathrm{crit}}\,\Omega_{\mathrm{b}} \frac{X_{\mathrm{H,const}}}{ m_{\mathrm{H}}}) \;,
\end{align} 

\noindent
and the final radiation field strength, including the density cutoff, is defined as 

\begin{equation}\label{eq:ISRFIcut}
    \log I_{\mathrm{ISRF}} \equiv \log \frac{n_{\mathrm{\gamma}}}{n_{\gamma,\mathrm{B87}}} = \log I_{\mathrm{ISRF}}' - \frac{\log I_{\mathrm{ISRF}}' - \log I_{\mathrm{ISRF,min}}}{1 +10 \left (\frac{n_{\mathrm{H}}}{n_{\mathrm{cut,ISRF}}}\right )^{2}}\; ,
\end{equation}

\noindent
with $I_{\mathrm{ISRF,min}} = n_{\gamma,\mathrm{min}}/n_{\gamma,\mathrm{B87}}$.
The ISRF strength relative to the \citet{Black1987} radiation field, $I_{\mathrm{ISRF}}$, approaches $I_{\mathrm{ISRF,min}}$ for $n_{\mathrm{H}}< n_{\mathrm{cut,ISRF}}$ and $I_{\mathrm{ISRF}}'$ for $n_{\mathrm{H}}> n_{\mathrm{cut,ISRF}}$. The value for $n_{\gamma,\mathrm{min}}$ is $n_{\gamma}'$ at the minimum column density, $N_{\mathrm{min}}$. For the fiducidal parameter values, $n_{\gamma,\mathrm{min}} = 19\, R_{\mathrm{ISRF}} \,(N_{\mathrm{min}}/N_{\mathrm{t,ISRF}})^{\alpha_{\mathrm{ISRF}}} = 1.5\times10^{-8}\,\mathrm{cm}^{-3}$ and therefore $I_{\mathrm{ISRF,min}} = 4\times10^{-5}$.

\begin{figure}
    \centering
    \includegraphics[width=\linewidth]{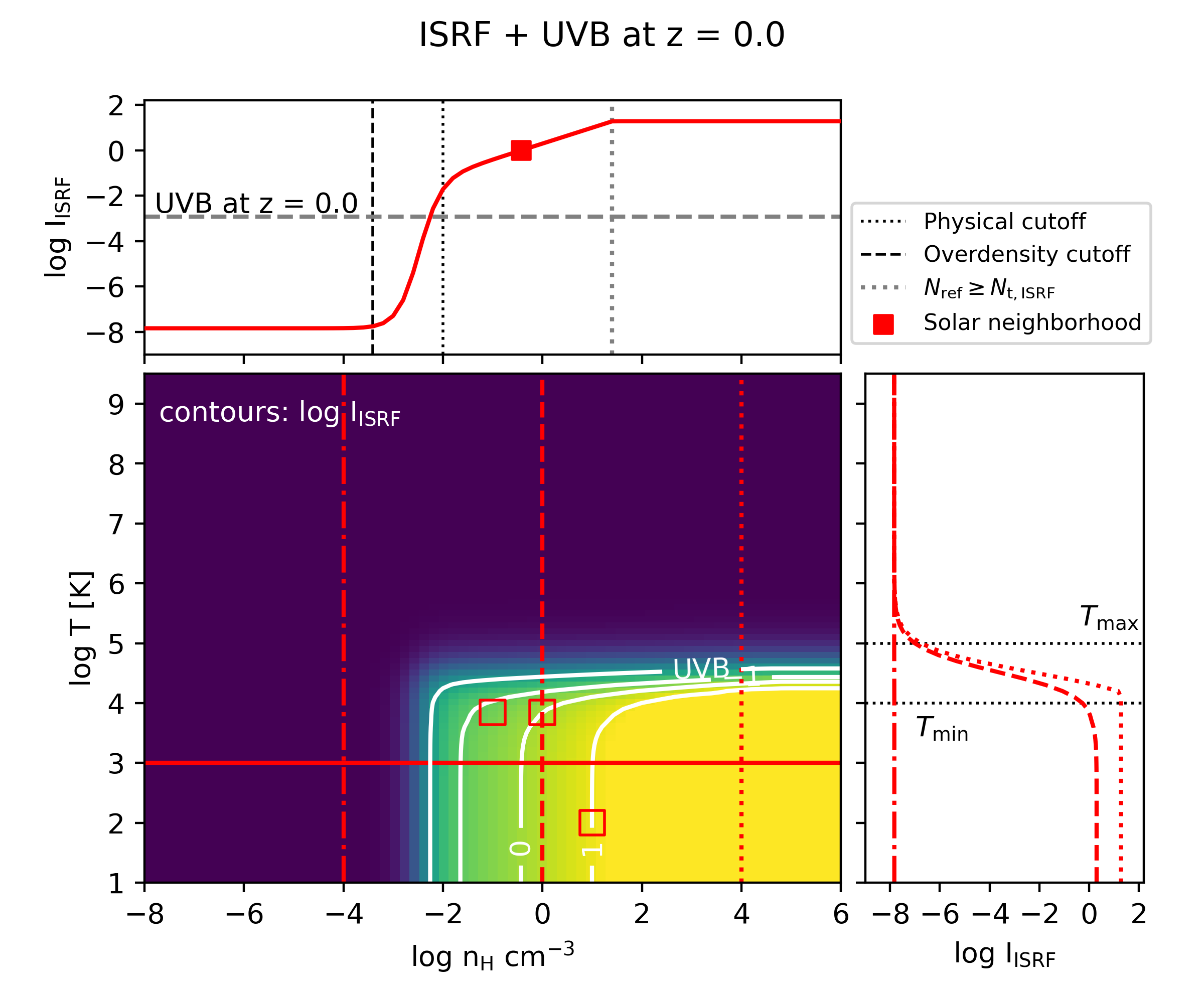}
    \caption{The strength of the interstellar radiation field and the UVB, relative to the \citet{Black1987} ISRF, $I_{\mathrm{ISRF}}$ (equation~\ref{eq:ISRFIcut}) at $z=0$ for the fiducial parameter values in Table~\ref{tab:fiducialparameter}. The colormap and white contours in the large panel show $\log I_{\mathrm{ISRF}}$. An additional white contour, labeled `UVB', indicates the strength of the UV/X-ray background at $z=0$. The small panels show $I_{\mathrm{ISRF}}$ for constant temperature (top panel) and constant densities (right panel) at a selected temperature ($1000\,\mathrm{K}$) and three selected densities ($10^{-4}$, $1$, and $10^{4}\,\mathrm{cm}^{-3}$). The line styles of the horizontal (vertical) red lines in the large panel indicate the values for $T$ ($n_{\mathrm{H}}$) used in the red lines in the small panels. The open squares indicate densities and temperatures typical for the thermally stable neutral phases of the ISM in the Galaxy (see Section~\ref{sec:neutral} and Table~\ref{tab:WNM}). The top panel includes vertical lines that show the physical cutoff density, $n_{\mathrm{cut,ISRF,phys}}$ (black dotted line) the over-density cutoff density at $z=0$, $n_{\mathrm{cut,ISRF,over}}$ (black dashed line) and the density at which $N_{\mathrm{ref}} = N_{\mathrm{t,ISRF}}$ (grey dotted line). The filled red square indicates solar neighborhood values ($N_{\mathrm{ref}} = N_{\odot} = 1.22\times10^{21}\,\mathrm{cm}^{-2}$, \citealp{McKee2015} and $I_{\mathrm{ISRF}} = 1$). The right panel shows the minimum ($T_{\mathrm{min}}$) and maximum ($T_{\mathrm{max}}$) temperatures for the transition to hot gas as horizontal dotted lines. }
    \label{fig:ISRF}
\end{figure}

An overview of $I_{\mathrm{ISRF}}$ for the fiducial parameter values at $z=0$ for the tabulated density and temperature range can be found in Fig.~\ref{fig:ISRF}.

\subsubsection{Cosmic rays}\label{sec:CR}

The cosmic ray (CR) rate is normalized with the CR ionization rate of atomic hydrogen, $\zeta_{\mathrm{H}}$. The ionization rate of all other species scales with $\zeta_{\mathrm{H}}$ following \citet{chimes2014optthin} (see also Section~\ref{sec:heatingelement}).
We describe first $\zeta_{\mathrm{H}}'$ with a similar double power law as the ISRF in equation~(\ref{eq:nISRF}):

\begin{equation}\label{eq:zetaCR}
    \zeta_{\mathrm{H}}' =\zeta_0  \,R_{\mathrm{CR}}
    \begin{cases}
      \left ( \frac{N_{\mathrm{ref}}}{N_{\mathrm{t,CR}}} \right )^{\alpha_{\mathrm{CR}}} & \text{if}\; N_{\mathrm{ref}} < N_{\mathrm{t,CR}}\\
      \left ( \frac{N_{\mathrm{ref}}}{N_{\mathrm{t,CR}}} \right )^{\beta_{\mathrm{CR}}} & \text{if}\; N_{\mathrm{ref}} \ge N_{\mathrm{t,CR}} \; ,
    \end{cases}
\end{equation}

\noindent
with the CR rate $\zeta_0 = 2\times10^{-16}\,\mathrm{s}^{-1}\,R_{\mathrm{CR}}$ at the transition column density, $N_{\mathrm{ref}} = N_{\mathrm{t,CR}}$, between two power laws with slopes $\alpha_{\mathrm{CR}}$ and $\beta_{\mathrm{CR}}$.

The fiducial model uses a constant ($\beta_{\mathrm{CR}}=0$) CR rate for $N_{\mathrm{ref}}>N_{\mathrm{t,CR}} = 10^{21}\,\mathrm{cm}^{-2}$
and a normalization of $R_{\mathrm{CR}} = 1$, which matches the CR rates for column densities of $N_{\mathrm{H}} > 10^{21}\,\mathrm{cm}^{-2}$ within the Galaxy from \citet{Indriolo2015}\footnote{Recent work by \citet{Obolentseva2024} report a lower CR ionization rate of $\zeta_{\mathrm{H2}} = 6\times10^{-17}\,\mathrm{s}^{-1}$ for H$_2$ and therefore $\zeta_{\mathrm{H}} = 3.6\times10^{-17}\,\mathrm{s}^{-1}$ for atomic hydrogen, following $\zeta_{\mathrm{H2}}=1.65 \zeta_{\mathrm{H}}$ from \citep{Glassgold1973}. We show results for different normalizations of the CR rate in Section~\ref{sec:neutral:var}.}. For $N_{\mathrm{ref}}<N_{\mathrm{t,CR}}$ we use the Kennicutt-Schmidt inspired power-law exponent of $\alpha_{\mathrm{CR}} = 1.4$, as for the ISRF.

A low-density threshold in both physical density, $n_{\mathrm{cut,CR,phys}}$ and over-density, $n_{\mathrm{cut,CR,over}}$ is combined into $n_{\mathrm{cut,CR}}$, defined as 

\begin{eqnarray}
    n_{\mathrm{cut,CR}} &=& \mathrm{max} ( n_{\mathrm{cut,CR,phys}}, n_{\mathrm{cut,CR,over}} ) \nonumber \\
    &=& \mathrm{max} ( n_{\mathrm{cut,CR,phys}}, \Delta_{\mathrm{cut,CR}} \,\rho_{\mathrm{crit}}  \Omega_{\mathrm{b}}\frac{X_{\mathrm{H}}}{ m_{\mathrm{H}}} ) \;, 
\end{eqnarray}

\noindent
with the fiducial parameters values, $n_{\mathrm{cut,CR,phys}} = 10^{-2}\,\mathrm{cm}^{-3}$, and $\Delta_{\mathrm{cut,CR}} = 100$, the over-density of the cutoff. The final CR rate, $\zeta_{\mathrm{H}}$, is defined as

\begin{equation}\label{eq:CRcut}
    \log \zeta_{\mathrm{H}} = \log \zeta_{\mathrm{H}}' - \frac{ \log \zeta_{\mathrm{H}}' - \log \zeta_{\mathrm{min}} }{1 + 10 \left ( \frac{n_{\mathrm{H}}}{n_{\mathrm{cut,CR}}}\right )^{2}} \; ,
\end{equation}

\noindent 
which asymptotically approaches a minimum value, $\zeta_{\mathrm{min}}$ for densities $n_{\mathrm{H}} < n_{\mathrm{cut,CR}}$. For higher densities ($n_{\mathrm{H}}>n_{\mathrm{cut,CR}}$), $\zeta_{\mathrm{H}}$ approaches $\zeta_{\mathrm{H}}'$ from equation~(\ref{eq:zetaCR}). As for the ISRF, the minimum value, $\zeta_{\mathrm{min}}$, matches that from other non-ISM regions in density-temperature space (i.e. gas with temperatures of $T>10^5\,\mathrm{K}$) by using the value of $\zeta_{\mathrm{min}}=\zeta_{\mathrm{H}}' \,(N_{\mathrm{ref}} = N_{\mathrm{min}})$. For the fiducial parameters $\zeta_{\mathrm{min}} = 2\times10^{-16} R_{\mathrm{CR}} \,(N_{\mathrm{min}}/N_{\mathrm{t,CR}})^{1.4} = 3.9\times10^{-24}\,\mathrm{s}^{-1}$.

\begin{figure}
    \centering
    \includegraphics[width=\linewidth]{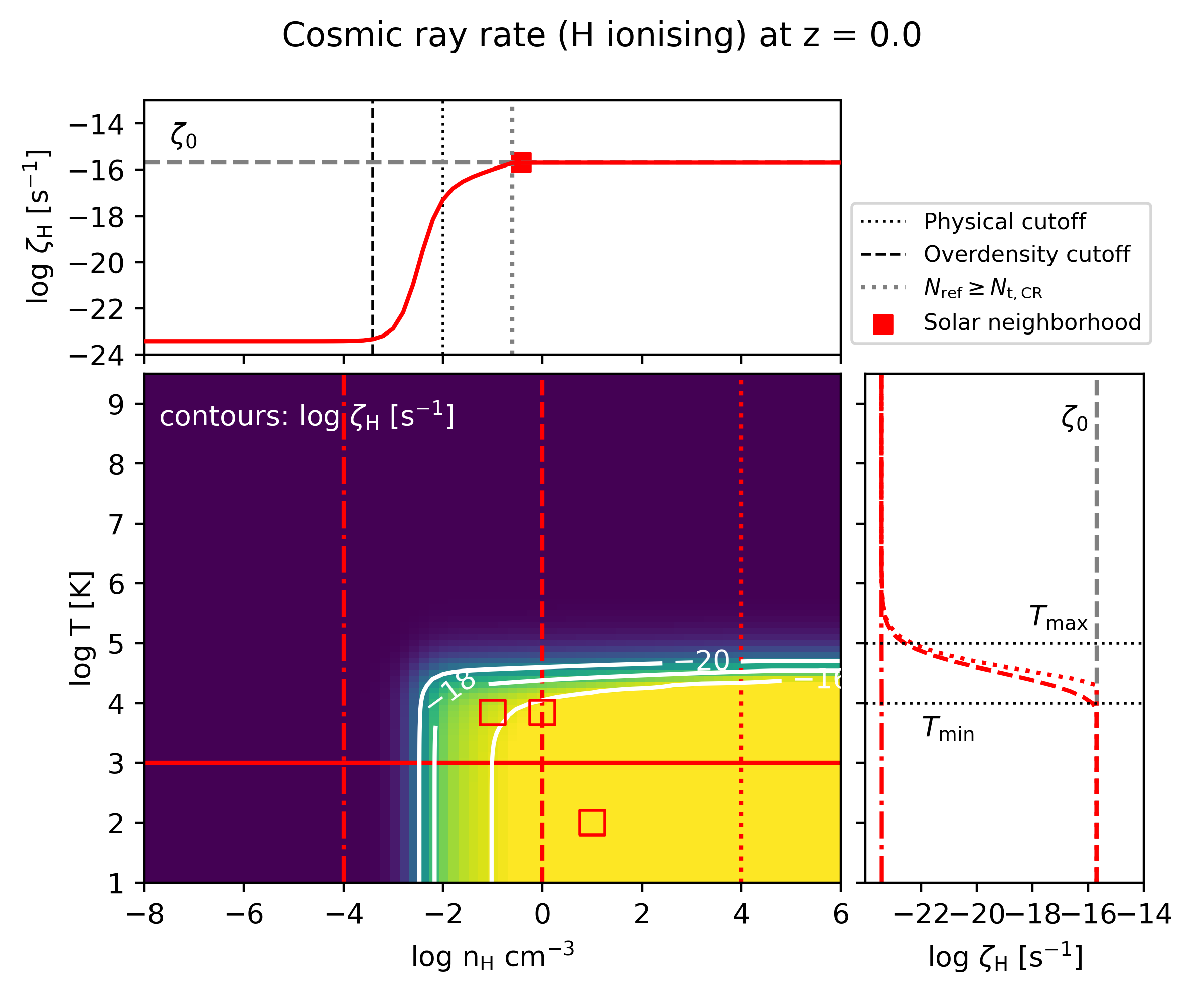}
    \caption{As Fig.~\ref{fig:ISRF} but for the hydrogen ionizing cosmic ray rate, $\zeta_{\mathrm{CR}}$ (equation~\ref{eq:CRcut}) at $z=0$ (large panel). The canonical value $\zeta_0 = 2\times10^{-16}\,\mathrm{s}^{-1}$ (e.g.~\citealp{Indriolo2015}) is added for reference as horizontal (top panel) or vertical (right panel) dashsed line. }
    \label{fig:CR}
\end{figure}

An overview of $\zeta_{\mathrm{CR}}$ for the fiducial parameter values at $z=0$ for the tabulated density and temperature range can be found in Fig.~\ref{fig:CR}.

\subsubsection{Shielding column density}\label{sec:Nsh}

In neutral gas, the diffuse radiation field can be efficiently absorbed before it reaches the individual gas resolution elements in the simulation. Following \citetalias{PS20}, we assume that the shielding column density, $N_{\mathrm{sh}}$, is proportional to the reference column density, $N_{\mathrm{ref}}$ (Section~\ref{sec:Nref}). If $N_{\mathrm{ref}}$ is the typical total column density of a symmetric self-gravitating structure, we first define $N_{\mathrm{sh}}'$ as

\begin{equation}\label{eq:Nsh}
    N_{\mathrm{sh}}' = R_{\mathrm{sh}} N_{\mathrm{ref}} \equiv l_{\mathrm{sh}} n_{\mathrm{H}}\;,
\end{equation}

\noindent
with the shielding length, $l_{\mathrm{sh}} = N_{\mathrm{sh}}'\,/\,n_{\mathrm{H}}$. Here, a parameter value of $R_{\mathrm{sh}} = 0.5$ may be interpreted as the shielding into the center of the gas cloud, while $R_{\mathrm{sh}} = 1$ is the shielding by the full gas cloud. We use $R_{\mathrm{sh}} = 0.5$ as the fiducial parameter value. 

\begin{figure}
    \centering
    \includegraphics[width=\linewidth]{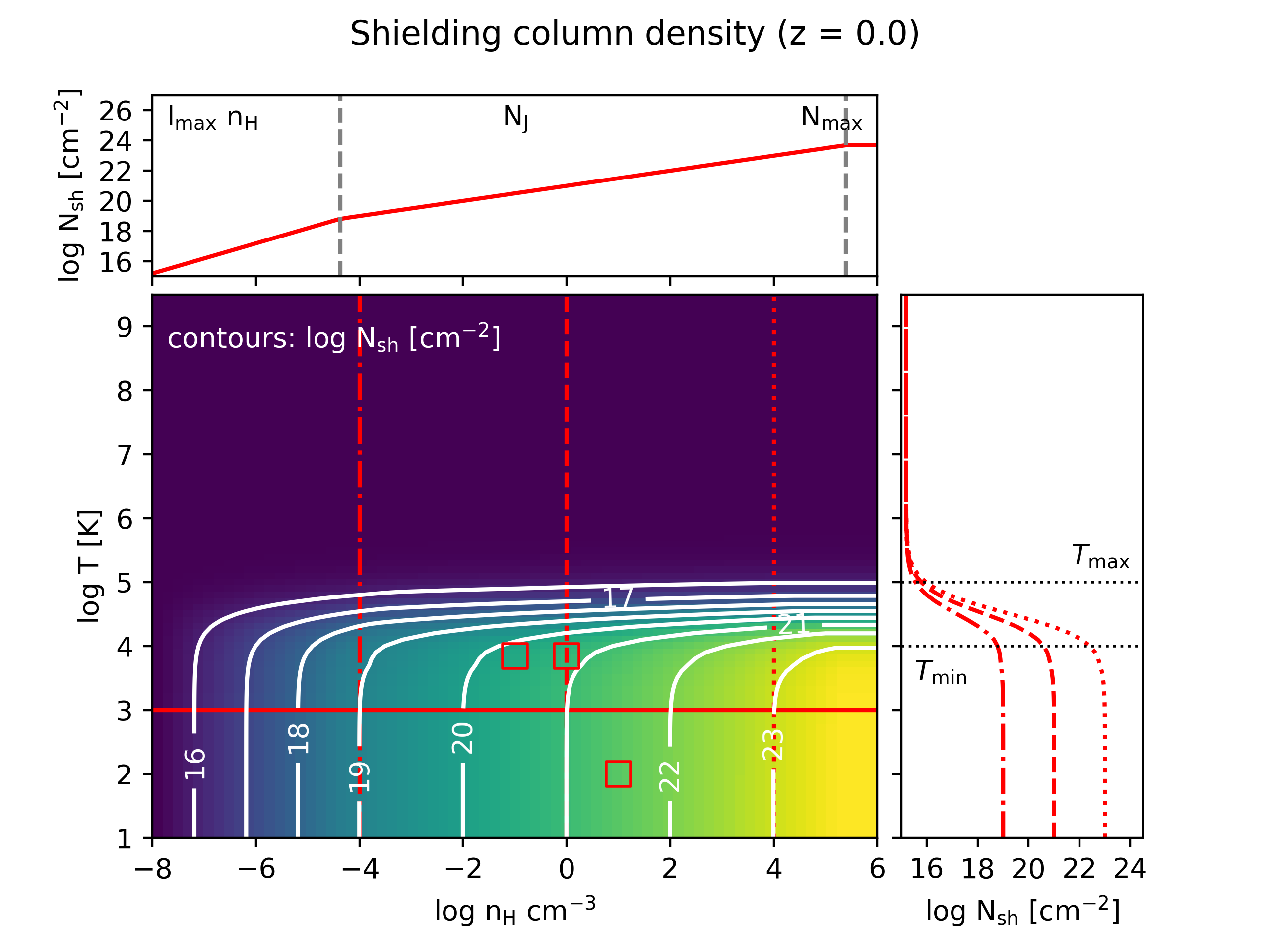}
    \caption{As Fig.~\ref{fig:ISRF} but for the shielding column density, $N_{\mathrm{sh}}$ (equation~\ref{eq:Nshcut}), for the fiducial parameter values at $z=0$. The constant temperature (top) panel illustrates the density ranges at which $N_{\mathrm{sh}}$ is limited by the redshift-dependent maximum coherence length ($l_{\mathrm{max}}n_{\mathrm{H}}$) and the maximum reference column density, $R_{\mathrm{sh}}\,N_{\mathrm{max}}$. For intermediate densities, between the vertical dashed lines, $N_{\mathrm{sh}} = R_{\mathrm{sh}} N_{\mathrm{J}}$ with the Jeans column density, $N_{\mathrm{J}}$ (equation~\ref{eq:NJ}).  }
    \label{fig:Nsh}
\end{figure}

We found that large shielding lengths, $l_{\mathrm{sh}}$, overestimate the shielding at high redshifts and delay the time of re-ionization in simulations with non-equilibrium chemistry. We therefore introduce a maximum length scale of coherent structures, $l_{\mathrm{max}}$, which limits $l_{\mathrm{sh}}$ at very low gas densities. This has no impact on the shielding within the ISM, but does affect the thermal evolution of gas at the cosmic mean density. Inspired by the rapid change of the mean free path of ionizing photons during the epoch of reionization (see e.g. \citealp{Gnedin2004, Rahmati2018}), we introduce a redshift-dependent function for $l_{\mathrm{max}}$ in which $l_{\mathrm{max}}$ transitions steeply from a smaller pre-reionization value ($l_{\mathrm{max,highz}}$ for $z>z_{\mathrm{t,lsh}}$) to a larger post-reionization value ($l_{\mathrm{max,lowz}}$ for $z\ll z_{\mathrm{t,lsh}}$):

\begin{eqnarray}
    l_{\mathrm{max}}(z)  &= l_{\mathrm{lowz}} - \frac{l_{\mathrm{lowz}} - l_{\mathrm{highz}}}{1 + \left ( \frac{z_{\mathrm{t,lsh}}}{z}\sqrt{1 - \frac{0.2}{z_{\mathrm{t,lsh}} - 0.2}}\right)^{100}} \;,
\end{eqnarray}\label{eq:lshmax}

\noindent 
with the fiducial parameter values $l_{\mathrm{max,lowz}}=50\,\mathrm{kpc}$, $l_{\mathrm{max,highz}}=10\,\mathrm{kpc}$, and $z_{\mathrm{t,lsh}} = 7$. All limits are in physical (proper) units. 
The final shielding column density, $N_{\mathrm{sh}}$, is therefore 

\begin{equation}\label{eq:Nshcut}
    N_{\mathrm{sh}} = \,
    \begin{cases}
      n_{\mathrm{H}} \;l_{\mathrm{max}}(z) & l_{\mathrm{sh}} \ge l_{\mathrm{max}}(z) \\
      n_{\mathrm{H}} \;l_{\mathrm{sh}} = N_{\mathrm{sh}}'&  l_{\mathrm{sh}} < l_{\mathrm{max}}(z) \; ,
    \end{cases}
\end{equation}

\noindent
and is shown for the fiducial parameter values at $z=0$ in Fig.~\ref{fig:Nsh}.

For the fiducial parameter values of the position ($z_{\mathrm{t,lsh}}=7$), width (0.2 in equation~\ref{eq:lshmax}), and steepness (set by the exponent of 100 in equation~\ref{eq:lshmax}) of the transition, the thermal evolution of gas at the cosmic mean baryon density matches observations and the desired reionization redshift (to be discussed in Section~\ref{sec:thermevol}).

\subsubsection{Dust}\label{sec:dust}

Dust is an important catalyst for various chemical reactions and photoelectric heating of dust grains is a critical heating process in the ISM. In addition, dust efficiently absorbs radiation and therefore contributes to self-shielding. For the pre-tabulated quantities, such as the species fractions, as well as heating and cooling rates, we need to assume both a dust-to-gas ratio, $\mathcal{DTG}$, and a dust composition (e.g. the type of grains and the grain size distribution). 

For a constant dust-to-metal ratio, $\mathcal{DTG}$ is proportional to the gas metallicity, $Z$, and we normalize $\mathcal{DTG}$ to $\mathcal{DTG}_{F_{\star} = 1}$ at solar metallicity ($Z=\mathrm{Z}_{\odot}$), with $\mathcal{DTG}_{F_{\star}=1} = 5.3\times10^{-3}$. This value matches the dust-to-gas ratio for the fiducial depletion model in this work ($F_{\star} = 1$ in \citealp{Jenkins2009}, see Section~\ref{sec:depletion} and Table~\ref{tab:depletion}). In hot gas ($T>T_{\mathrm{max}} = 10^{5}\,\mathrm{K}$), we assume instantaneous destruction of dust and set $\mathcal{DTG} = 0$. At low column densities ($N_{\mathrm{min}}<N_{\mathrm{ref}}<N_{\mathrm{t,dust}}$, with $ N_{\mathrm{t,dust}} = 10^{20}\,\mathrm{cm}^{-2}$ in the fiducial model) we link the dust content to the star formation rate surface density, with a power law exponent of $\alpha_{\mathrm{dust}}=1.4$. Taking these ingredients together, $\mathcal{DTG}$ is defined as 

\begin{align}
    \mathcal{DTG} = & \,\mathcal{DTG}_{F_{\star}=1}\, \frac{Z}{\mathrm{Z}_{\odot}}  \,\begin{cases}
             0 & N_{\mathrm{ref}} = N_{\mathrm{min}} \\
              \left ( \frac{N_{\mathrm{ref}}}{N_{\mathrm{t,dust}}} \right )^{\alpha_{\mathrm{dust}}} &  N_{\mathrm{min}}<N_{\mathrm{ref}} < N_{\mathrm{t,dust}}\\
     1 &  N_{\mathrm{ref}} \ge N_{\mathrm{t,dust}} \; .
    \end{cases} \label{eq:dusttogas}
\end{align}

\noindent
The visual extinction, $A_{\mathrm{V}}$, is given by

\begin{equation}\label{eq:AV}
    A_{\mathrm{V}} = \mathcal{D} \; N_{\mathrm{sh}} \; \mathcal{DTG} \; .
\end{equation}

\noindent
It is proportional to the dust-to-gas ratio, $\mathcal{DTG}$, and the shielding column density, $N_{\mathrm{sh}}$ (from Section~\ref{sec:Nsh}). The proportionality constant, $\mathcal{D}$, depends on the assumed dust composition. We use \textsc{cloudy} to calculate $A_{\mathrm{V}}$ for both their \textsc{orion} ($\mathcal{DTG}_{\mathrm{Orion}} = 5.6\times10^{-3}$) and \textsc{ism} ($\mathcal{DTG}_{\mathrm{ISM}} = 6.6\times10^{-3}$) grain sets\footnote{In \citetalias{PS20} we used the \textsc{orion} grain set, see the discussion in Appendix~\ref{sec:app:PS20comp}.} and find $\mathcal{D}_{\mathrm{Orion}} = 6.1\times 10^{-20} \,\mathrm{mag\,cm}^{2}$ and $\mathcal{D}_{\mathrm{ISM}} = 7.1\times 10^{-20} \,\mathrm{mag\,cm}^{2}$. The resulting visual extinction per unit hydrogen column density, $(A_{\mathrm{V}}/N_{\mathrm{H}})$, is nearly the same for both grain sets (within 0.2 per cent) while their dust-to-gas ratios differ by a factor of 0.85. This means that dust shielding is identical for both grain sets but rates that scale with $\mathcal{DTG}$, such as the photoelectric heating rate, are lower in the \textsc{orion} grain set than in the \textsc{ism} grain set by a factor of 0.85. 

In \textsc{chimes}, the dust properties are parametrized with the effective grain size, $d_{\mathrm{eff}}$, and the dust-to-gas ratio relative to that of the MW, $[\mathcal{DTG}/\mathcal{DTG}_{\mathrm{ISM}}]$. The visual extinction in \textsc{chimes}, $A_{\mathrm{V,chimes}}$ is given by 

\begin{equation}\label{eq:AVchimes}
    A_{\mathrm{V,Chimes}} = d_{\mathrm{eff}} \; \left [\frac{\mathcal{DTG}}{\mathcal{DTG}_{\mathrm{ISM}}} \right ]\; N_{\mathrm{sh}} \;.
\end{equation}

\noindent
Comparing equations~(\ref{eq:AV}) and (\ref{eq:AVchimes}), we find $d_{\mathrm{eff,Orion}} = \mathcal{D}_{\mathrm{Orion}} \,\mathcal{DTG}_{\mathrm{ISM}} = 4.72\times10^{-22}\,\mathrm{mag\,cm}^{-2}$ and $d_{\mathrm{eff,ISM}} = \mathcal{D}_{\mathrm{ISM}} \,\mathcal{DTG}_{\mathrm{ISM}} = 4\times10^{-22}\,\mathrm{mag\,cm}^{-2}$ (see Table~\ref{tab:dustmodel} for an overview). 

The fiducial dust model assumes the same grain composition as the \textsc{ism} grain model in \textsc{cloudy}, but with a slightly lower dust-to-gas ratio, $[\mathcal{DTG}_{F_{\star}=1}/\mathcal{DTG}_{\mathrm{ISM}}] = 0.8$, consistent with the strongest depletion model from \citet{Jenkins2009} (see Section~\ref{sec:depletion}).

\begin{table}
    \caption{Overview of selected dust models (column 1). In this work, the fiducial model assumes the same dust grain composition as in the \textsc{cloudy} \textsc{ism} grain model with an effective grain size of $d_{\mathrm{eff}}$ (column 3), but re-normalized to a dust-to-gas ratio (column 2) that is consistent with the depletion from \citet{Jenkins2009}. The properties of the dust models used in \citetalias{PS20} (\textsc{cloudy orion} grain model) and in the \textsc{colibre} project (\textsc{cloudy ism} grain model) are listed for comparison.  }
    \label{tab:dustmodel}
    \begin{tabular}{lrrcc}
         \hline
         Dust model &  $\mathcal{DTG}$  &  $d_{\mathrm{eff}}\,[\mathrm{mag\,cm}^{-2}]$ & $\left [\frac{\mathcal{DTG}}{\mathcal{DTG}_{\mathrm{ISM}}}\right ]$\\
         \hline
         This work ($F_{\star}=1$)   & $5.3\times10^{-3}$  & $4\times 10^{-22}$ & 0.8 \\
         \textsc{cloudy orion} (\citetalias{PS20})&  $5.6\times10^{-3}$ &  $4.72\times10^{-22}$& 0.85\\
         \textsc{cloudy ism} (\textsc{colibre})& $6.6\times10^{-3}$ & $4\times10^{-22}$ & 1\\
         \hline
    \end{tabular}
\end{table}

\subsubsection{Dust boost}\label{sec:dustboost}
The formation rate of H$_{2}$ on dust grains increases with gas density (see e.g. equation 18 in \citealp{Cazaux2002}). For simulations that do not resolve the full distribution of gas densities in the ISM, the total formation rate of H$_{2}$ may be underestimated. To enable correcting for unresolved gas clumping, 
we implement the option to boost the rate of reactions on the surface of dust grains, including the formation of H$_{2}$, with a density-dependent boost factor, $b_{\mathrm{dust}}$, that relates to unresolved gas clumping and is defined as 

\begin{align}\label{eq:dustboost}
    &\log b_{\mathrm{dust}} = \nonumber \\
    &\begin{cases}
        0 &  n_{\mathrm{H}} \le n_{\mathrm{b,min}} \\
        \log b_{\mathrm{dust,max}}\,\frac{\log n_{\mathrm{H}} - \log n_{\mathrm{b,min}}}{\log n_{\mathrm{b,max}} - \log n_{\mathrm{b,min}}} &  n_{\mathrm{b,min}}< n_{\mathrm{H}} < n_{\mathrm{b,max}}\\
       \log  b_{\mathrm{dust,max}} &  n_{\mathrm{H}} \ge n_{\mathrm{b,max}} \\
    \end{cases}
\end{align}

\noindent
and we multiply the formation rate of H$_{2}$ on dust grains by $b_{\mathrm{dust}}$. The boost factor is parametrized by the maximum boost factor, $b_{\mathrm{dust,max}}$, for densities $n_{\mathrm{H}} > n_{\mathrm{b,max}}$, a minimum density below which no dust boost is applied (i.e. $b_{\mathrm{dust}} = 1$ for $n_{\mathrm{H}} < n_{\mathrm{b,min}}$) and a linear interpolation in $\log-\log$ space between these limits. 

Depending on the resolution of the simulation, the maximum boost factor, $b_{\mathrm{dust,max}}$, can be set to unity or calibrated. Note that $b_{\mathrm{dust}}$ only boosts the formation rate of H$_{2}$ on dust grains and the grain recombination reactions but does not affect the visual extinction (equation~\ref{eq:AV}) nor the PE heating rates (equation~\ref{eq:PER+14}), because the total $\mathcal{DTG}$ ratio is unaffected by the dust boost. 

We use a fiducial maximum boost factor of unity (i.e. no boost) in Sections~\ref{sec:neutral} and ~\ref{sec:galaxy:eq}, and we explore varying this factor in Section~\ref{sec:galaxy:H2}.

\subsubsection{Dust depletion} \label{sec:depletion}

The tables with cooling and heating rates in this work do not include any depletion, because we will use these tables in combination with a live dust model \citep{Trayford2025arXiv} in large-scale simulations. Therefore, we correct for dust depletion on the fly and on a gas-particle-by-gas-particle basis. 

For the exploration of the thermal equilibrium in this work, we correct the undepleted tables following the depletion patterns from \citet{Jenkins2009}\footnote{We provide the routine to apply different depletion strengths to the tables, see data availability section.}. In \citet{Jenkins2009}, the depletion strength is parametrized by $F_{\star}$, normalized so that $F_{\star}=0$ ($F_{\star}=1$) represents the weakest (strongest) depletion in their sample. Table~\ref{tab:depletion} lists the gas fractions, $f_{\mathrm{gas}}$, of each element included in this work that are depleted on dust grains for low (column 2), moderate (column 3), and high (column 4) values of $F_{\star}$. The fiducial depletion strength in this work is a strong depletion of $F_{\star}=1$, which corresponds to a total dust-to-gas ratio of $\mathcal{DTG}_{F_{\star}=1} = 5.3\times10^{-3}$ in the ISM. 

Because we decrease $\mathcal{DTG}$ in our model towards low densities and high temperatures (see equation~\ref{eq:dusttogas}), we vary the depletion of each element on dust grains, $f_{\mathrm{dust}} = 1- f_{\mathrm{gas}}$, accordingly.

\begin{table}
    \caption{Fraction of each depleted element (column 1) that remains in the gas phase for different depletion models from \citet{Jenkins2009}: a low level of depletion on dust grains with $F_{\star} = 0$ (column 2), a moderate depletion level with $F_{\star} = 0.5$ (column 3) and a high depletion level ($F_{\star} = 1$, column 4). The bottom row list the total dust-to-gas ratio, $\mathcal{DTG}$, for each depletion model for solar metallicity. }
    \centering
    \begin{tabular}{llll}
        \hline
        Element & $f_{\mathrm{gas,low}}$ ($F_{\mathrm{\star}} = 0$) & $f_{\mathrm{gas,mod}}$ ($F_{\mathrm{\star}} = 0.5$) & $f_{\mathrm{gas,high}}$ ($F_{\mathrm{\star}} = 1$)  \\
        \hline
         C  & 0.828 & 0.737& 0.656\\
         N  & 0.914 & 0.914& 0.914\\
         O  & 1.000 & 0.885& 0.683\\
         Mg & 0.562 & 0.178& 0.057\\
         Si & 0.752 & 0.203& 0.055\\
         Fe & 0.123 & 0.028& 0.006\\
         \hline
         $\mathcal{DTG}$ & $2.08\times10^{-3}$ &$3.72\times10^{-3}$&  $5.30\times10^{-3}$ \\
         \hline
    \end{tabular}
    \label{tab:depletion}
\end{table}

\subsection{Equilibrium abundances}\label{sec:equilibriumabundances}

We use an updated version of the wrapper script \textsc{chimes-driver} (publicly available, see data availability section) which uses the stand-alone version of the chemical network \textsc{chimes} \citep{chimes2014optthin, chimes2014shielded} to calculate the abundances of 157 species, assuming chemical equilibrium, for redshifts between $z=0$ and $9$, gas temperatures between $T=10~\mathrm{and}\,10^{9.5}\,\mathrm{K}$, gas densities between $n_{\mathrm{H}}=10^{-8}\,\mathrm{and}\,10^6\,\mathrm{cm}^{-3}$, and metallicities between $Z/\mathrm{Z}_{\odot} = 0\,\mathrm{and}\,10^{0.5}$ with $Z_{\odot} = 0.0134$ and solar relative element abundance ratios from \citet{Asplund2009}.

The equilibrium abundances of species $x$, $n_{\mathrm{x}}/n_{\mathrm{H}}$, of gas with properties spanning the full grid ($z$, $T$, $n_{\mathrm{H}}$, $Z/\mathrm{Z}_{\odot}$) are calculated for the turbulent velocity dispersion described in Section~\ref{sec:Nref}, the radiation fields from Section~\ref{sec:ISRF}, the cosmic ray rate from Section~\ref{sec:CR}, the shielding column density from Section~\ref{sec:Nsh}, the dust-to-gas ratio from Section~\ref{sec:dust} and the optional dust boost from Section~\ref{sec:dustboost}.
We calculate the equilibrium species abundances for the undepleted element abundances and add an option to later adjust the cooling and heating rates of the individual elements (Section~\ref{sec:processes}) for their gas-phase fractions ($f_{\mathrm{gas}}$, see Section~\ref{sec:depletion}) to account for depletion. 

\subsubsection{Reduced chemical networks}\label{sec:equilibriumabundances:red}

The chemical reactions included in \textsc{chimes} for the full network are listed in table~B1 of \citet{chimes2014optthin}. Naturally, networks that include only a subset of elements also include only a subset of reactions. 
We discuss here the differences in the equilibrium species abundances of H and He for a few different chemical networks, described in Table~\ref{tab:reducednetworks}.

\begin{figure}
    \centering
    \includegraphics[width=\linewidth]{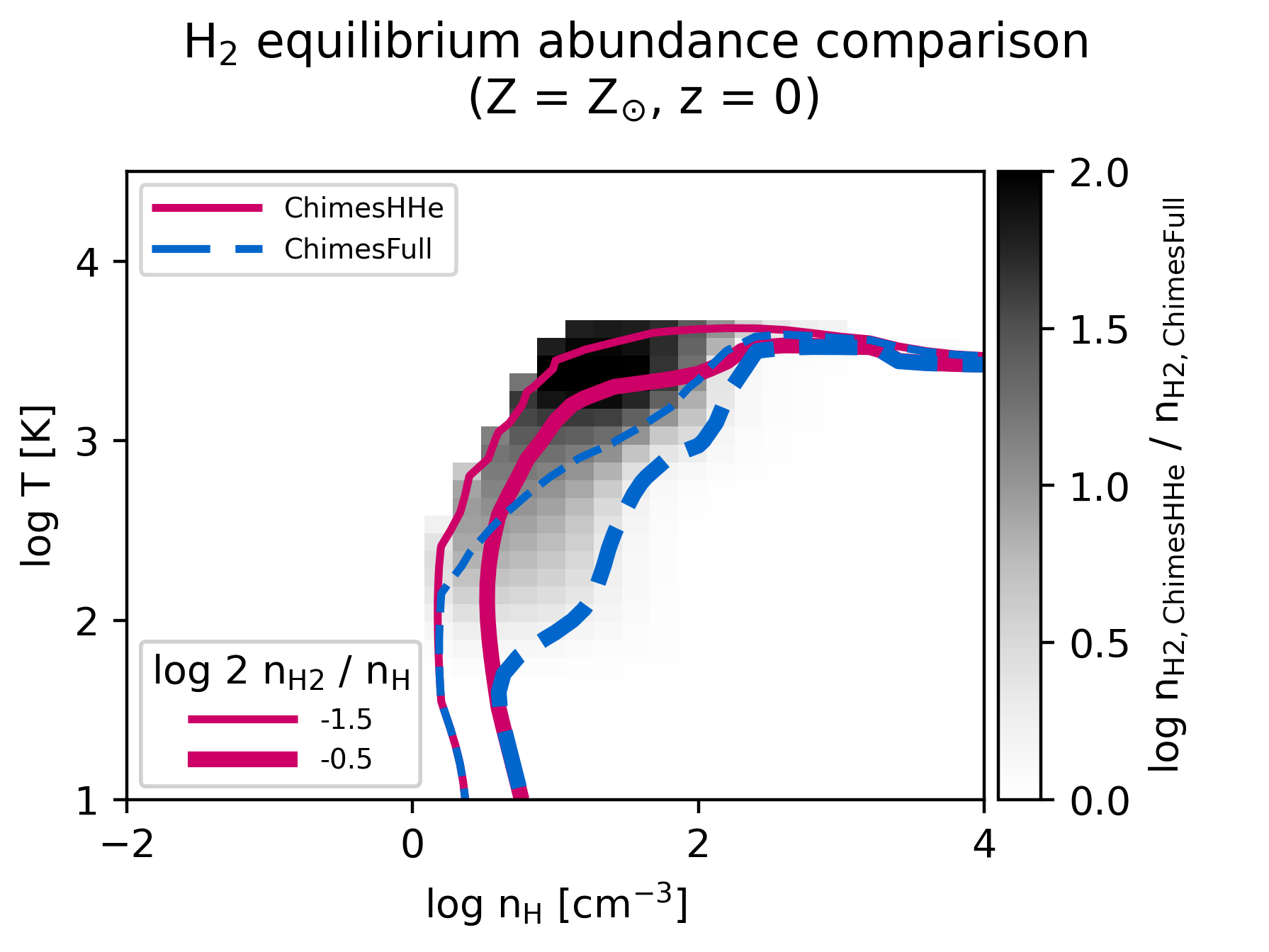}
    \caption{Comparison of the H$_2$ abundances between the full network (ChimesFull, 157 species) and a network that only includes H and He (ChimesHHe, 10 species). The molecular hydrogen abundances are evolved to chemical equilibrium for solar metallicity gas at $z=0$. The ratio $n_{\mathrm{H2,ChimesHHe}}/n_{\mathrm{H2,ChimesFull}}$ is shown as 2D histogram (see colorbar) and the absolute values are represented by red solid (ChimesHHe) and blue dashed (ChimesFull) contours at $\log 2 n_{\mathrm{H2}}/ n_{\mathrm{H}} = -1.5$ (thin lines) and $-0.5$ (thick lines). The H and He network underestimates the H$_2$ abundances compared to full network ($\log n_{\mathrm{H2,ChimesHHe}} / n_{\mathrm{H2,ChimesFull}} \ge 0$) by up to 2 orders of magnitude. }
    \label{fig:equabundancecomparison}
\end{figure}

In Fig.~\ref{fig:equabundancecomparison}, we compare the equilibrium abundances for molecular hydrogen for a chemical network including all 157 species (`ChimesFull' in Table~\ref{tab:reducednetworks}) and for a reduced chemical network that only follows H and He (`ChimesHHe' in Table~\ref{tab:reducednetworks}) for solar metallicity and $z=0$. In a narrow range in density and temperature, at the transition from atomic to molecular gas, the H$_2$ fraction in the ChimesHHe network is overestimated by up to $2\,\mathrm{dex}$, compared to ChimesFull. For example, at a density of $\log n_{\mathrm{H}}\,[\mathrm{cm}^{-3}] = 1$ and a temperature of $\log T \,[\mathrm{K}] = 3$, the H$_2$ fraction $\log 2 n_{\mathrm{H2}}/n_{\mathrm{H}}$ is $\lesssim -1.5$ with ChimesHHeO (blue dashed contours) and $\gtrsim-0.5$ when using the ChimesHHe network (red solid contours). The helium species, as well as the electron abundances, are unaffected. 

\begin{figure}
    \centering
    \includegraphics[width=\linewidth]{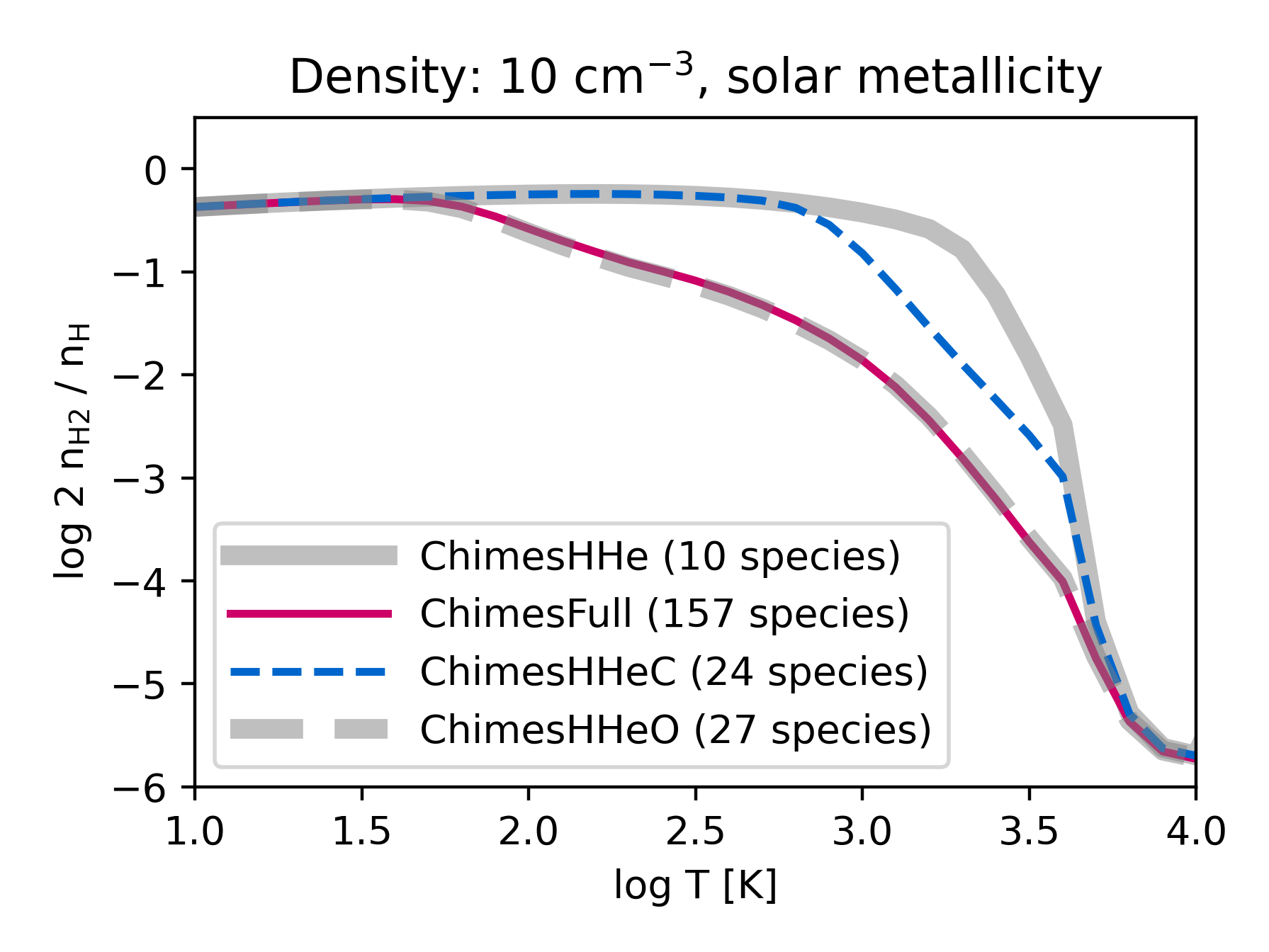}
    \caption{Equilibrium abundance of H$_2$ for a gas density of $n_{\mathrm{H}} = 10\,\mathrm{cm}^{-3}$ and solar metallicity for different chemical networks (see Table~\ref{tab:reducednetworks} for details). The smallest chemical network, ChimesHHe (thick solid line), overestimates the H$_2$ fractions compared to the full network, ChimesFull (thick dashed line). In the network ChimesHHeC (thin blue dashed line), the H$_2$ abundance is reduced at temperatures of $\approx 1000~\mathrm{K}$ but still largely follows the H$_2$ fraction from the minimal network ChimesHHe. On the other hand, when including O (ChimesHHeO, thin red solid line), the resulting H$_2$ abundances are indistinguishable from those of the full network. The H$_2$ fraction from the network ChimesHHeCO overlaps with the lines from ChimesHHeO and ChimesFull and is not shown for clarity. }
    \label{fig:rednetworks}
\end{figure}

Further investigation reveals that the difference in the H$_2$ abundance is mainly caused by reactions between H$_2$ and oxygen species, and also to a smaller extent with carbon species, which are missing from the reduced network. Fig.~\ref{fig:rednetworks} shows the H$_2$ abundance for solar metallicity gas with a density of $n_{\mathrm{H}} = 10\,\mathrm{cm}^{-3}$. At this gas density, the differences in the H$_2$ abundances between ChimesFull (thick grey dashed line) and ChimesHHe (thick grey solid line) are up to $2~\mathrm{dex}$, in agreement with Fig.~\ref{fig:equabundancecomparison}. Including carbon (ChimesHHeC, blue dashed line) reduces the H$_2$ abundance, but a $>1\,\mathrm{dex}$ difference to ChimesFull remains. In contrast, any network that includes oxygen (e.g. ChimesHHeO, red solid line) recovers the H$_2$ fractions from the full chemical network. 

We identify seven chemical reactions in ChimesHHeO that destroy H$_2$ and are not present in the ChimesHHe network: 

\begin{tabular}{l@{\hspace{0.5em}}l@{\hspace{0.5em}}l@{\hspace{0.5em}}l@{\hspace{0.5em}}l@{\hspace{1em}}l}
    O &+ H$_2$ & $\rightarrow$ OH &+ H & \citep{Natarajan1987}\\
    O$^+$ &+ H$_2$ & $\rightarrow$ OH$^+$ &+ H & \citep{Smith1978}  \\
    O$^-$ &+ H$_2$ & $\rightarrow$ H$_2$O &+ e$^-$ & \citep{LeTeuff2000} \\
    OH &+ H$_2$ & $\rightarrow$ H$_2$O &+ H & \citep{Oldenborg1992} \\
    O$_2$ &+ H$_2$ & $\rightarrow$ OH &+ OH & \citep{azatyan1975rate} \\
    OH$^+$ &+ H$_2$ & $\rightarrow$ H$_2$O$^+$ &+ H & \citep{Jones1981} \\
    H$_2$O$^+$ &+ H$_2$ & $\rightarrow$ H$_3$O$^+$ &+ H & \citep{Rakshit1980}
\end{tabular}

\noindent

The references for the reaction rates are added in parentheses. For a full list of reactions included in \textsc{chimes} and their references, see table~B1 in \citet{chimes2014optthin}. We limit our analysis here to the reactions that are directly involved in destroying H$_2$ but caution that additional reactions may influence the effect of oxygen on the H$_2$ fractions\footnote{For example, a higher rate for the dissociative recombination of OH$^+$ (OH$^+$ + e$^-$ $\rightarrow$ O + H), as suggested by \citet{Kalosi2023}, compared to the rate assumed in \textsc{chimes} (from \citealp{Mitchell1990}) may decrease the OH$^+$ abundance and hence decrease the destruction rate of H$_2$ by OH$^+$.}.

\begin{figure}
    \centering
    \includegraphics[width=\linewidth]{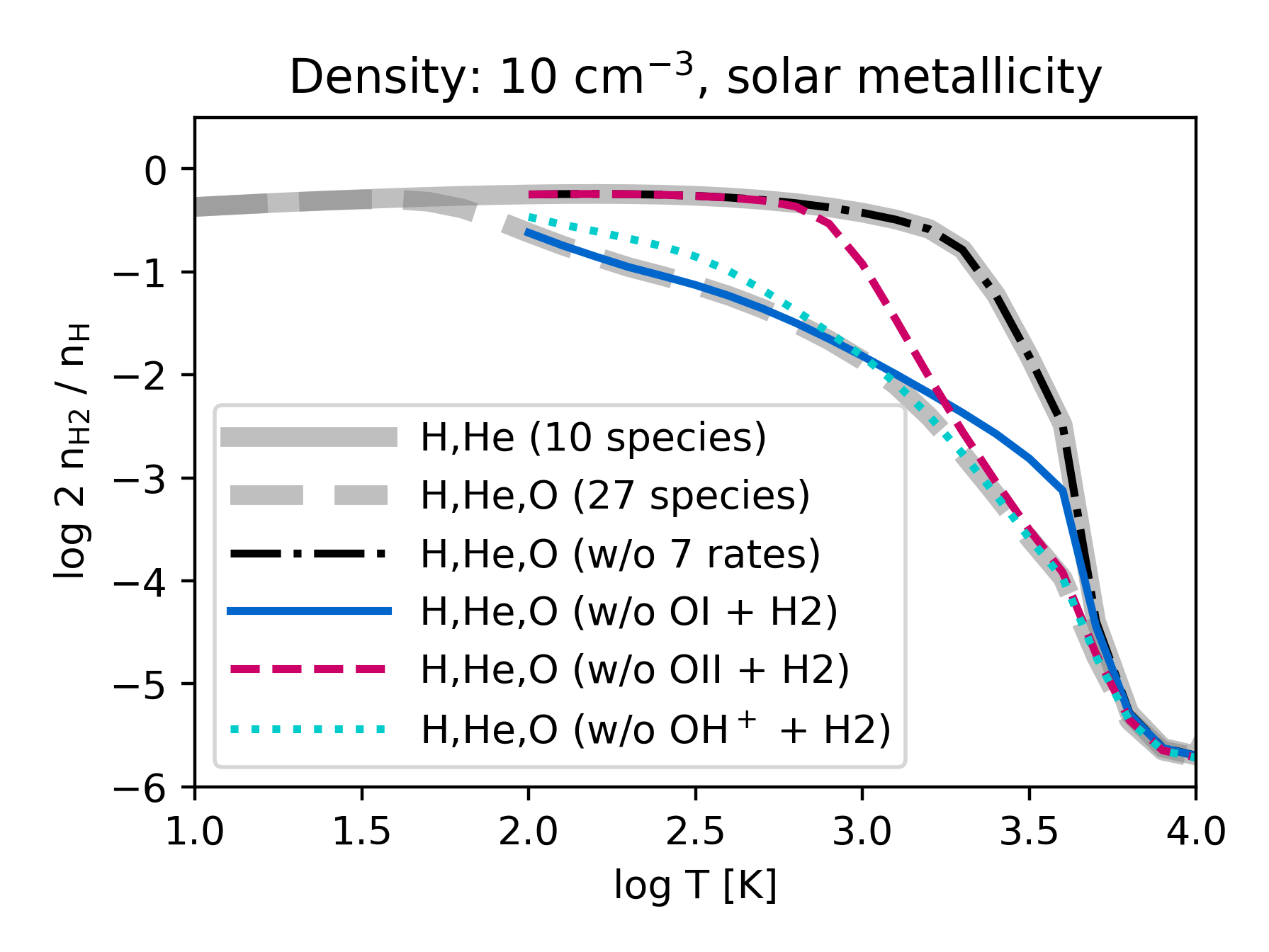}
    \caption{Equilibrium abundance of H$_2$ for a gas density of $n_{\mathrm{H}} = 10\,\mathrm{cm}^{-3}$ and solar metallicity for the ChimesHHe (thick grey solid line) and the ChimesHHeO (thick grey dashed line) chemical network. The thin lines show the recalculated H$_2$ fractions of ChimesHHeO when excluding individual reactions between H$_2$ and different oxygen species (\ion{O}{I}: blue solid line, \ion{O}{II}: red dashed line, OH$^+$: green dotted line, all seven reactions listed in the text: black dash-dotted line).}
    \label{fig:Oreactions}
\end{figure}

In order to quantify the individual contributions of these reactions in reducing the H$_2$ fraction within the ChimesHHeO network, we manually remove individual reactions by reducing their reaction coefficients by a factor of 10,000 and recalculate the H$_2$ fractions for $2\le\log T \,[\mathrm{K}]\le 4$. The results are shown in Fig.~\ref{fig:Oreactions}. The H$_2$ fractions for the ChimesHHe and ChimesHHeO networks are as in Fig.~\ref{fig:rednetworks} for reference and, as expected, removing all seven reactions (black dash-dotted line) recovers the results from ChimesHHe. The main reactions that involve oxygen species and destroy H$_2$ are with \ion{O}{I} (blue solid line), \ion{O}{II} (red dashed line), and OH$^+$ (green dotted line). Each of these reactions triggers further reactions from the list above that destroy additional hydrogen molecules. For clarity, we do not show the H$_2$ fractions after removing the reactions between H$_2$ and O$^-$, OH, O$_2$, and H$_2$O$^+$ because they overlap with the ChimesHHeO results, indicating that their contribution is negligible. 

We will show and discuss the impact of excluding oxygen from the non-equilibrium calculations on the H$_2$ fractions in simulated galaxies in Section~\ref{sec:galaxysimulations}.

\begin{table*}
    \caption{List of included species in chemical networks of different sizes.}
    \centering
    \begin{tabular}{lll}
        \hline
        Name &  Nr. of species  & List of species \\
        \hline
        ChimesHHe & 10 & e$^{-}$, \ion{H}{I}, \ion{H}{II}, $\mathrm{H}^{-}$, \ion{He}{I}, \ion{He}{II}, \ion{He}{III}, $\mathrm{H}_{2}$, $\mathrm{H}_{2}^{+}$, and $\mathrm{H}_{3}^{+}$\\ 
        ChimesHHeC & 24 & as ChimesHHe; C atom and ions; C$^-$, C$_2$, CH, CH$_2$, CH$_3^+$, CH$^+$, CH$_2^+$\\
        ChimesHHeO &  27& as ChimesHHe; O atom and ions; O$^-$, OH, H$_2$O, O$_2$, OH$^+$, H$_2$O$^+$, H$_3$O$^+$, O$_2^+$  \\
        ChimesFull & 157& e$^-$; atoms and ions from H, He, C, N, O, Ne, Si, Mg, S, Ca, Fe; \\
        &&all molecules listed in ChimesHHe, ChimesHHeC, ChimesHHeO, as well as HCO$^+$, CO, CO$^+$, HOC$^+$ \\
        \hline 
    \end{tabular}
    \label{tab:reducednetworks}
\end{table*}

\subsection{The \textsc{hybrid-chimes} model}\label{sec:hybrid}

We introduce the \textsc{hybrid-chimes} model, which combines equilibrium and non-equilibrium chemistry calculations, both done with the \textsc{chimes} chemical network, to balance computational efficiency with physical accuracy. In this work, \textsc{hybrid-chimes} is implemented in the hydrodynamic code \textsc{swift} \citep{swift2024} but it can be included as cooling module in any hydrodynamic code, especially in those that already have a \textsc{chimes} integration, such as \textsc{gizmo} \citep{Hopkins2015Gizmo, Richings2022}. In this work we parametrize the strength of the ISRF, the shielding column density, and the cosmic ray rate as discussed in Section~\ref{sec:inputparameter}, but \textsc{hybrid-chimes} supports all parameter choices from the general \textsc{chimes} package.

\textsc{Chimes} can be used to follow the non-equilibrium species abundances of hydrogen, helium and any selection from the elements C, N, O, Ne, Si, Mg, S, Ca, and Fe. The different chemical networks include between 10 (electrons, hydrogen and helium species) and 157 (electrons and all 11 elements) species. For simulations in which following the non-equilibrium chemistry of all 157 species is prohibitively expensive, the species abundances of the remaining elements are pre-tabulated with \textsc{chimes-driver} and the respective cooling and heating rates are pre-tabulated with \textsc{chimes-rates} (see data availability statement for access to these packages). Each software package uses the same input parameters as described in Section~\ref{sec:inputparameter} for consistency.

\begin{figure}
    \centering
    \includegraphics[width=\linewidth]{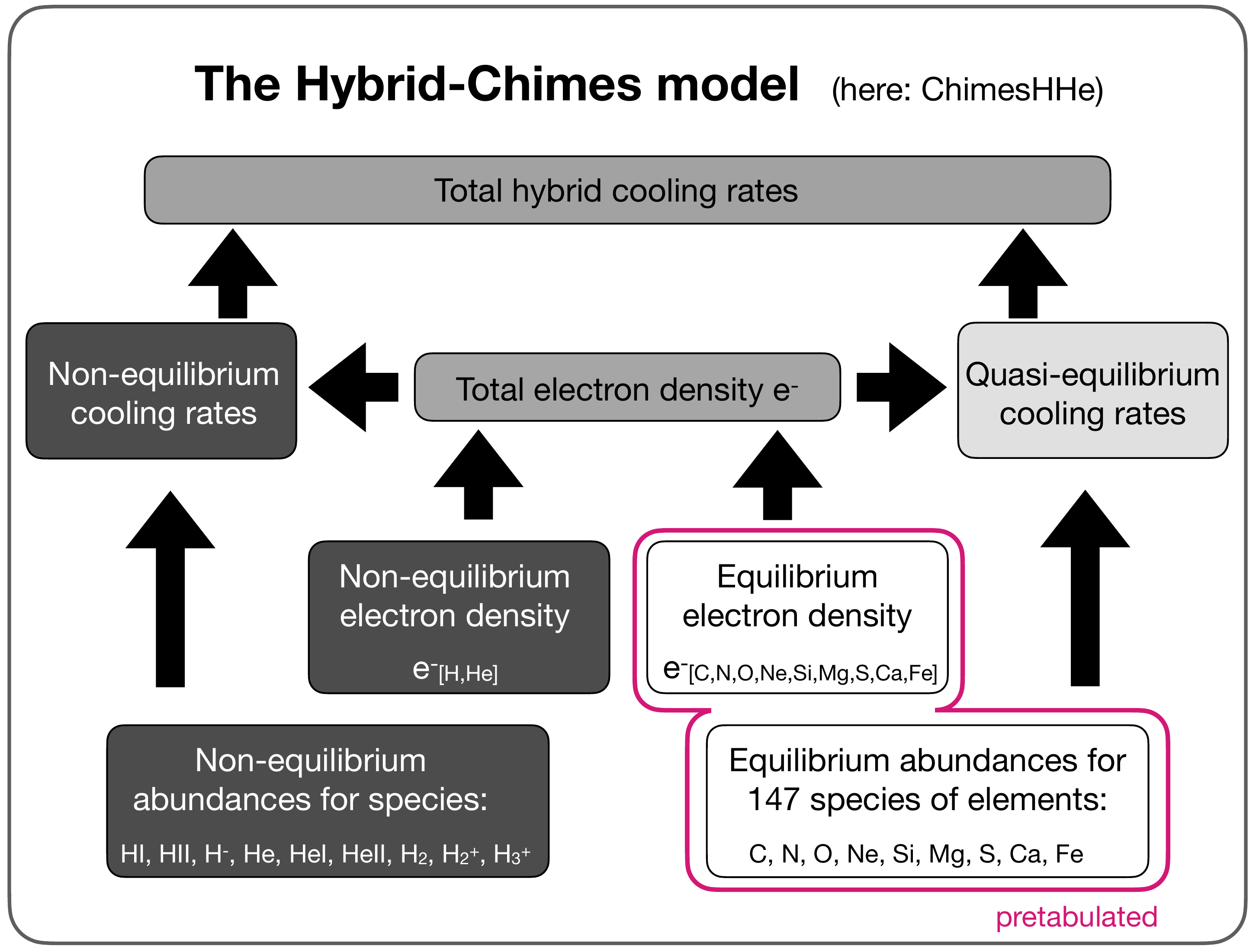}
    \caption{An overview of the structure of the \textsc{hybrid-chimes} cooling model, using the reduced network ChimesHHe as an example. The pre-tabulated properties (equilibrium species abundances and electron fractions from equilibrium elements) are highlighted. During the simulation, the non-equilibrium species abundances of nine hydrogen and helium species and their electrons are evolved on the fly. The total electron density (from equilibrium and non-equilibrium calculations) is used to scale both the non-equilibrium and the (quasi-)equilibrium cooling rates. Grey shades illustrate the contributions of non-equilibrium and equilibrium calculations, transitioning smoothly from white (completely in equilibrium) to dark grey (completely in non-equilibrium). }
    \label{fig:hybrid}
\end{figure}

Combining (1) the cooling and heating rates from a non-equilibrium chemical network for a subset of species and (2) tabulated rates, calculated under the assumption of chemical equilibrium, together represent an established paradigm, exemplified by e.g. \textsc{grackle} \citep{Grackle2017}. For \textsc{hybrid-chimes} we place special emphasis on consistency between the non-equilibrium and the equilibrium calculations. Both parts of \textsc{hybrid-chimes} rely on the same chemical network (\textsc{chimes}) and the same assumptions on radiation fields, cosmic rays, and shielding column densities. Furthermore, as outlined in Fig.~\ref{fig:hybrid}, \textsc{hybrid-chimes} includes important non-equilibrium effects even in the cooling rates of elements that are assumed to be in equilibrium: in the hybrid cooling module, the total electron density is the sum of the electrons from the non-equilibrium species abundances in the chemical network and from the equilibrium species abundances of the pre-tabulated elements. Because of the much higher number density of hydrogen and helium, compared to the number density of metal species, non-equilibrium effects are well represented in the total electron density. The cooling rates of all elements are proportional to the electron density (see Eq.~\ref{eq:coolingion}). By using the total electron density for all cooling rates, the non-equilibrium effects of hydrogen and helium impact the cooling rates of the elements that are assumed to be in chemical equilibrium, resulting in `quasi-equilibrium' cooling rates for metals. In turn, the electrons from equilibrium species are also included in the calculation of the non-equilibrium cooling rates.

In contrast, \textsc{grackle} computes primordial cooling rates using its own chemical network, while metal cooling rates are determined independently with \textsc{cloudy}, with electron densities dominated by hydrogen and helium electrons from the equilibrium \textsc{cloudy} calculations.

Note, that while this hybrid approach has the clear advantage of drastically reducing computational cost compared to evolving the full network in non-equilibrium, it may introduce the errors discussed in Section~\ref{sec:equilibriumabundances:red}, depending on the included elements.

\subsection{Cooling and heating rates}\label{sec:processes}
The cooling and heating rates for the equilibrium species abundances from Section~\ref{sec:equilibriumabundances} are calculated with the python package \textsc{chimes-rates}, which is made publicly available with this work (see data availability statement). For a given redshift\footnote{While in \citetalias{PS20} all redshifts were combined within one \texttt{hdf5} file, here we provide one \texttt{hdf5} file per redshift.}, the structure of the \texttt{hdf5} files produced with \textsc{chimes-rates} is very close to that of \citetalias{PS20} for backward compatibility. However, in this update, some differences are introduced, mainly due to the change from \textsc{cloudy} to \textsc{chimes}, which will be discussed in this section. 

The main references for the included cooling and heating rates are the \textsc{chimes} publications, \citet{chimes2014optthin} and \citet{chimes2014shielded}. The package \textsc{chimes-rates} closely follows the processes included in \textsc{chimes}. For completeness and to highlight some differences from both the originally published \textsc{chimes} version and from the tables in \citetalias{PS20}, we briefly summarize all the thermal processes included. An overview of all individually tabulated cooling and heating channels is provided in Tables~\ref{tab:cooloverview} and \ref{tab:heatoverview}, respectively.

The \textsc{chimes} code includes a large data file that tabulates various rate coefficients for all included species. These data sets are read by \textsc{chimes-rates} and therefore any update in the \textsc{chimes} data file would propagate to the results from \textsc{chimes-rates}. For this work, we use commit 1a837f8 from 2023-07-18 of the main \textsc{chimes} datafile. For details, see the data availability statement.

\subsubsection{Cooling from ions per element}\label{sec:coolingelement}

The cooling rates from hydrogen, helium and the metal species in \textsc{chimes} are based on \citet{OppenheimerSchaye2013} and include the cooling rates from radiative and
di-electronic recombination, collisional ionization and excitation,
and Bremsstrahlung, tabulated for each ion individually. 

\citet{OppenheimerSchaye2013} used \textsc{cloudy} \citep{Cloudy1998PASP} versions 10.00 and 13.01 (updated on their project webpage \href{https://noneq.strw.leidenuniv.nl}{https://noneq.strw.leidenuniv.nl}) to calculate the ion-by-ion cooling rates, $\Lambda_{\mathrm{x_i}} / (n_{\mathrm{e}} n_{\mathrm{x_i}}) \, [\mathrm{erg\,cm^3\,s^{-1}}]$ for elements $x =$ H, He, C, N, O, Ne, Si, Mg, S, Ca, and Fe and their respective ions, $x_{\mathrm{i}}$. If collisions between electrons with number density $n_{\mathrm{e}}$ and ions with number density $n_{\mathrm{x_i}}$ dominate the cooling rate, the term $\Lambda_{\mathrm{x_i}} / (n_{\mathrm{e}} n_{\mathrm{x_i}}) \, [\mathrm{erg\,cm^3\,s^{-1}}]$ only depends on temperature and is tabulated by \citet{OppenheimerSchaye2013} for temperatures between $\log T [K] = 2$ and $9$ (see Table~\ref{tab:thermalprocesses} for exceptions). 

For each element $x$, we tabulate the total cooling rate as
\begin{eqnarray}
    \Lambda_{\mathrm{x}}\,/ \,n_{\mathrm{H}}^2 \, [\mathrm{erg\,cm^3\,s^{-1}}] &=&\frac{1}{n_{\mathrm{H}}^2}\sum_{i} \Lambda_{\mathrm{x}_i}  \, [\mathrm{erg\,cm^3\,s^{-1}}] \nonumber \\
    &=& \sum_{i} \underbrace{\frac{\Lambda_{\mathrm{x}_i}}{n_{\mathrm{e}} n_{\mathrm{x}_i}}}_{\mathrm{OS2013+}} \cdot \underbrace{\frac{n_{\mathrm{e}}}{n_{\mathrm{H}}} \cdot \frac{n_{\mathrm{x}_i}}{n_{\mathrm{H}}}}_{\mathrm{CHIMES}}\;, \label{eq:coolingion}
\end{eqnarray}

\noindent
with $\Lambda_{\mathrm{x}_i} / (n_{\mathrm{e}} n_{\mathrm{x}_i})$ tabulated in the main \textsc{chimes} data file based on \citet{OppenheimerSchaye2013} and the references in Table~\ref{tab:thermalprocesses}. The electron and ion abundances are from the equilibrium abundances tabulated with \textsc{chimes-driver}, as a function of redshift, gas density, gas temperature, and metallicity\footnote{For the hybrid chemistry case discussed in Section~\ref{sec:hybrid}, the electron abundance uses the non-equilibrium contribution for H and He.} (see Section~\ref{sec:equilibriumabundances}).

\paragraph*{Caution when upgrading from \citetalias{PS20}:} In \citetalias{PS20}, the cooling from bremsstrahlung is split into three separate categories: (1) `NetFFH', the bremsstrahlung from H and He ions, (2) `NetFFM', the bremsstrahlung from metal ions, and (3), `eeBrems', the electron - electron bremsstrahlung (see their table 7). In this work, the bremsstrahlung from all ions of element $x$ is included in the total cooling rate of element $x$ instead. We do not include electron - electron bremsstrahlung, which is only important at temperatures of $\log T \,[\mathrm{K}] \gtrsim 10^9$ (see for example figure 9 in \citetalias{PS20}). 

The tables of \citetalias{PS20} also containes a separate cooling channel for all metal elements that were not tabulated individually (`OtherA', see their table 7). This minor cooling contribution is not included in \textsc{chimes}. For an easy upgrade, we keep the dimensions of the cooling dataset from \citetalias{PS20} and set the unused fields to 0  (see also Table~\ref{tab:cooloverview}).

\begin{table}
    \caption{Overview of all individually tabulated cooling channels. }
    \centering    
    \begin{tabular}{llll}
    \hline
    Index & Identifier &  Comment\\
    \hline
    0 - 10  & Elements (H to Fe)  &  see Section~\ref{sec:coolingelement}\\
   11 & OtherA          &  $= 0$; not included, see Section~\ref{sec:coolingelement}\\
   12 & H2              & see Section~\ref{sec:coolingH2}  \\
   13 & Molecules       & see Section~\ref{sec:coolingmol} \\
   14 & HD              & see Section~\ref{sec:coolingHD}  \\
   15 & NetFFH          & $=0$; included in Elements,  see Section~\ref{sec:coolingelement}\\
   16 & ComptonUI       & was `NetFFM' in \citetalias{PS20}, see Section~\ref{sec:coolingCompton} \\
   17 & eeBrems         & $=0$; included in Elements,  see Section~\ref{sec:coolingelement} \\
   18 & ComptonCMB      & was `Compton' in \citetalias{PS20}, see Section~\ref{sec:coolingCompton} \\
   19 & Dust            & see Section~\ref{sec:coolingdust}   \\
   20 & TotalPrim       &  Sum of (1), (2), (12), (14), (16), (18)  \\
   21 & TotalMetal      &  Sum of (3-10), (13), (19)   \\
    \hline
    \end{tabular}
    \label{tab:cooloverview}
\end{table}

\subsubsection{Molecular hydrogen cooling}\label{sec:coolingH2}

The H$_2$ cooling rates in \textsc{chimes} include ro-vibrational cooling caused by collisional excitation of H$_2$ and cooling due to collisional dissociation of H$_2$.

The ro-vibrational cooling rates are from \citet{GloverAbel2008} for an ortho-to-para ratio of 3:1 (i.e. $x_{\mathrm{p}} = 0.25$ and $x_{\mathrm{o}} = 0.75$ in their notation). The rates from collisions with \ion{H}{I}, \ion{H}{II}, and \ion{He}{I} are described in their equation 30 and for collisions with H$_2$ in their equation 32. For clarity, we repeat the ro-vibrational cooling rates from collisional excitation by electrons, because we believe their equation 36 is missing the `$+$' symbol and should be

\begin{eqnarray}
   \log \frac{\Lambda_{\mathrm{pH_2},e^-}}{\mathrm{erg\,cm^{3}\,s^{-1}}} &=& \log e^{-509.85\,\mathrm{K}/T} + \sum_{i=0}^5 a_i\,\log(T_3)^i \\
    \log \frac{\Lambda_{\mathrm{oH_2},e^-}}{\mathrm{erg\,cm^{3}\,s^{-1}}} &=& \log e^{-845\,\mathrm{K}/T} + \sum_{i=0}^5 a_i\,\log(T_3)^i \;,
\end{eqnarray}

\noindent
for para-H$_{\mathrm{2}}$ and ortho-H$_{\mathrm{2}}$, respectively, and with the fitting coefficients, $a_i$, from their table 7.

In \citet{chimes2014optthin} the cooling rate from collisional dissociation of H$_2$ is referenced as table 7 from \citet{GloverJappsen2007}. The rate coefficient $k_9$ has since been updated in \textsc{chimes} to that from \citet{GloverAbel2008}, which includes a transition from low-density gas with all H$_2$ in the vibrational ground state to high-density gas with all H$_2$ level populations in local thermodynamic equilibrium (LTE). 

\begin{table}
    \caption{Overview of all individually tabulated heating channels.}
    \centering    
    \begin{tabular}{llll}
    \hline
    Index & Identifier &  Comment\\
    \hline
    0 - 10  & Elements (H to Fe)  &  see Section~\ref{sec:heatingelement}\\
   11 & OtherA          &  $= 0$; not included, see Section~\ref{sec:heatingelement}\\
   12 & H2              & see Section~\ref{sec:heatingH2}  \\
   13 & COdiss          & $=0$; not included, see Section~\ref{sec:heatingobsolete} \\
   14 & CosmicRay       & $= 0$; included in Elements, see Section~\ref{sec:heatingelement}  \\
   15 & UTA             & $=0$; not included, see Section~\ref{sec:heatingobsolete}\\
   16 & line            & $=0$; not included, see Section~\ref{sec:heatingobsolete}\\
   17 & Hlin            & $=0$; not included, see Section~\ref{sec:heatingobsolete} \\
   18 & ComptonUI       & was `ChaT' in \citetalias{PS20}, see Section~\ref{sec:heatingCompton} \\
   19 & HFF              & $=0$; not included, see Section~\ref{sec:heatingobsolete} \\
   20 & ComptonCMB      & was `Compton' in \citetalias{PS20}, see Section~\ref{sec:heatingCompton} \\
   21 & Dust            & see Section~\ref{sec:PEheating}   \\
   22 & TotalPrim       &  Sum of (1), (2), (12), (18), (20)  \\
   23 & TotalMetal      &  Sum of (3-10), (19)   \\
    \hline
    \end{tabular}
    \label{tab:heatoverview}
\end{table}

\subsubsection{Cooling from molecules including metal species}\label{sec:coolingmol}

The cooling channel `molecules' is the sum of the cooling rates from CO, H$_2$O, and OH.

The CO and H$_2$O rotational and vibrational cooling rates are from \citet{Glover2010} and include cooling through collisions with H$_2$, \ion{H}{I}, and electrons. The fits to their cooling rates transition from the low-density limit to the LTE limit and depend on gas temperature, gas density, and the effective optical depth, parametrized as the effective column density per unit velocity, $\tilde{N}_x$, for molecule $x$. As in \citet{chimes2014optthin}, we use the approximation for static gas 

\begin{eqnarray}
    \tilde{N}_{\mathrm{CO}} &=& \left(\frac{n_{\mathrm{CO}}}{n_{\mathrm{H}}}\right )\frac{N_{\mathrm{sh}}}{\sigma} \nonumber \\
     \tilde{N}_{\mathrm{H_2O}} &=& \left(\frac{n_{\mathrm{H_2O}}}{n_{\mathrm{H}}}\right ) \frac{N_{\mathrm{sh}}}{\sigma} \;,  
\end{eqnarray}

\noindent
with the thermal velocity dispersion, $\sigma = \sqrt{3k_{\mathrm{B}}T / (\mu m_{\mathrm{H}})}$, the shielding column density, $N_{\mathrm{sh}}$, from Section~\ref{sec:Nsh}, and the CO and H$_2$O abundances from the equilibrium abundance tables (Section~\ref{sec:equilibriumabundances}).

The rotational cooling rates of OH are from \citet{HollenbachMcKee1979} and depend on temperature, density, OH column density, $N_{\mathrm{OH}} = (n_{\mathrm{OH}}/n_{\mathrm{H}})\,N_{\mathrm{sh}}$, and on the dust extinction, $A_{\mathrm{V}}$, from equation~(\ref{eq:AV}).

\begin{table}
    \caption{The metal-line cooling rates per ion in \textsc{chimes} are from \citet{OppenheimerSchaye2013} and depend solely on gas temperature, $T$. The rates of a few individual species (column 1) depend on additional gas properties (column 2) for low temperatures (column 3) and have therefore been updated (see column 4 for reference). }
    \begin{center}
    \begin{tabular}{llcl}
        \hline
        Species & Dependence  & Temperatures & Reference \\
        \hline
        \ion{C}{I}  & $T$, $n_{\mathrm{e}}$, $n_{\mathrm{HI}}$, $n_{\mathrm{HII}}$ &$<10^4\,\mathrm{K}$   &\citet{chimes2014optthin}\footnotemark[1]\\
        \ion{C}{II} & $T$, $n_{\mathrm{e}}$ &$<10^5\,\mathrm{K}$   & \citet{chimes2014optthin}\footnotemark[2]\\
        \ion{N}{II} & $T$, $n_{\mathrm{e}}$  &$<10^5\,\mathrm{K}$   & \citet{chimes2014optthin}\footnotemark[2]\\
        \ion{O}{I}  & $T$, $n_{\mathrm{e}}$, $n_{\mathrm{HI}}$, $n_{\mathrm{HII}}$ &$<10^4\,\mathrm{K}$   & \citet{GloverJappsen2007}\\
        \ion{Si}{II}  & $T$, $n_{\mathrm{e}}$  &$<10^5\,\mathrm{K}$ &\citet{chimes2014optthin}\footnotemark[2]\\
        \ion{Fe}{II}  & $T$, $n_{\mathrm{e}}$  &$<10^5\,\mathrm{K}$ &\citet{chimes2014optthin}\footnotemark[2]\\
        \hline
    \end{tabular}
    \end{center}
    {\footnotesize{$^1$Based on \citet{GloverJappsen2007} but including additional energy levels.\\ $^2$Based on the \textsc{CHIANTI} database version 7.1 \citep{DereChianti1997, LandiChianti2013}}}
    \label{tab:thermalprocesses}
\end{table}

\subsubsection{HD cooling}\label{sec:coolingHD}

\textsc{chimes} does not include deuterium chemistry. As in \citetalias{PS20}, we use the HD cooling rate based on the fit by \citet{Flower2000} and assume a constant HD abundance relative to H$_2$ of $n_{\mathrm{HD}}/n_{\mathrm{H_2}} = 1.65\times10^{-5}$.

\subsubsection{Compton processes} \label{sec:coolingCompton}
In \citetalias{PS20} the cooling (and heating) channel `Compton' is the sum of contributions from all radiation fields. In this work, we divide the total rates from Compton scattering processes into two cooling entries: (1) the energy transfer between photons from the cosmic microwave background (CMB) and electrons (`ComptonCMB') and  (2) the energy transfer between electrons and photons from both the UVB and the ISRF (`ComptonUI').  The contribution from the CMB has a simple analytic form with a steep redshift dependence (see e.g. equation A3 in \citetalias{PS20}) and is therefore typically directly added in the hydro code. We calculate the Compton cooling (and heating) rates separately as follows: The non-relativistic cooling and heating rate through Compton processes in units of $\mathrm{erg\,cm^3\,s^{-1}}$ is

\begin{equation}\label{eq:Comptonnonrel}
    \frac{\Lambda_{\mathrm{Compton,non-rel}}}{n_{\mathrm{H}}^2} \,= \frac{4\pi n_{\mathrm{e}}}{n_{\mathrm{H}}^2}\frac{\sigma_{\mathrm{T}}}{m_{\mathrm{e}}c^2} \int_0^{\infty} \mathrm{d}\nu J_{\nu} (h\nu - 4k_{\mathrm{B}}T)
\end{equation}

\noindent
(see e.g. equation 3 in \citealp{ME99}), with the specific intensity of the radiation field, $J_{\nu}$, at frequency, $\nu$, the electron mass, $m_{\mathrm{e}}$, the speed of light, $c$, the Planck constant, $h$, and the Thompson cross section, $\sigma_{\mathrm{T}}$. Because our radiation fields include photons with energies above $100\,\mathrm{keV}$, we generalize the above equation by using the relativistic Klein-Nishina cross-section, $\sigma_{\mathrm{KN}}$, which approaches the classical Thompson cross section, $\sigma_{\mathrm{T}}$, for small photon energies ($\eta = h\nu / (m_{\mathrm{e}}c^2) \ll 1$) and is given by \citet{Blumenthal1974}:

\begin{align}\label{eq:sigmaKN}
    \sigma_{\mathrm{KN}} &= \frac{3\sigma_{\mathrm{T}}}{8\eta} f_{\mathrm{KN}}(\eta)\\
    f_{\mathrm{KN}}(\eta) & =\left [\frac{2\eta(1+\eta)}{(1+2\eta)^2} + \left (1- \frac{2}{\eta} - \frac{2}{\eta^2}\right)\mathrm{ln}(1+2\eta) + \frac{4}{\eta}\right ] \; .
\end{align}

\noindent
Because $\sigma_{\mathrm{KN}}$ depends on the dimensionless photon energy $\eta$, we express equation~(\ref{eq:Comptonnonrel}) in terms of $\eta$ with $J_{\eta} = J_{\nu} m_{\mathrm{e}} c^2 / h$ and use the general cross section $\sigma_{\mathrm{KN}}$ instead of $\sigma_{\mathrm{T}}$ to account for relativistic effects:

\begin{align}
    \frac{\Lambda_{\mathrm{Compton,rel}}}{n_{\mathrm{H}}^2} = 3\pi\sigma_{\mathrm{T}} \frac{n_{\mathrm{e}}}{n_{\mathrm{H}}^2} & \left [ \frac{1}{2} \int_0^{\infty} J_{\mathrm{\eta}} f_{\mathrm{KN}} (\eta) \mathrm{d}\eta -  \right . \\
    & \left . 2\frac{k_{\mathrm{B}}T}{m_{\mathrm{e}}c^{2}} \int_0^{\infty} \frac{J_{\mathrm{\eta}}}{\eta} f_{\mathrm{KN}} (\eta) \mathrm{d}\eta \right ]
\end{align}

\noindent
The integrals $\int_0^{\infty} J_{\mathrm{\eta}} f_{\mathrm{KN}} (\eta) \mathrm{d}\eta$ and $\int_0^{\infty} \frac{J_{\mathrm{\eta}}}{\eta} f_{\mathrm{KN}} (\eta) \mathrm{d}\eta$ are tabulated for each included radiation field.

If the energy transfer between photons and charged particles leads to a net cooling of the gas, the rate is added to the respective Compton cooling channels, while for a net gas heating, the rate is added to the respective Compton heating channels (Section~\ref{sec:heatingCompton}). 

\paragraph*{Caution when upgrading from \citetalias{PS20}:} We replace the now redundant cooling channel `NetFFM' (see Section~\ref{sec:coolingelement}) with the new `ComptonUI' cooling rates. The original cooling channel `Compton' has been renamed to `ComptonCMB' and only includes the contribution from CMB photons. 

\subsubsection{Dust cooling}\label{sec:coolingdust}

We include cooling by dust grains as in \textsc{chimes}. Gas can transfer energy to dust grains by inelastic gas-grain collisions (table 7 of \citealp{GloverJappsen2007}, important only at densities $n_{\mathrm{H}}\gtrsim 10^4\,\mathrm{cm}^{-3}$). The rates for this process depend on the dust grain temperature, which we assume to be constant ($T_{\mathrm{dust}} = 10\,\mathrm{K}$). 
In addition, gas may cool through the recombination of free electrons and ions on the surface of dust grains (\citealp{GloverJappsen2007}, originally from \citealp{Wolfire1995}). This process is mainly important for high temperatures ($T\gtrsim10^4\,\mathrm{K}$).

\subsubsection{Heating from atoms and ions per element}\label{sec:heatingelement}

Each element, $x$, with $x =$ H, He, C, N, O, Ne, Si, Mg, S, Ca, and Fe is heated through photoheating by both the UVB and the ISRF radiation fields (see Section~\ref{sec:ISRF}). Both radiation fields are split into extreme ultraviolet (EUV, hydrogen ionizing radiation, $E>13.6\,\mathrm{eV}$) and far ultraviolet (FUV, $E<13.6\,\mathrm{eV}$) radiation. Per element, we add their respective photo-heating rates plus the heating from cosmic rays for all their atom and ion species.

For each individual species $i$ (e.g. $i$ = \ion{C}{II}), the photoheating rate  is 

\begin{equation}
    \Lambda_{\mathrm{heat,}i}\,[\mathrm{erg\,cm^{-3}\,s^{-1}}] = \Gamma_i\, \langle \epsilon_{i} \rangle  \, n_{i}\;,
\end{equation}
with the photoionization rate $\Gamma_i\,[\mathrm{s}^{-1}]$, the average excess energy of the ionizing photons, $\langle \epsilon_{i} \rangle $, and the species number density, $n_{i}\,[\mathrm{cm}^{-3}]$. 

\paragraph*{EUV:}

For species that are ionized by EUV radiation, the intrinsic photoionization rate may be reduced due to shielding of \ion{H}{I}, H$_2$, \ion{He}{I}, and \ion{He}{II}. In addition, the average excess energy, $\langle \epsilon_{i} \rangle $, can increase due to the change of the shape of the partially absorbed UV spectrum. 

The reduction of the photoionization rate due to shielding of \ion{H}{I}, H$_2$, \ion{He}{I}, and \ion{He}{II} is $\Gamma_{i,\mathrm{thick}}/\Gamma_{i,\mathrm{thin}}$ and follows equation~3.8 in \citet{chimes2014shielded}. The shielded average excess energy, $\langle \epsilon_{\mathrm{i,thick}} \rangle $, is from equation 3.22 in \citet{chimes2014shielded}. The integrals in these equations depend on the shielding column densities of \ion{H}{I}, H$_2$, \ion{He}{I}, and \ion{He}{II} and are pre-tabulated within the \textsc{chimes} datafiles for each species, $i$. The shielding column densities are the product of $N_{\mathrm{sh}}$ from equation~(\ref{eq:Nshcut}) and the abundance of species $i$, $n_{i}\,/\,n_{\mathrm{H}}$, either in chemical equilibrium (Section~\ref{sec:equilibriumabundances}) or non-equilibrium (Section~\ref{sec:hybrid}).

\paragraph*{FUV:}
The photoionization rate for FUV radiation of species $i$ is reduced by dust shielding, 

\begin{equation}\label{eq:dustshielding}
   \frac{ \Gamma_{i,\mathrm{thick}}}{\Gamma _{i,\mathrm{thin}}} = e^{-\gamma_i A_{\mathrm{V}}} \; ,
\end{equation}

\noindent
and depends on the dust extinction, $A_{\mathrm{V}}$ (equation~\ref{eq:AV}), and a species dependent prefactor, $\gamma_i$, that is stored in the main \textsc{chimes} data file (see section 3.1 in \citealp{chimes2014shielded} for details). The average excess energies for the FUV range do not depend on the shielding column density because the spectral shape is assumed to remain constant with the FUV energy range (`grey approximation'). For FUV, $\langle \epsilon_{i} \rangle$ only depends on the intrinsic spectral shape of the radiation field and is tabulated in the cross section file of each radiation field within \textsc{chimes}.

\paragraph*{Cosmic rays:}
The primary ionization rate of atomic hydrogen due to cosmic rays, $\zeta_{\mathrm{H}}$, is an input parameter of the model and is described in Section~\ref{sec:CR}. The ionization rate of all other species scales linearly with $\zeta_{\mathrm{H}}$ and depends on the effective number of electrons in the outer shell and the ionization energy of the species. Secondary ionization, i.e. ionization by electrons that were released through the primary CR ionization process, is included for \ion{H}{I} and \ion{He}{I} ionization (see section 2.3 in \citealp{chimes2014optthin} for details).  

We add the heating rates due to CR ionization from all species of an element to the heating rate per element. For example, the `Carbon' heating channel includes the photoheating rates of all C species (atomic carbon and carbon ions) from both EUV and FUV radiation of all included radiation fields and the heating rate from CR ionization.

\paragraph*{Caution when upgrading from \citetalias{PS20}:} 
The heating from cosmic ray ionization is a separate heating channel in \citetalias{PS20} while in this work the total heating from cosmic rays is distributed into the heating channels of each element. This simplifies the scaling of the individual heating rates for different element abundances. As for the cooling, no heating contributions from elements other than the individually traced elements are included. The heating channel `OtherA', which was used in \citetalias{PS20} for all other elements, is therefore here set to zero.

\subsubsection{H$_2$ heating}\label{sec:heatingH2}

Molecular hydrogen may heat the gas through photo-dissociation by the included radiation fields,  UV pumping, and the formation of H$_2$ (gas phase and on dust grains). Radiation at the relevant energies is absorbed mainly by dust and through self-shielding of H$_2$. Dust shielding is included following equation~(\ref{eq:dustshielding}) with tabulated values of $\gamma_{\mathrm{i}}$ for the photo-dissociation rates. The self-shielding of H$_2$ is approximated by the fitting function in \citet{chimes2014shielded} (their equation 3.12), which depends on the gas temperature, the H$_2$ column density and a parameter that describes the Doppler broadening of the Lyman-Werner bands. The Doppler broadening parameter, $b$, with 

\begin{equation}
    b = \sqrt{b^2_{\mathrm{therm}} + b^2_{\mathrm{turb}}} = \sqrt{\frac{2k_{\mathrm{B}}T}{m_{\mathrm{H_2}}} + b^2_{\mathrm{turb}}} \;,
\end{equation}

\noindent
includes a thermal ($b_{\mathrm{therm}}$) and a turbulent ($b_{\mathrm{turb}}$) component. Unresolved turbulence is an input parameter and in our model expressed as the 1D turbulent velocity dispersion, $v_{\mathrm{turb}}$ (see Section~\ref{sec:Nref}) with $b_{\mathrm{turb}} = \sqrt{2} v_{\mathrm{turb}}$.

\subsubsection{Compton heating}\label{sec:heatingCompton}

This heating channel includes the rates of Compton processes described in Section~\ref{sec:coolingCompton} for net heating. As for cooling, the contribution from the CMB is tabulated separately in the channel `ComptonCMB' while the contributions of photons from both the UVB and the ISRF are combined in the channel `ComptonUI'. 

\paragraph*{Caution when upgrading from \citetalias{PS20}:} The new channel `ComptonUI' replaces the no longer used charge transfer (`ChaT') heating channel.

\subsubsection{Photo-electric heating by dust}\label{sec:PEheating}

Photo-electric (PE) heating, i.e. the absorption of FUV photons by dust grains and the subsequent electron ejection, is an import heating mechanism in the neutral (CNM, WNM) phases of the ISM. Different functional forms exist in the literature that estimate the PE heating rate as a function of the local FUV intensity, assuming a specific radiation field spectrum and dust grain size and shape distribution.

The PE heating rate in \textsc{chimes}\footnote{In \citet{chimes2014optthin}, the PE heating rate in \textsc{chimes} is described as using equations 1 and 2 from \citet{Wolfire1995}, but this rate has since been updated in \textsc{chimes} (see section 2.2 in \citealp{Richings2018ChimesPEupdate}).} follows equations 19 and 20 from \citet{Wolfire2003} and is defined as

\begin{equation}\label{eq:PER+14}
    \Gamma_{\mathrm{PE}} = 1.3 \times 10^{-24}\,\mathrm{erg\,s^{-1}} \, \epsilon \, G_0 \, \mathrm{Z_d'}\;,
\end{equation}

\noindent
with the heating efficiency (i.e. the fraction of FUV radiation absorbed by the dust grains and converted to gas heating), $\epsilon$,

\begin{equation}\label{eq:epsilonBT}
    \epsilon = \frac{4.9\times10^{-2}}{\left [ 1 + 4\times10^{-3} \,\psi^{0.73} \right]} + \frac{3.7\times10^{-2} \left ( \frac{T}{10^4\,\mathrm{K}}\right )^{0.7}}{\left [ 1+ 2\times 10^{-4}\,\psi \right ]}
\end{equation}

\noindent
the grain charging parameter, $\psi$, 

\begin{equation}\label{eq:psiBT}
\psi = \frac{G_0 \left ( \frac{T}{1\,\mathrm{K}}\right)^{1/2}}{\left ( \frac{n_{\mathrm{e}}}{1\,\mathrm{cm^{-3}}}\right ) \phi_{\mathrm{PAH}}}\;,
\end{equation}

\noindent
and a parameter $\phi_{\mathrm{PAH}}=0.5$, which scales the electron-PAH (polycyclic aromatic hydrocarbons) collision rates (see \citealp{Wolfire2003} for details). 
The strength of the UV radiation field, $G_0$, is expressed as the integrated energy flux, $u\,[\mathrm{erg\,cm^{-2}\,s^{-1}}]$, in the energy range between $6\,\mathrm{eV}$ and $13.6\,\mathrm{eV}$ relative to the radiation field from \citet{Habing1968} ($G_0 = u / [1.6\times10^{-3}\,\mathrm{erg\,cm}^{-2}\,\mathrm{s}^{-1}]$).
The photoelectric heating rate is furthermore assumed to scale linearly with the dust content relative to the local ISM value ($Z_{\mathrm{d}}' = [\mathcal{DTG} / \mathcal{DTG}_{\mathrm{ISM}}]$, see Section~\ref{sec:dust}).

\paragraph*{Dust extinction}

The extinction of the FUV flux through a shielding column (see Section~\ref{sec:Nsh}) is taken into account by reducing the incident radiation field by a factor that depends on the visual extinction, $A_{\mathrm{V,Chimes}}$ (equation~\ref{eq:AVchimes}), with 

\begin{equation}\label{eq:dustextinction}
    G_0 = G_{\mathrm{0,incident }} \times e^{-2.77 A_{\mathrm{V,Chimes}}} \; 
\end{equation}

\noindent
in the calculation of the PE heating rate. 
For reference, for the dust model in this work, the radiation field strength is $0.92 \times G_{\mathrm{0,incident}}$ and $0.41 \times G_{\mathrm{0,incident}}$ for total hydrogen shielding column densities of $N_{\mathrm{sh}} = 10^{20}\,\mathrm{cm}^{-2}$ and $10^{21}\,\mathrm{cm}^{-2}$, respectively. For $N_{\mathrm{sh}} = 5\times10^{21}\,\mathrm{cm}^{-2}$, only 1 per cent of the incident FUV radiation remains.

\subsubsection{No longer included heating processes}\label{sec:heatingobsolete}

Compared to \citetalias{PS20}, we no longer include heating from the dissociation of CO (`COdiss'), unresolved transition array heating (`UTA'), heating due to induced line absorption of the continuum (`Hlin'), the iso-sequence line heating (`Hlin;'), and the heating from Bremsstrahlung (`HFF'), if Bremsstrahlung has a net heating effect. These heating channels were included in \citetalias{PS20} for completeness, because they were a standard output in \textsc{cloudy}, but none contribute significantly to the total heating rate. 

\begin{figure*}
    \centering
    \includegraphics[width=\linewidth]{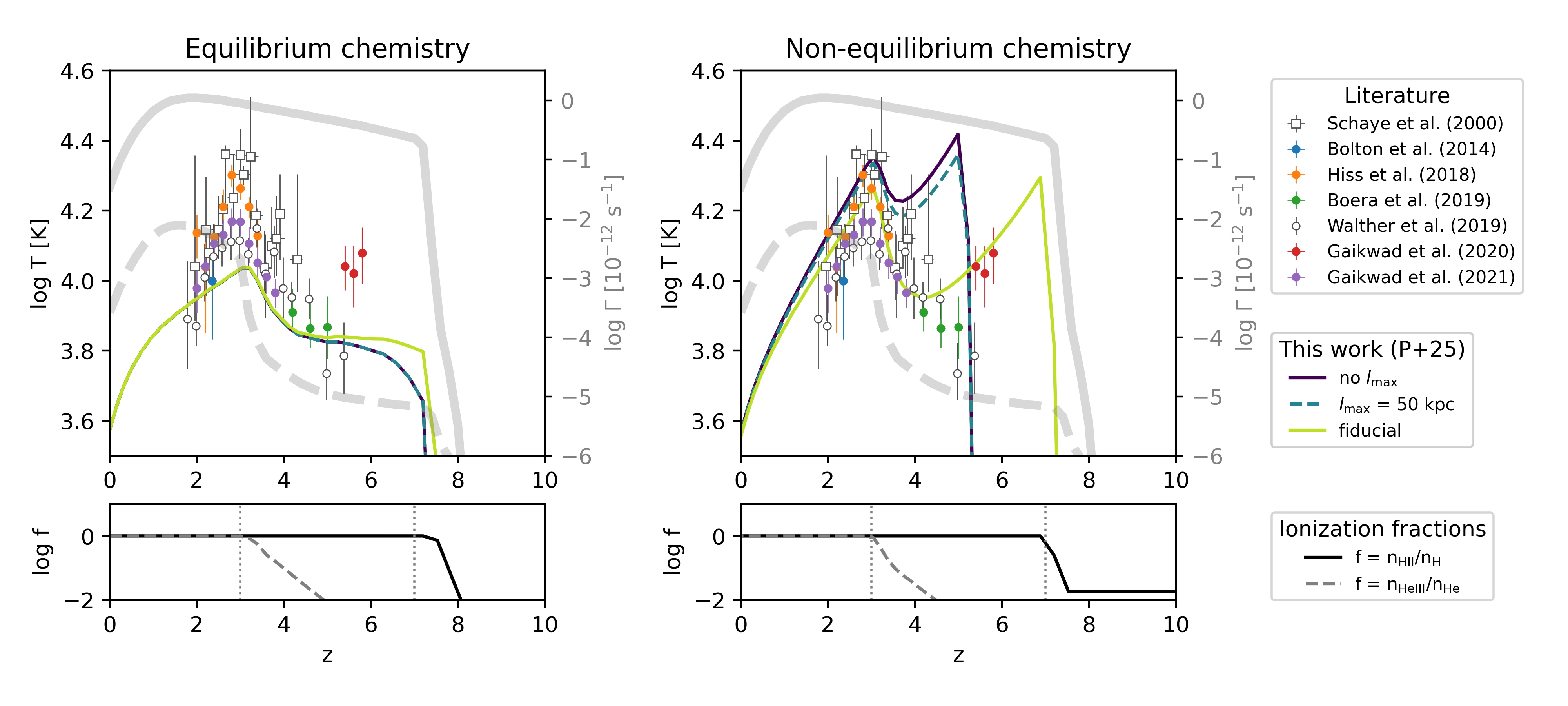}
    \caption{Thermal evolution for gas at the cosmic mean baryon density (large panels) for equilibrium (left plot) and non-equilibrium chemistry (right plot), and for different prescriptions of the maximum shielding length, $l_{\mathrm{max}}$ (constant $l_{\mathrm{max}}$ of $50\,\mathrm{kpc}$, the fiducial redshift-dependent $l_{\mathrm{max}}$ from equation~\ref{eq:lshmax}, and no limit on the shielding length, see middle legend). The HI (thick grey solid line) and HeII ionization rates (thick grey dashed line) from the background radiation field (figure B2 from \citetalias{PS20}) are over-plotted (right y-axis) in the large panels. Data from \citet{Schaye2000thermalhistory}, \citet{Walther2019}, \citet{Boera2019}, \citet{Bolton2014}, \citet{Hiss2018}, \citet{Gaikwad2020}, and \citet{Gaikwad2021} are added with various symbols (see top legend). The small panels show the species fractions $n_{\mathrm{HII}}/n_{\mathrm{H}}$ and $n_{\mathrm{HeIII}}/n_{\mathrm{He}}$ for the model with the fiducial parameter values. The vertical, dotted lines indicate redshifts $z=3$ and $z=7$, at which helium and hydrogen are completely ionized at the cosmic mean density, respectively.}
    \label{fig:thermevol}
\end{figure*}

\section{Results}\label{sec:results}

The model for radiative cooling described in Section~\ref{sec:method} was developed to represent gas in any environment that is present in a cosmological galaxy formation and evolution simulation, from the densities and temperatures typical for the ISM to those characteristic of the circum- and intergalactic medium. Here, we evaluate the model for a few selected conditions: (1) the thermal evolution of primordial gas at the cosmic mean density, (2) the warm and cold neutral medium at solar metallicity in MW-mass galaxies, and (3) the transition from atomic to molecular gas in solar metallicity gas.

\subsection{Thermal evolution at the cosmic mean baryon density}\label{sec:thermevol}

Following the thermal evolution and ionization levels of primordial gas with the cosmic mean baryon density provides insight into the efficacy of hydrogen and helium reionization. We use the \textsc{ConstantCosmoTempEvolution} example within \textsc{swift} that evolves a cosmological box filled with homogeneous gas at the mean baryon density to $z=0$ without gravity. This setup includes adiabatic cooling from the expansion of the universe and radiative cooling and heating as described in Section~\ref{sec:method}.

We use the same cosmological parameters from the Dark Energy Survey \citep{Desy2022} as in the FLAMINGO simulations \citep{Flamingo2023}, with the cold dark matter fraction, $\Omega_{\mathrm{CDM}}=0.256011$, the baryon fraction, $\Omega_{\mathrm{b}} = 0.0486$, the dark energy fraction, $\Omega_{\Lambda} = 0.693922$, and the Hubble constant, $H_{\mathrm{0}} = 68.1\,\mathrm{km\,s}^{-1}\,\mathrm{Mpc}^{-1}$. For this cosmology, the cosmic mean baryon density evolves from $\log n_{\mathrm{H,mean}} [\mathrm{cm}^{-3}]= -3.7$ at $z=9$ to $\log n_{\mathrm{H,mean}} [\mathrm{cm}^{-3}] = -6.7$ at $z=0$ (for $X_{\mathrm{H}} = 0.756$). 

The shielding length estimate used in this work is based on the Jeans length, which is derived from assuming an equality between the free-fall time and the sound-crossing time. For the cosmic mean baryon density and a temperature of $10^4\,\mathrm{K}$, the Jeans length is $32\,\mathrm{kpc}$ (physical),  $320\,\mathrm{kpc}$ (comoving) at $z =9$ and $1.02\,\mathrm{Mpc}$ at $z=0$, each with a sound crossing time that is larger than the corresponding Hubble times. The equilibrium assumption is therefore invalid for these low densities which is the reason for limiting the shielding length, $l_{\mathrm{sh}} = N_{\mathrm{sh}}/n_{\mathrm{H}}$ to the redshift-dependent maximum value, $l_{\mathrm{max}}$, from equation~\ref{eq:lshmax}.

Fig.~\ref{fig:thermevol} shows the evolution of the gas temperature (large panels) and ion fractions (black solid line: $n_{\mathrm{HII}}\,/\,n_{\mathrm{H}}$, grey dashed line: $n_{\mathrm{HeIII}} \,/\,n_{\mathrm{He}}$) at the cosmic mean baryon density for equilibrium (left plot) and non-equilibrium chemistry (right plot). The temperature evolution is shown for different models (see legend) while the ion fractions are only shown for the fiducial model. 

In the case of equilibrium chemistry (left plot of Fig.~\ref{fig:thermevol}) the background radiation field ionizes hydrogen at the mean baryonic density around $z=7.5$ (solid line, bottom panel) as expected by the sharp increase of the \ion{H}{I} ionization rate (thick grey solid line, right y-axis) of the assumed background radiation field. Because the photoheating rates of hydrogen depend on the abundance of neutral hydrogen, the unrealistic instantaneous increase of the ionized fraction implies an underestimated photo-heating rate (e.g. \citealp{Puchwein2015}). As a result, the gas temperature does not increase to temperatures above $10^4\,\mathrm{K}$ for any of the considered models (colored lines) at hydrogen reionization, in disagreement with data from \citet{Gaikwad2020}. Similarly, during \ion{He}{II} reionization (\ion{He}{II} ionization rate, thick grey dashed line, right y-axis), the gas temperature is underestimated in all models compared to observations. This is a known issue when using equilibrium chemistry models (see also \citealp{Gaikwad2021}) and was solved in e.g. the \textsc{eagle} \citep{EagleSchaye2015} simulations by injecting additional heat during the reionization epochs of \ion{H}{I} and \ion{He}{II}. 

For non-equilibrium chemistry (right plot of Fig.~\ref{fig:thermevol}), the temperature is sensitive to the assumed maximum shielding length. Larger shielding columns decrease the photoheating and the photoionization rates and without an additional limit on the shielding length (purple, solid line) or a constant maximum shielding length of $l_{\mathrm{max}} = 50\,\mathrm{kpc}$: blue dashed line) reionization is noticeably delayed, from $z\approx7.5$ (as expected from the background radiation field) to $z\approx5$. Limiting the shielding length to smaller values before and during reionization to $l_{\mathrm{max}} = 10\,\mathrm{kpc}$, as in the fiducial model, largely avoids this delay. Here, hydrogen reionization is complete at $z\approx7$ and helium reionization is complete at $z\approx3$ (vertical lines in the bottom panels of Fig.~\ref{fig:thermevol}).

\subsection{Thermal equilibrium in the ISM}\label{sec:neutral}

\subsubsection{Thermal equilibrium in the fiducial model}\label{sec:neutral:fid}

The neutral atomic ISM can be described by a thermally stable warm and a cold phase that co-exist in pressure equilibrium \citep{Field1969, Wolfire1995, Wolfire2003}. The physical properties (i.e. the gas temperatures and densities) of the warm neutral medium (WNM) and the cold neutral medium (CNM) can be estimated by exploring the thermal equilibrium pressure, $P_{\mathrm{eq}}$, as a function of the gas density, $n_{\mathrm{H}}$, defined as  

\begin{equation}\label{eq:Peq}
    P_{\mathrm{eq}}/k_{\mathrm{B}}  = n\,T_{\mathrm{eq}} = T_{\mathrm{eq}} \,n_{\mathrm{H}} / (X_{\mathrm{H}}\,\mu_{\mathrm{eq}})\;,
\end{equation}

\noindent
for the thermal equilibrium temperature, $T_{\mathrm{eq}}$,
at which radiative heating and cooling processes (e.g. those described in Section~\ref{sec:processes}) are in balance, i.e. $\Lambda_{\mathrm{cool}} \,(T_{\mathrm{eq}}) = \Lambda_{\mathrm{heat}} \,(T_{\mathrm{eq}})$. The factor $\mu_{\mathrm{eq}}$ is the mean particle mass in units of the proton mass at thermal equilibrium, .

\begin{figure}
    \centering
    \includegraphics[width=\linewidth]{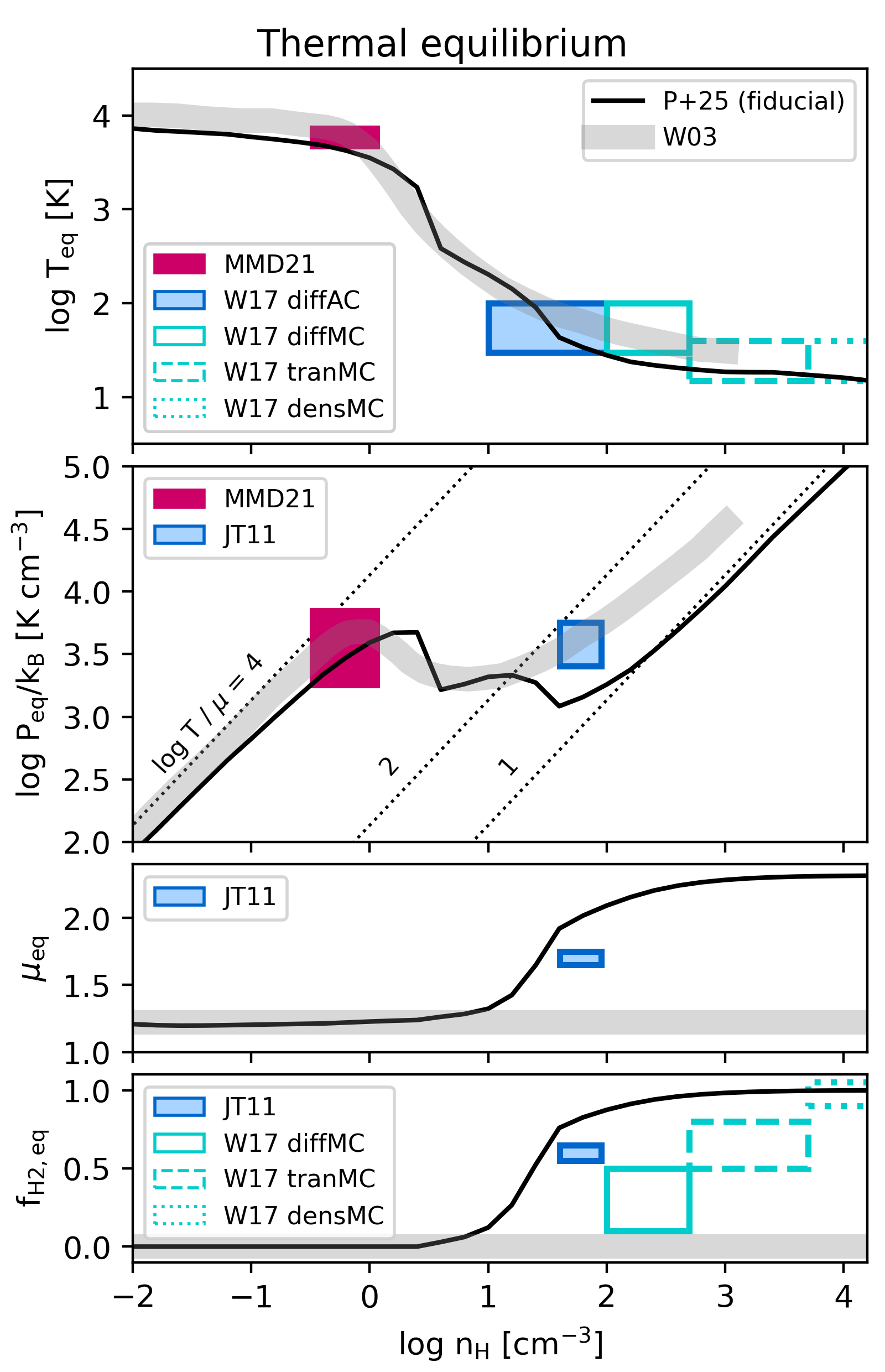}
    \caption{Physical properties of gas in thermal equilibrium for the fiducial model with the parameter values from Table~\ref{tab:fiducialparameter} (solid black line) compared to selected data and models from the literature. The individual panels show the gas temperature, $T_{\mathrm{eq}}$ (first panel), the thermal pressure, $P_{\mathrm{eq}}$ (second panel), the mean particle mass, $\mu_{\mathrm{eq}}$ (third panel), and the molecular hydrogen fraction, $f_{\mathrm{H2}}$ (fourth panel), each for thermal equilibrium. The fiducial model from \citet{Wolfire2003} for the solar neighborhood and for a constant shielding column density of $10^{19}\,\mathrm{cm}^{-2}$ is shown as a thick grey band. The colored rectangles refer to measurements or definitions by \citet{Marchal2021} (MMD21) for the WNM, \citet{Jenkins2011} (JT11) and \citet{Wakelam2017} (W17 diffAC, diffuse atomic clouds) for the CNM, and \citet{Wakelam2017} for diffuse molecular clouds (W17 diffMC), translucent molecular clouds (W17 tranMC), and dense molecular clouds (W17 densMC) in the Milky Way Galaxy. }
    \label{fig:peqfiducial}
\end{figure}

The thick grey line in the second panel of Fig.~\ref{fig:peqfiducial} is the $P_{\mathrm{eq}}\,(n_{\mathrm{H}})$ curve for the fiducial model in \citet{Wolfire2003} for the solar neighborhood and a constant shielding column density of $10^{19}\,\mathrm{cm}^{-2}$. This function consists of two parts with $\mathrm{d}P_{\mathrm{eq}}\,/\,\mathrm{d}n_{\mathrm{H}}>0$: one at low densities ($n_{\mathrm{H}}\lesssim 1\,\mathrm{cm}^{-3}$) and one at high densities ($n_{\mathrm{H}}\gtrsim 10\,\mathrm{cm}^{-3}$), both of which describe a thermally stable density regime. At these densities, gas with $P<P_{\mathrm{eq}}$ is heated, increasing the thermal pressure which causes expansion and therefore a density decrease. On the other hand, gas with $P>P_{\mathrm{eq}}$ cools and increases its density. In both cases, equilibrium can be restored ($P\rightarrow P_{\mathrm{eq}}$). At low densities, the dynamically stable track corresponds to temperatures of $T/\mu \approx 10^4\,\mathrm{K}$ (thin black dotted lines show constant $T/\mu$), while at high densities, the second thermally stable density range corresponds to temperatures of $T/\mu \approx 50\,\mathrm{K}$ (CNM). For densities in between, we see $\mathrm{d}P_{\mathrm{eq}}\,/\,\mathrm{d}n_{\mathrm{H}}<0$, such that gas is thermally unstable. The pressures of the two thermally stable phases overlap and in this pressure range both phases may be in thermal pressure equilibrium. This theoretical result agrees with the observed general properties for the WNM (\citealp{Marchal2021}, MMD21) and CNM (\citealp{Jenkins2011}, JT11) in the Milky Way Galaxy.

\begin{figure*}
    \centering
    \includegraphics[width=\linewidth]{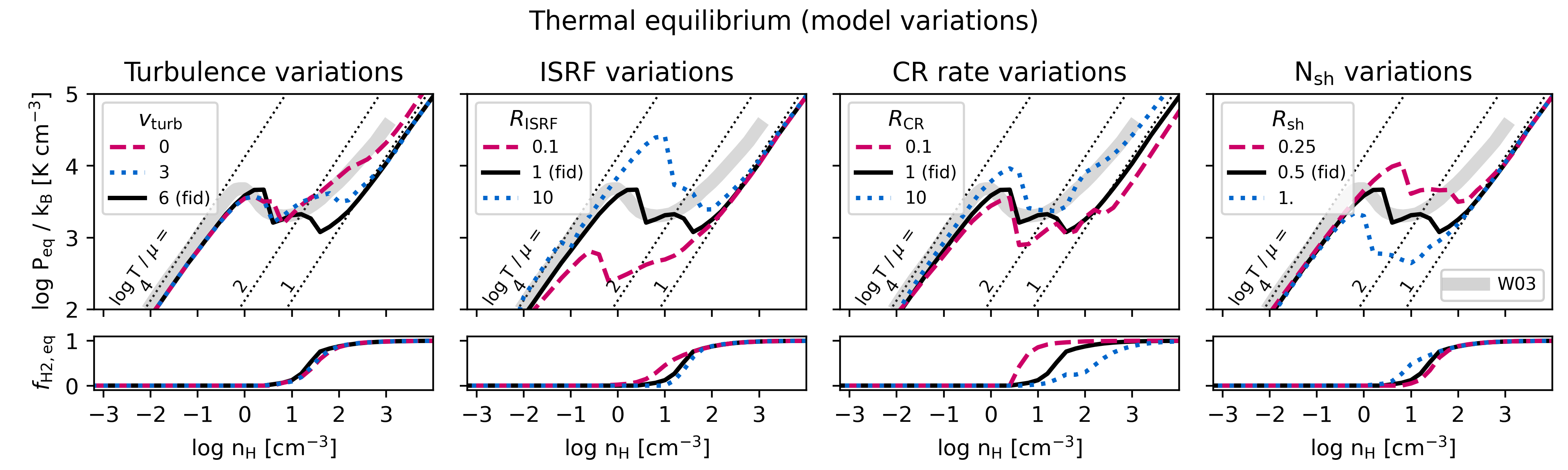}
    \caption{Thermal equilibrium pressure, $P_{\mathrm{eq}}$ (big panels, see equation~\ref{eq:Peq}), and thermal equilibrium molecular hydrogen fraction, $f_{\mathrm{H2,eq}}$ (small panels), of solar metallicity gas at $z=0$ for selected model parameter variations (see also Table~\ref{tab:WNM}). We show $P_{\mathrm{eq}}$ and $f_{\mathrm{H2,eq}}$ for a turbulent velocity dispersion of $v_{\mathrm{turb}} = 0\,,3\,,6\,\mathrm{km\,s}^{-1}$ (first column), an ISRF strength (equation~\ref{eq:ISRFIcut}) with normalizations $R_{\mathrm{ISRF}}=0.1\,,1\,, 10$ (second column), a CR rate (equation~\ref{eq:CRcut}) with normalizations $R_{\mathrm{CR}} = 0.1\,,1\,, 10$ (third column), and shielding column densities (equation~\ref{eq:Nshcut}) with normalizations of $R_{\mathrm{sh}} = 0.25\,,0.5\,, 1$ (fourth column). The fiducial model (black solid line, as in Fig.~\ref{fig:peqfiducial}) is repeated in each variation set. The thick grey line in the background is the pressure equilibrium line from \citet{Wolfire2003} for the solar neighborhood and for a constant shielding column density of $10^{19}\,\mathrm{cm}^{-2}$, as in Fig.~\ref{fig:peqfiducial}. The dotted lines show $\log T \,[\mathrm{K}]\,/\,\mu = 4$, $2$, and $1$ for reference.}
    \label{fig:pressurequilibrium}
\end{figure*}

For the WNM, \citet{Marchal2021} find values of $P_{\mathrm{WNM}}/k_{\mathrm{B}} = (4.4\pm 2.6)\times10^3\,\mathrm{K\,cm}^{-3}$, $n_{\mathrm{WNM}} = 0.74\pm0.41\,\mathrm{cm}^{-3}$, and $T_{\mathrm{WNM}} = (6.0\pm1.3)\times 10^3\,\mathrm{K}$ for the pressure, density, and temperature of WNM gas near the Local Bubble, based on 21cm data from the GHIGLS HI survey \citep{Martin2015}. The local CNM pressure is measured by \citet{Jenkins2011} from \ion{C}{I} absorption features of nearby stars as $\log P_{\mathrm{CNM}}/k_{\mathrm{B}} = 3.58\pm0.175\,\mathrm{K\,cm}^{-3}$. For their assumed H$_{\mathrm{2}}$ fraction ($f_{\mathrm{H2}} =0.6$, $\mu = 1.7$, third and fourth panels) and their reported mean CNM temperature of $80\,\mathrm{K}$, their CNM densities are between $n_{\mathrm{H}} = 40$ and $90\,\mathrm{cm}^{-3}$. In the review by \citet{Wakelam2017}, diffuse atomic clouds are characterized by densities of $n_{\mathrm{H}} = 10-100\,\mathrm{cm}^{-3}$, temperatures of $T = 30 - 100\,\mathrm{K}$, and H$_2$ fractions of $f_{\mathrm{H2}}<0.1$ (W17 diffAC, first panel).

We convert the equilibrium pressure from \citet{Wolfire2003} to $T_{\mathrm{eq}}$ (thick grey line in the first panel of Fig.~\ref{fig:peqfiducial}) with their factor $X_{\mathrm{H}}\,\mu = 0.9$. For $X_{\mathrm{H}} = 0.75$, the mean particle mass $\mu = 1.21$ (thick grey line in the third panel), which corresponds to a molecular fraction of $f_{\mathrm{H2}} = 0$ (thick grey line in the fourth panel).  

Temperatures, pressures, mean particle masses and molecular hydrogen fractions in thermal equilibrium with the fiducial parameter values from Table~\ref{tab:fiducialparameter} and for solar metallicity at $z=0$ are shown as black lines in Fig.~\ref{fig:peqfiducial}. The transition from the WNM to the CNM loosely follows the \citet{Wolfire2003} model but allows higher maximum WNM densities, by $\approx 0.5\,\mathrm{dex}$. Based on $P_{\mathrm{eq}}$, the minimum density of the CNM is $\log n_{\mathrm{CNM,min}} [\mathrm{cm}^{-3}]= 0.5$, which is minimally (by $\approx0.2\,\mathrm{dex}$) lower than in \citet{Wolfire2003}. The minimum and maximum pressures for pressure equilibrium between WNM and CNM are as in \citet{Wolfire2003} and the thermal equilibrium temperature in our model (first panel) matches both the WNM (\citealp{Marchal2021}, MMD21) and CNM (\citealp{Wakelam2017}, W17 diffAC) temperatures.

It is important to note that the classical S-shaped function $P_{\mathrm{eq}}$ from \citet{Wolfire2003} describes the gas conditions at the WNM-CNM interface, and therefore their CNM is representative only of the outermost layer of cold gas clouds. Deeper into the gas cloud, as the shielding column density increases, gas can cool to lower temperatures and eventually form molecular hydrogen, H$_{2}$, which is not included in the \citet{Wolfire2003} model. In our model, the increase in the molecular hydrogen fraction, $f_{\mathrm{H2}}$ (fourth panel), which increases the mean particle mass, $\mu$ (third panel), appears as a reduced pressure for $\log n_{\mathrm{H}}\,[\mathrm{cm}^{-3}]\gtrsim 1.2$ compared to \citet{Wolfire2003} because $P_{\mathrm{eq}}\propto T_{\mathrm{eq}}\,/\mu_{\mathrm{eq}}$. The deviation of our fiducial model from \citet{Wolfire2003} for $\log n_{\mathrm{H}}\, [\mathrm{cm}^{-3}] \gtrsim 1$ is partly explained by the difference between modeling the exterior of the cloud and the interior of the cloud. Furthermore, the $P_{\mathrm{eq}}$ line from \citet{Wolfire2003} in Fig.~\ref{fig:peqfiducial} is for a specific assumed parametrization of the radiation field (here: typical for a Galactocentric radius of 8.5~kpc, varied in their figure 7) and shielding column density (here: $10^{19}\,{\mathrm{cm}}^{-2}$, varied in their figure 9).

\begin{table*}
    \caption{An overview of the shielding column densities, $N_{\mathrm{sh}}$ (column 3), incident ISRF strengths in Habing units, $G_{\mathrm{0,i}}$ (column 4), the shielded ISRF strengths in Habing units, $G_{\mathrm{0,sh}}$ (equation~\ref{eq:dustextinction}, column 5), and cosmic ray rates, $\zeta_{\mathrm{CR}}/\zeta_0$ with $\zeta_0=2\times10^{-16}\,\mathrm{s}^{-1}$, (column 5) for the model variations (column 1) discussed in Section~\ref{sec:neutral} for densities and temperatures typical for the WNM and CNM in the Galaxy. Column 2 lists which input parameter value was changed relative to the fiducial values in Table~(\ref{tab:fiducialparameter}). }
    \centering
    \begin{tabular}{llllll}
\hline
Model & Parameter & $N_{\mathrm{sh}}$ [cm$^{-2}$]& $(G/G_0)_i$ & $(G/G_0)_{\mathrm{sh}}$ &$\zeta_{\mathrm{CR}}$ \\
\hline
\multicolumn{6}{c}{Low WNM density ($n_{\mathrm{H}} = 0.1\,\mathrm{cm}^{-3}$) and $T=7000\,\mathrm{K}$} \\
\hline
Fiducial                         & -                             & 	$2 \times 10^{20}$& 	0.21	 & 	0.18	 & 	0.28 \\
No turbulence                    & $v_{\mathrm{turb}}=0$         & 	$2 \times 10^{20}$& 	0.21	 & 	0.18	 & 	0.28 \\
Low turbulence                   & $v_{\mathrm{turb}}/2$         & 	$2 \times 10^{20}$& 	0.21	 & 	0.18	 & 	0.28 \\
Weak ISRF                        & $G/10$                        & 	$2 \times 10^{20}$& 	0.02	 & 	0.02	 & 	0.28 \\
Strong ISRF                      & $G\times10$                   & 	$2 \times 10^{20}$& 	2.10	 & 	1.76	 & 	0.28 \\
Low CR rate                      & $\zeta_{\mathrm{CR}}/10$      & 	$2 \times 10^{20}$& 	0.21	 & 	0.18	 & 	0.03 \\
High CR rate                     & $\zeta_\mathrm{CR}\times10$   & 	$2 \times 10^{20}$& 	0.21	 & 	0.18	 & 	2.78 \\
Small $N_{\mathrm{sh}}$          & $N_{\mathrm{sh}}/2$           & 	$1 \times 10^{20}$& 	0.21	 & 	0.19	 & 	0.28 \\
Large $N_{\mathrm{sh}}$          & $N_{\mathrm{sh}}\times2$      & 	$4.1 \times 10^{20}$& 	0.21	 & 	0.15	 & 	0.28 \\
\hline
\multicolumn{6}{c}{High WNM density ($n_{\mathrm{H}} = 1\,\mathrm{cm}^{-3}$) and $T=7000\,\mathrm{K}$} \\
\hline
Fiducial                         & -                             & 	$6.1 \times 10^{20}$& 	0.99	 & 	0.58	 & 	1.00 \\
No turbulence                    & $v_{\mathrm{turb}}=0$         & 	$6.1 \times 10^{20}$& 	0.99	 & 	0.58	 & 	1.00 \\
Low turbulence                   & $v_{\mathrm{turb}}/2$         & 	$6.1 \times 10^{20}$& 	0.99	 & 	0.58	 & 	1.00 \\
Weak ISRF                        & $G/10$                        & 	$6.1 \times 10^{20}$& 	0.10	 & 	0.06	 & 	1.00 \\
Strong ISRF                      & $G\times10$                   & 	$6.1 \times 10^{20}$& 	9.93	 & 	5.80	 & 	1.00 \\
Low CR rate                      & $\zeta_{\mathrm{CR}}/10$      & 	$6.1 \times 10^{20}$& 	0.99	 & 	0.58	 & 	0.10 \\
High CR rate                     & $\zeta_\mathrm{CR}\times10$   & 	$6.1 \times 10^{20}$& 	0.99	 & 	0.58	 & 	10.00 \\
Small $N_{\mathrm{sh}}$          & $N_{\mathrm{sh}}/2$           & 	$3 \times 10^{20}$& 	0.99	 & 	0.76	 & 	1.00 \\
Large $N_{\mathrm{sh}}$          & $N_{\mathrm{sh}}\times2$      & 	$1.2 \times 10^{21}$& 	0.99	 & 	0.34	 & 	1.00 \\
\hline
\multicolumn{6}{c}{CNM density ($n_{\mathrm{H}} = 10\,\mathrm{cm}^{-3}$) and $T=100\,\mathrm{K}$} \\
\hline
Fiducial                         & -                             & 	$3.2 \times 10^{21}$& 	10.06	 & 	0.60	 & 	1.00 \\
No turbulence                    & $v_{\mathrm{turb}}=0$         & 	$4.3 \times 10^{20}$& 	0.62	 & 	0.42	 & 	0.81 \\
Low turbulence                   & $v_{\mathrm{turb}}/2$         & 	$1.6 \times 10^{21}$& 	3.81	 & 	0.93	 & 	1.00 \\
Weak ISRF                        & $G/10$                        & 	$3.2 \times 10^{21}$& 	1.01	 & 	0.06	 & 	1.00 \\
Strong ISRF                      & $G\times10$                   & 	$3.2 \times 10^{21}$& 	100.57	 & 	6.03	 & 	1.00 \\
Low CR rate                      & $\zeta_{\mathrm{CR}}/10$      & 	$3.2 \times 10^{21}$& 	10.06	 & 	0.60	 & 	0.10 \\
High CR rate                     & $\zeta_\mathrm{CR}\times10$   & 	$3.2 \times 10^{21}$& 	10.06	 & 	0.60	 & 	10.00 \\
Small $N_{\mathrm{sh}}$          & $N_{\mathrm{sh}}/2$           & 	$1.6 \times 10^{21}$& 	10.06	 & 	2.46	 & 	1.00 \\
Large $N_{\mathrm{sh}}$          & $N_{\mathrm{sh}}\times2$      & 	$6.3 \times 10^{21}$& 	10.06	 & 	0.04	 & 	1.00 \\
\hline

    \end{tabular}
    \label{tab:WNM}
\end{table*}

In their review, \citet{Wakelam2017} categorize molecular gas into diffuse ($0.2\le A_{\mathrm{V}}\le1$,  $0.1\leq f_{\mathrm{H2}} \lesssim 0.5$, W17 diffMC), translucent ($1\le A_{\mathrm{V}}\le2$, $f_{\mathrm{H2}} \approx 0.5$, W17 tranMC), and dense ($A_{\mathrm{V}}>2$, $f_{\mathrm{H2}} \approx 1$, W17 densMC) molecular clouds with densities  $n_{\mathrm{H}}\approx 100-500\,\mathrm{cm}^{-3}$, $500-5000\,\mathrm{cm}^{-3}$, and $>5000\,\mathrm{cm}^{-3}$, respectively. The gas temperature range of diffuse molecular clouds is equal to that of diffuse CNM atomic clouds ($T = 30 - 100\,\mathrm{K}$), and decreases for translucent and dense molecular clouds to $15-40\,\mathrm{K}$. The molecular hydrogen fractions in our model are higher than those in the \citet{Wakelam2017} categories (fourth panel), which is the main reason for the low CNM pressure compared to that measured by \citet{Jenkins2011} (second panel, JT11).

In Section~\ref{sec:galaxysimulations}, we discuss the temperatures, pressures, mean particle masses, and molecular hydrogen fraction of ISM gas in a galaxy simulation that uses the cooling and heating rates from the fiducial model. We show that the median gas properties in the ISM may deviate strongly from the thermal equilibrium properties. Furthermore, we will show that galaxy simulations with particle masses of $\approx 10^5\,\mathrm{M}_{\odot}$ underestimate the molecular hydrogen mass surface densities, which motivates the choice for the fiducial model to have the atomic to molecular hydrogen transition at lower densities compared to observations of the Milky Way Galaxy\footnote{Note, that we use the total molecular hydrogen mass surface densities in the simulations, which may include CO-dark molecular hydrogen (see e.g. \citealp{Seifried2020}). For a more detailed comparison with observations, \textsc{chimes} can be used to produce mock CO maps (see e.g. \citealp{Thompson2024}).}.
Because in highly shielded regions H$_2$ is mainly dissociated by CRs, $f_{\mathrm{H2}}$ is very sensitive to variations in the CR rate. A CR rate that does not saturate at high column densities, e.g. $\beta_{\mathrm{CR}}>0$ in equation~\ref{eq:zetaCR}, shifts the \ion{H}{I}-H$_2$ transition to higher densities. This may be appropriate for very high resolution simulations that include enough high-density gas.

\subsubsection{Thermal equilibrium variations}\label{sec:neutral:var}

The pressures and temperatures of the transitions from the WNM ($T/\mu\approx 10^4\,\mathrm{K}$) to the thermally unstable medium and further to the CNM ($T/\mu\approx 50\,\mathrm{K}$) have been shown to be sensitive to the assumed ISRF strengths, shielding column densities and $\mathcal{DTG}$ ratios \citep{Wolfire1995}, to the Galactocentric distance within the Milky Way \citep{Wolfire2003}, and to the CR rate and gas metallicity \citep{BialySternberg2019}. 

In Fig.~\ref{fig:pressurequilibrium}, we demonstrate the sensitivity of $P_{\mathrm{eq}}\,(n_{\mathrm{H}})$ (big panels) and $f_{\mathrm{H2}}$ (small panels) in our model to parameters related to unresolved turbulence, parametrized by $v_{\mathrm{turb}} = 0,\,3,\,6\,\mathrm{km\,s}^{-1}$, (first column), the strength of the ISRF ($R_{\mathrm{ISRF}}$, second column), the CR rate ($R_{\mathrm{CR}}$, third column), and the shielding column density, $N_{\mathrm{sh}}$ ($R_{\mathrm{sh}}$, fourth column). For reference, we list the values for $N_{\mathrm{sh}}$, the incident ISRF strength in Habing units, $G_{\mathrm{0,i}} = I_{\mathrm{ISRF}}$, the shielded ISRF strength in Habing units, $G_{\mathrm{0,sh}}$, and the CR rate relative to $\zeta_0=2\times10^{-16}\,\mathrm{s}^{-1}$ for solar metallicity and typical WNM and CNM densities and temperatures in Table~\ref{tab:WNM}.

The maximum WNM density as well as the minimum CNM density are very sensitive to the radiation field used in the PE heating, which depends on the ISRF strength (second column) and $N_{\mathrm{sh}}$ (fourth column). The CR rate (third column) has only a small effect on the transition densities but does impact the pressure (i.e. temperature) of gas close to the transition density, confirming that heating from ionization by CRs is a relevant heating mechanism in the WNM (e.g. \citealp{BakesTielens1994}). 

Decreasing the level of turbulence (first panel) does not affect the properties of the WNM because at temperatures of $T/\mu\approx 10^4\,\mathrm{K}$ the thermal pressure rather than the turbulent pressure sets the Jeans column density (equation~\ref{eq:NJ}) for all models and therefore the ISRF strength, the CR rate and $N_{\mathrm{sh}}$ do not depend on $v_{\mathrm{turb}}$ at $T/\mu\approx 10^4$. For cold gas, a smaller value of $v_{\mathrm{turb}}$ decreases the turbulent pressure which decreases the Jeans column density, $N_{\mathrm{J}}$. For cold ($T<T_{\mathrm{min}}=10^4\,\mathrm{K}$) gas, the reference column density $N_{\mathrm{ref}} = N_{\mathrm{J}}$ (equation~\ref{eq:Nref}), and the ISRF strength (equation~\ref{eq:nISRF}), CR rate (equation~\ref{eq:zetaCR}), and the shielding column density (equation~\ref{eq:Nsh}) depend on $N_{\mathrm{ref}}$. 
The interplay of decreasing the incident radiation field and CR rates but also decreasing the shielding column density leads to a shielded radiation field in Habing units that is lowest for $v_{\mathrm{turb}} = 0\,\mathrm{km\,s}^{-1}$ ($G_{\mathrm{0,sh}} = 0.42$) and highest for $v_{\mathrm{turb}} = 3\,\mathrm{km\,s}^{-1}$ ($G_{\mathrm{0,sh}} = 0.93$), with the fiducial model ($v_{\mathrm{turb}} = 6\,\mathrm{km\,s}^{-1}$) in between with $G_{\mathrm{0,sh}} = 0.6$ (see Table~\ref{tab:WNM}). The resulting differences in the PE heating rates are responsible for the CNM temperature differences between models with different values for $v_{\mathrm{turb}}$. 

The molecular hydrogen fraction, $f_{\mathrm{H2,eq}}$ (small panels) as well as $P_{\mathrm{eq}}$ of high-density gas ($n_{\mathrm{H}}\gtrsim 100\,\mathrm{cm}^{-3}$) are mainly sensitive to the CR rate because both FUV and EUV radiation are efficiently attenuated and CRs are the main source of both heating through CR ionization and the dissociation of H$_2$ in highly shielded gas. 

As discussed at the end of Section~\ref{sec:neutral:fid}, gas in the fiducial model is predominantly molecular ($f_{\mathrm{H2}}>0.5$) at lower gas densities than e.g. in the review of molecular gas in \citet{Wakelam2017}. We see in Fig.~\ref{fig:pressurequilibrium} (blue dotted line, third column, bottom panel), that increasing the CR rate at CNM densities by a factor of 10 results in a better agreement of $f_{\mathrm{H2,eq}}$ but we will see in Section~\ref{sec:galaxysimulations} that for simulations that do not resolve molecular clouds, a transition to molecular gas at lower densities, as in the fiducial model, may be desirable.

\subsection{Isolated disc galaxy simulations}\label{sec:galaxysimulations}

The assumption of thermal equilibrium that was made in Section~\ref{sec:neutral} provides a reasonable estimate of the physical conditions of the neutral ISM but radiative processes are not the only cooling and heating mechanisms in the ISM. Adiabatic processes, but also shock heating as well as the thermalization of turbulent motions driven by supernova explosions, can impact the temperature of ISM gas. We use our tabulated cooling rates in simulations of an isolated galaxy and compare the thermal equilibrium curves (see Section~\ref{sec:neutral}) to the densities and temperatures of gas particles in the galaxy simulated with the \textsc{swift} code \citep{swift2024}, using the energy-density \textsc{sphenix} smoothed particle hydrodynamics scheme from \citet{Borrow2022sphenix}.

The initial conditions for the isolated galaxy are the same as in \citet{Ploeckinger2024} and available within the `IsolatedGalaxy' example in \textsc{swift}. In short, a disc galaxy with a total disc gas mass of $1.6\times10^{10}\,\mathrm{M}_{\odot}$ and a stellar disc of $3.8\times10^{10}\,\mathrm{M}_{\odot}$ is set up in equilibrium with an analytic \citet{Navarro1997} dark matter halo of $M_{\mathrm{200}} = 1.37\times10^{12}\,\mathrm{M}_{\odot}$ and a concentration of $c=9$. The galaxy is evolved for 500~Myr, at $z=0$. The particle mass of each gas and star particle initially is $10^5\,\mathrm{M}_{\odot}$ and the gravitational force softening length is $\epsilon=265\,\mathrm{pc}$.

Gas may cool down to an internal energy floor that corresponds to a temperature of $10\,\mathrm{K}$ for neutral atomic gas (for $\mu = 1.22$) and $16\,\mathrm{K}$ for molecular gas with $\mu = 2$. We include star formation with the star formation criterion described in \citet{Nobels2024} with $\alpha_{\mathrm{crit}} = 1$, which limits star formation to gravitationally unstable gas. Eligible gas particles are converted into star particles stochastically with a star formation efficiency of 1 per cent per free-fall time. The gas metallicity is initially set to solar abundance ratios and solar metallicity \citep{Asplund2009} and for simplicity, we do not include metal enrichment nor any form of early feedback (i.e. stellar winds, radiation pressure, \ion{H}{II} regions; to be discussed in Ben\'itez-Llambay et al., in preparation), nor supermassive black holes. The energy feedback of core-collapse (CC) supernovae (SN) follows \citet{Chaikin2023} and here we use the \textsc{colibre} implementation, described in detail in Schaye et al. (in preparation). In short, the total injected energy per CC-SN is between $0.8 \times 10^{51}\,\mathrm{erg}$ and $4 \times 10^{51}\,\mathrm{erg}$, depending on the thermal pressure of the star particle at the time of its formation (`stellar birth pressure'), with a pivot point of $P/k_{\mathrm{B}} = 1.5\times10^4\,\mathrm{K\,cm}^{-3}$. The majority of SN energy (90 per cent) is used to stochastically heat individual gas particles to very high temperatures \citep{DallaVecchia2012}, between $10^7\,\mathrm{K}$ and $10^8\,\mathrm{K}$, depending on the birth density with a pivot point of $n_{\mathrm{H}} = 1\,\mathrm{cm}^{-3}$. For the remaining energy (10 per cent), kinetic feedback with a kick velocity of $50\,\mathrm{km\,s}^{-1}$ is applied to neighboring gas particles. This combination of individual large thermal explosions and frequent smaller kicks allows for efficient SN feedback in massive galaxies as well as sufficient sampling of SN events in galaxies of lower masses or low star formation rates. Additionally, we include energy feedback from type-Ia supernovae. The implementation of this process follows that of CC SN feedback, but all energy is injected thermally, with no kinetic component, and where the energy per type-Ia SN is fixed at $10^{51}\,\mathrm{erg}$ (i.e., independent of stellar birth pressure).

The simulations here do not include a live dust model. We demonstrate in \citet{Trayford2025arXiv} how this cooling model can be combined with a live dust model and introduce in Schaye et al. (in preparation) the \textsc{colibre} cosmological simulation project that applies this combined model.

\subsubsection{Thermal pressure distribution}\label{sec:galaxy:eq}

\begin{figure*}
    \centering
    \includegraphics[width=0.45\linewidth]{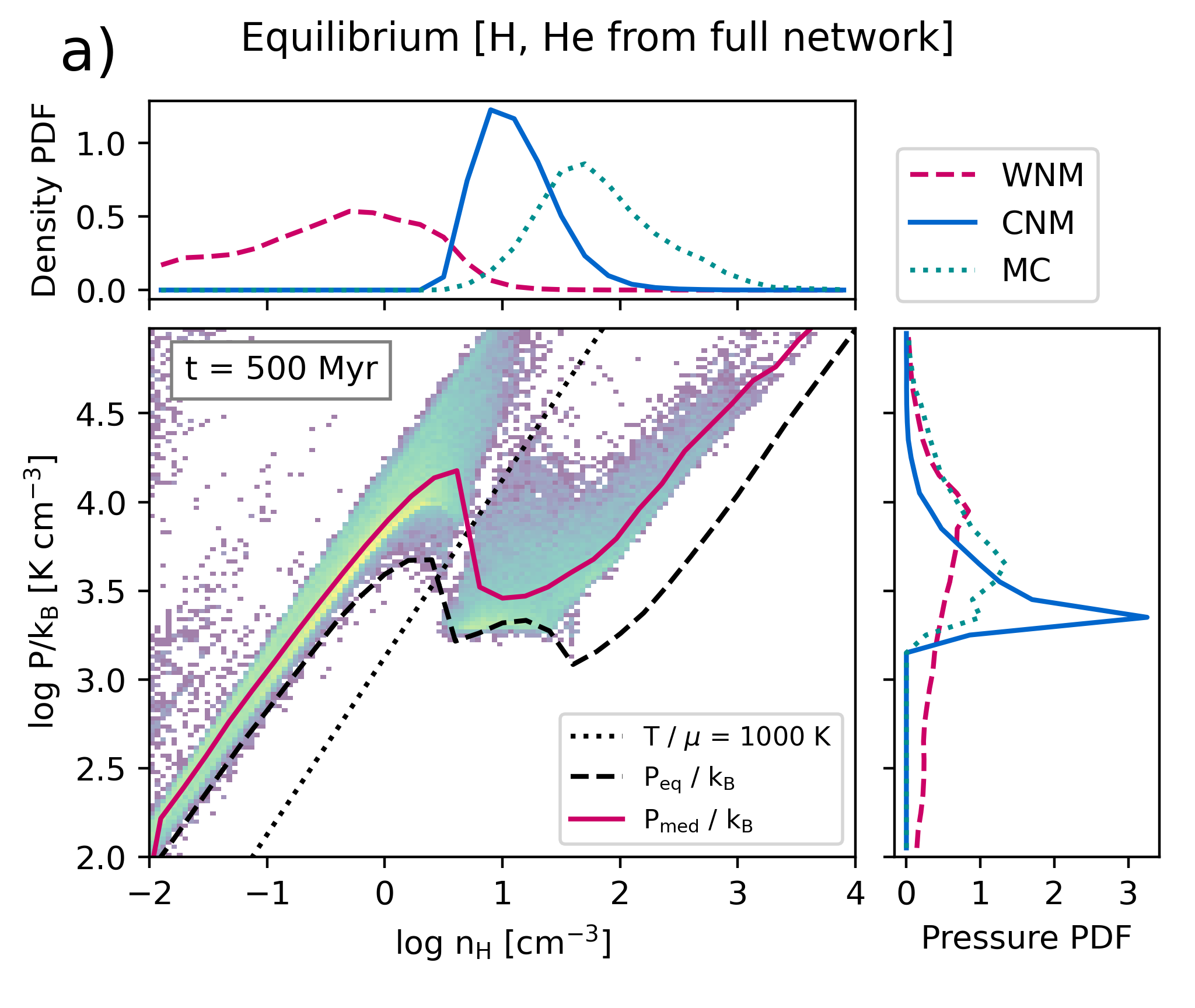}
    \includegraphics[width=0.45\linewidth]{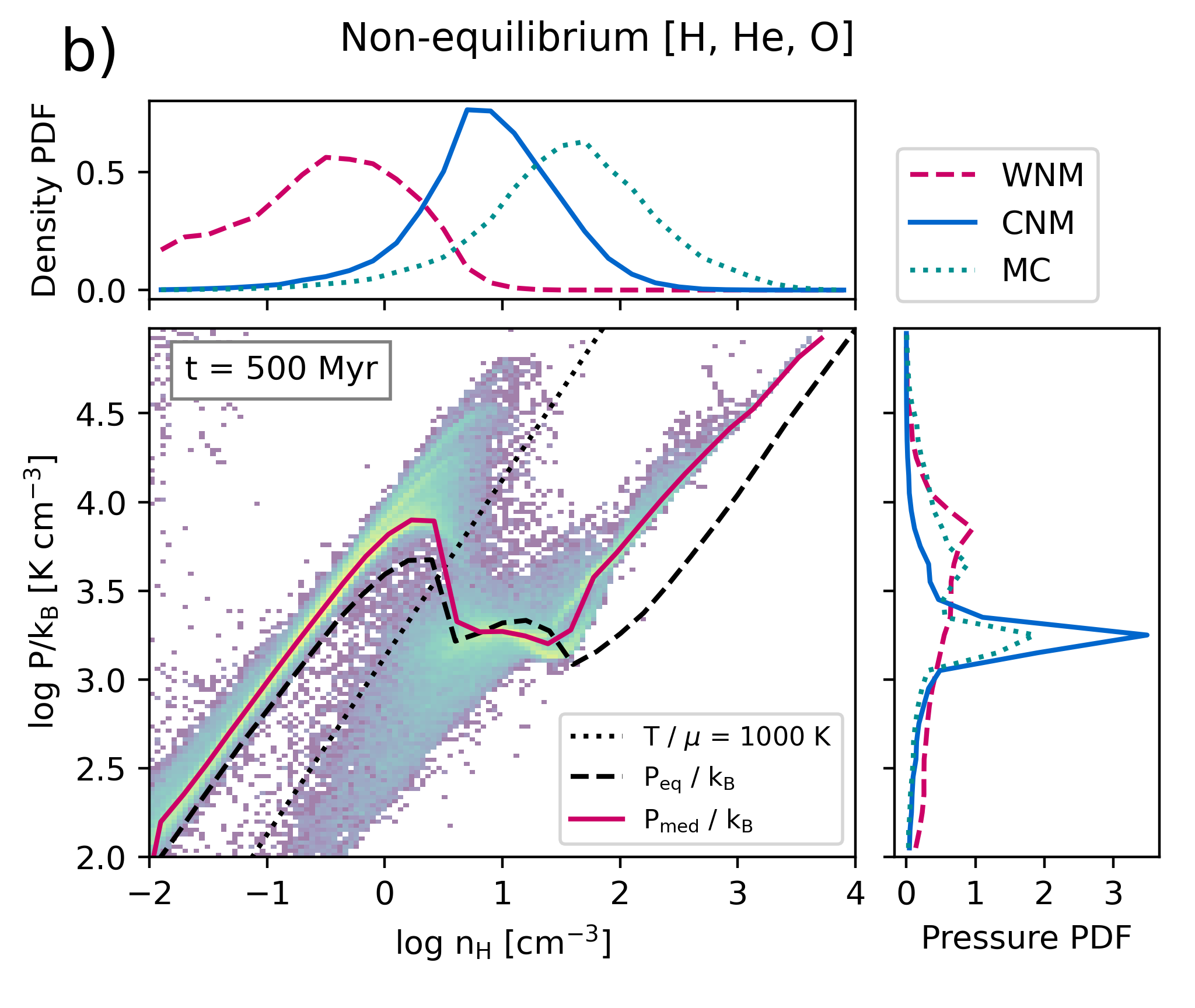}
    \includegraphics[width=0.45\linewidth]{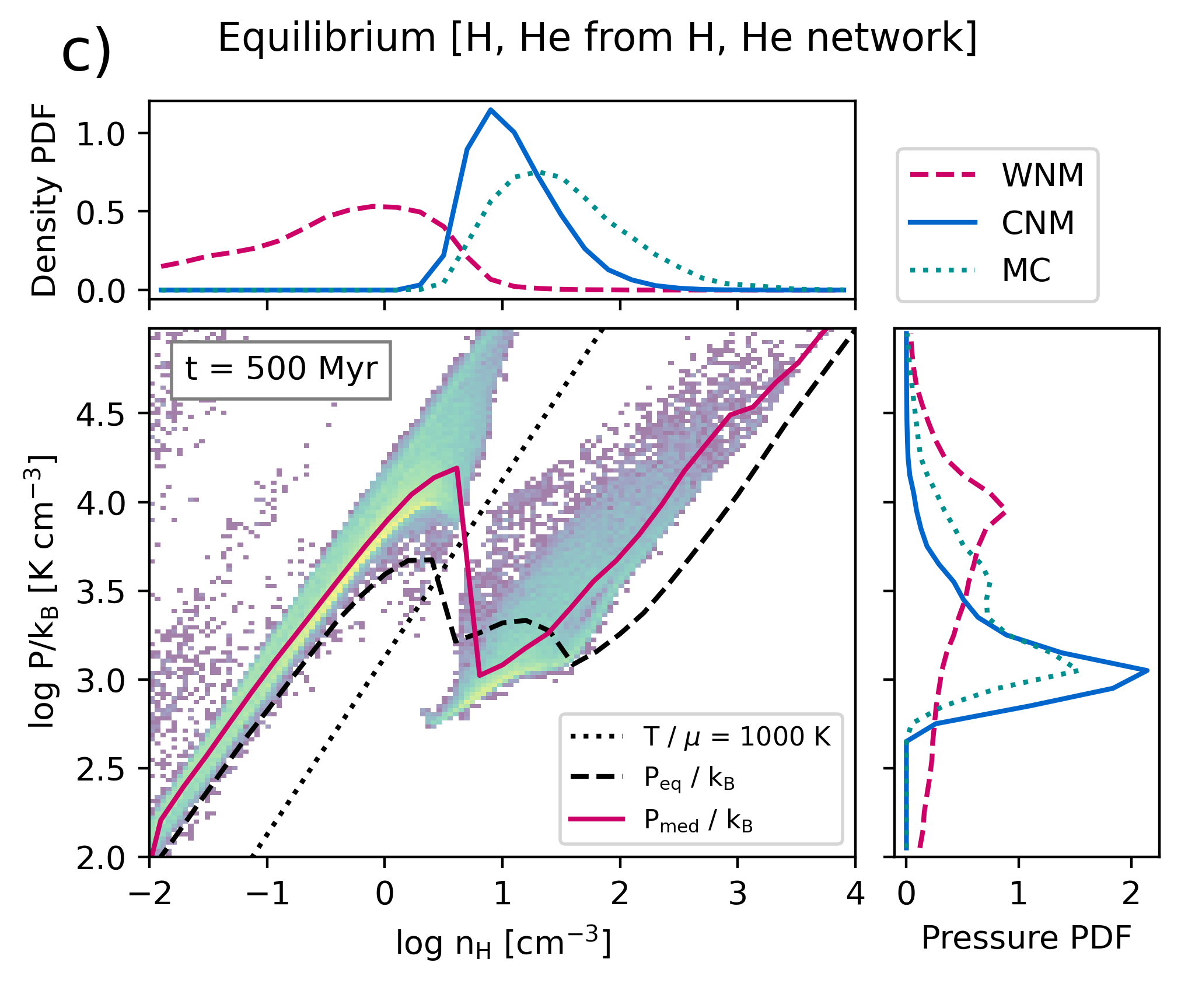}
    \includegraphics[width=0.45\linewidth]{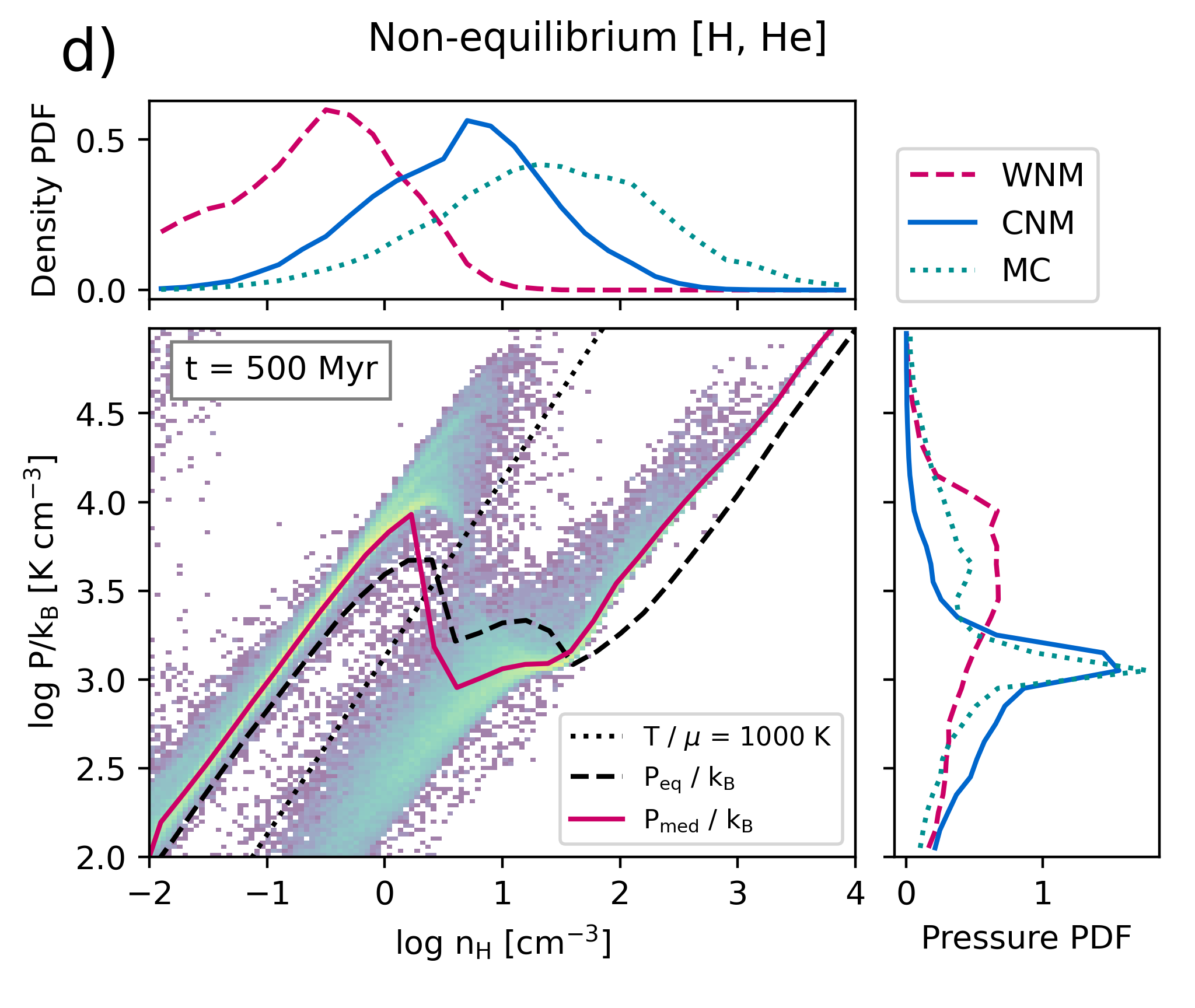}
    \caption{Distribution of thermal pressures, $P/k_{\mathrm{B}}$, and densities, $n_{\mathrm{H}}$, of gas particles within an isolated disc galaxy at $t=500\,\mathrm{Myr}$ using the fiducial parameter values from Table~\ref{tab:fiducialparameter}. The abundances of H and He species are pre-calculated with \textsc{Chimes-Driver} assuming chemical equilibrium in (a) and (c), while they are evolved `on-the-fly' by \textsc{Swift-Chimes} for each gas particle in non-equilibrium in (b) and (d). In the top row (a and b) the H and He species abundances are calculated with a chemical network including O (`ChimesFull' and `ChimesHHeO' in Table~\ref{tab:reducednetworks}) and in the bottom row with a smaller network of only H and He (`ChimesHHe'). The species abundances from the remaining elements are identical in all four plots and calculated with \textsc{Chimes-Driver} using the full chemical network (`ChimesFull" in Table~\ref{tab:reducednetworks}).
    The large panel in each figure shows the thermal pressures and densities of gas particles as 2D histogram where the colormap from purple to yellow is proportional to the log of the number of particles per pixel. The median pressures as a function of density is indicated as red solid line and the constant $T/\mu = 1000\,\mathrm{K}$ with a black dotted line. The thermal equilibrium pressure, $P_{\mathrm{eq}}$ (see Fig~\ref{fig:peqfiducial}), for the tabulated rates using the full chemical network is repeated as a black dashed line for reference. The right (top) panels in each figure show the individual pressure (density) probability density functions of warm neutral gas (WNM, $T/\mu > 1000\,\mathrm{K}$, red dashed line), cold neutral gas (CNM, $T/\mu < 1000\,\mathrm{K}$, blue solid line) and molecular gas (MC, green dotted line). The PDFs of the individual phases are calculated by weighting the contribution of the individual gas particles with their \ion{H}{I} (WNM, CNM) or H$_2$ (MC) mass fractions.}
    \label{fig:simpressureP25}
\end{figure*}

The hydrodynamic properties of gas, such as pressure, internal energy (or temperature), and density, respond to changes in the radiative gas cooling and heating rates, especially for gas with short cooling timescales, as in the ISM. These rates, in turn, depend on the abundance of individual species. If chemical equilibrium is assumed, the equilibrium species abundances are pre-tabulated with \textsc{Chimes-Driver}, as described in Section~\ref{sec:equilibriumabundances}.

For simulations that assume chemical equilibrium, there are two equivalent ways to calculate the cooling and heating rates from the model presented in this work: (1) the rates are read from the provided tables and during the simulation interpolated in redshift, gas density, metallicity and temperature. This method is independent of the hydrodynamic code and provides a trivial update for any simulation project that has previously used the tabulated rates of \citetalias{PS20} (although note the differences outlined in Section~\ref{sec:processes}); (2) the rates are calculated with the \textsc{chimes} cooling module within \textsc{swift}, in equilibrium mode, for the provided equilibrium species abundances and with the \textsc{chimes} parameters set to the values from Table~\ref{tab:fiducialparameter}. By construction, these two methods are expected to return the same rates for the same species abundances and input parameters and we have verified that this is indeed the case. In the following, we focus on simulations for which the cooling and heating rates are calculated by the \textsc{chimes} cooling module within \textsc{swift}.

In the hybrid non-equilibrium model, we follow the non-equilibrium species abundances of a subset of species with \textsc{Swift-Chimes} and assume chemical equilibrium for the remaining individually traced metal species (see Section~\ref{sec:hybrid}). Due to the timescales associated with formation and dissociation (for H$_2$) as well as ionization and recombination (for H and He atoms and ions), the abundances of species may differ from those calculated under the assumption of chemical equilibrium. This affects the cooling and heating rates and therefore the thermal pressures of gas particles. 

As discussed in Section~\ref{sec:equilibriumabundances:red}, the abundances of individual species (here: H$_2$) may depend on the completeness of the reaction network. This affects both the pre-calculated equilibrium abundances (as shown in Section~\ref{sec:equilibriumabundances:red}) as well as the non-equilibrium abundances. Fig.~\ref{fig:simpressureP25} summarizes the impact of different chemical network sizes as well as non-equilibrium effects in the ISM of the simulated galaxy. The left column of plots (a and c) in Fig.~\ref{fig:simpressureP25} shows the thermal pressure and density distribution of the ISM, assuming chemical equilibrium, while the right column (b and d) shows the hybrid model where the species of a subset of elements may be out of chemical equilibrium. Furthermore, the species abundances of H and He are calculated with a chemical network that includes oxygen in the top row (a:~ChimesFull and b:~ChimesHHeO\footnote{We showed in Section~\ref{sec:equilibriumabundances:red} that including O results in H and He species abundances that are identical to those from the full network.}) and with a network of only H and He (ChimesHHe) in the bottom row (c and d). 

Each figure shows the thermal pressures and densities of gas in a simulated galaxy $500\,\mathrm{Myr}$ after the simulation started. Fig.~\ref{fig:simpressureP25}a uses the abundances and rates from the full network with the fiducial parameter values from Table~\ref{tab:fiducialparameter}. The black dashed line in the large panel is the thermal equilibrium pressure, $P_{\mathrm{eq}}\,/\,k_{\mathrm{B}}$, as in Fig.~\ref{fig:pressurequilibrium}. The median thermal pressures of the gas particles in the simulation ($P_{\mathrm{sim}}\,/\,k_{\mathrm{B}}$, red solid line, large panel) are higher than $P_{\mathrm{eq}}\,/\,k_{\mathrm{B}}$, which shows that even when assuming equilibrium abundances, typical WNM pressures and temperatures deviate systematically from those of the thermal equilibrium model. At densities of $0.5 \lesssim n_{\mathrm{H}}\,[\mathrm{cm}^{-3}] \lesssim 5$ and a temperature of $T\approx 5000\,\mathrm{K}$, the cooling rate is $\Lambda/n_{\mathrm{H}}^2\,[\mathrm{cm}^{-3}]\approx 10^{-26}\,\mathrm{erg\,cm^{3}\,s^{-1}}$ in the fiducial equilibrium model. For these values, the cooling time, $\tau_{\mathrm{cool}}=(3/2)\,k_{\mathrm{B}}\,T/\,[n_{\mathrm{H}} \, (\Lambda/n_{\mathrm{H}}^2)]$, much smaller than the free-fall time, $\tau_{\mathrm{ff}} = [3\pi/ (32\,G\,n_{\mathrm{H}}\,m_{\mathrm{H}}\,\mu)]^{1/2}$, with a ratio of $\tau_{\mathrm{cool}}/\tau_{\mathrm{ff}} = 0.07 \, (n_{\mathrm{H}}/[1\,\mathrm{cm}^{-3}])^{-1/2}$. We conclude that the cause of the deviation of the median pressures from $P_{\mathrm{eq}}$ are mechanical heating processes, such as shock heating, that are missing from the radiative heating and cooling balance assumed for $P_{\mathrm{eq}}$.  Adiabatic cooling is subdominant at the densities and pressures displayed, and the thermal equilibrium line, therefore, describes the lower limit of thermal pressures in the simulation. At high densities, the temperatures of the gas particles are limited by the imposed minimum temperature within \textsc{Swift-Chimes} of $T_{\mathrm{min}} = 10\,\mathrm{K}\,(\mu/1.224)$, which results in a constant offset to the dashed line for which the minimum temperature is 10~K, independent of the value for $\mu$. 

The pressure and density probability density functions (PDF) for the neutral phases in the ISM (WNM, CNM, MC) are shown in the right and top panels, respectively. Here, we weigh the contributions of each gas particle by their \ion{H}{I} mass fraction for WNM and CNM and by their H$_2$ mass fraction for molecular gas (MC). The atomic neutral medium is separated into gas that is warmer (WNM) and colder (CNM) than $T/\mu = 1000\,\mathrm{K}$ (black dotted line, large panel). The lack of gas close to $T/\mu = 1000\,\mathrm{K}$ is explained by thermal instability (see the discussion in Section~\ref{sec:neutral}).

At thermal pressures of $\log P\,/\,k_{\mathrm{B}}\,[\mathrm{K\,cm^{-3}}]\approx 3.4$ or $P\,/\,k_{\mathrm{B}} \approx 2.5\times10^3\,\mathrm{K\,cm^{-3}}$, the peak of the CNM PDF, all three phases (WNM, CNM, MCs) are in pressure equilibrium (right panel, Fig~\ref{fig:simpressureP25}a). The density PDFs (top panel) peak at $\log n_{\mathrm{H}} [\mathrm{cm}^{-3}]\approx -0.3,\,1$, and $1.7$ for WNM, CNM and MCs, respectively. Note, that the MC distribution will extend to higher densities in simulations with higher resolution. 

The results for an isolated galaxy with the hybrid non-equilibrium model `ChimesHHeO' (H, He, and O in non-equilibrium, Table~\ref{tab:reducednetworks}) are shown in Fig.~\ref{fig:simpressureP25}b. In agreement with previous work (e.g. \citealp{Richings2016}), the density distributions (top panels) of cold neutral and molecular gas (CNM, MC) are wider in non-equilibrium chemistry because of the lag associated with each phase transition. Interestingly, compared to the simulation with the corresponding equilibrium abundances (Fig.~\ref{fig:simpressureP25}a), the median thermal pressures are closer to the thermal equilibrium pressure. 

A chemical network that does not include reactions between hydrogen and oxygen species (ChimesHHe) overestimates the H$_2$ fractions (see Section~\ref{sec:equilibriumabundances:red}). The increased H$_2$ cooling leads to a lower thermal pressure in the CNM by $\approx 0.3\,\mathrm{dex}$, both when using equilibrium abundances (compare solid blue lines in the right panels of Figs.~\ref{fig:simpressureP25}a and c) and in the hybrid model with only H and He in non-equilibrium (compare solid blue lines in the right panels of Figs.~\ref{fig:simpressureP25}b and d).

\paragraph*{Computational expense: } The simulation with H and He in non-equilibrium is a factor of 2 slower than the simulation that assumes chemical equilibrium and interpolates the pre-tabulated species abundances. Adding oxygen species to the non-equilibrium network increases the CPU time by another factor of 1.85. 

\subsubsection{Atomic to molecular hydrogen transition}\label{sec:galaxy:H2}

\begin{figure}
    \centering
    \includegraphics[width=\linewidth]{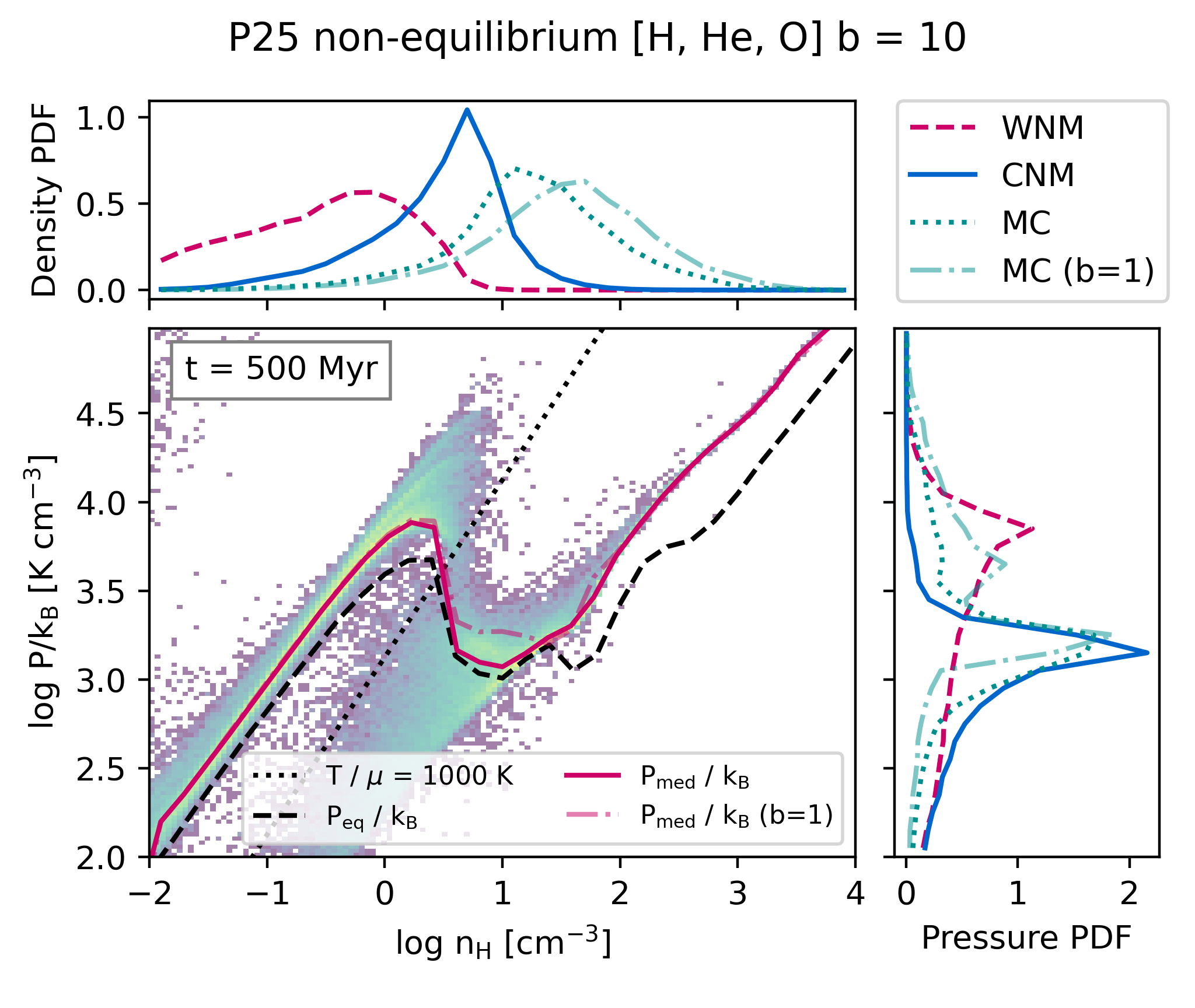}
    \caption{As Fig.~\ref{fig:simpressureP25}b, but with a maximum boost factor of $b_{\mathrm{dust,max}} = 10$. For reference, we add the median pressure line (large panel) and the molecular PDFs (small panels) from Fig.~\ref{fig:simpressureP25}b (same model but without a boost factor, $b=1$) as dash-dotted lines.
    Comparing the median pressures lines, we see that the WNM-CNM transition is unaffected by the boost factor and the CNM pressures are decreased only very slightly (by $\approx 0.15~\mathrm{dex}$ at $\log n_{\mathrm{H}} \,[\mathrm{cm}^{-3}]\approx 1$). As expected, the CNM to MC transition is shifted to lower densities (upper panel), compared to Fig.~\ref{fig:simpressureP25}b (repeated here as dash-dotted line).}
    \label{fig:sim:peq:boost10}
\end{figure} 

The simulations shown here use a mass of $10^5\,\mathrm{M}_{\odot}$ per baryon resolution element and a gravitational force softening length of $\epsilon = 265~\mathrm{pc}$, which is a competitive resolution for simulations of cosmological volumes. For this resolution, gravitational collapse is slowed down (i.e. follows the softened free-fall time, see \citealp{Ploeckinger2024}) for gas clouds with a Jeans length, $\lambda_{\mathrm{J}}$, smaller than the softening length ($\lambda_{\mathrm{J}}\le\epsilon$), a condition that is fulfilled for densities of $n_{\mathrm{H}} \ge 3.2\,\mathrm{cm}^{-3}\,(\,T/10^4\,\mathrm{K})$. In addition, star formation reduces the amount of unresolved high-density gas, which is critical for computational efficiency. Acknowledging that the molecular phase remains poorly resolved in cosmological volumes, we nevertheless demonstrate the transition from atomic to molecular gas phase, which will be better resolved than the internal properties of molecular clouds, in the simulations of isolated galaxies. 

In Section~\ref{sec:dustboost}, we introduce a boost factor that increases the H$_2$ formation rate on dust grains by a factor, $b_{\mathrm{dust}}$. In the fiducial model we use a maximum boost factor of unity (i.e. no boost) but in the following figures we compare the results from the fiducial model to those that use a boost factor that increases from $b_{\mathrm{dust}}=1$ at $\log n_{\mathrm{b,min}}\,[\mathrm{cm}^{-3}] = 0.5$ to $b_{\mathrm{dust}}=b_{\mathrm{dust,max}} =10$ at $\log n_{\mathrm{b,max}}\,[\mathrm{cm}^{-3}]  = 1$ (see equation~\ref{eq:dustboost}). Increasing $b_{\mathrm{dust,max}}$ allows us to increase the H$_2$ fraction for a specific resolution in a controlled way\footnote{Note, that a different normalization or exponent for the cosmic ray rate has a comparable effect (see the bottom panel in the third column of Fig.~\ref{fig:pressurequilibrium}).} in an effort to compensate the above mentioned resolution effects.

The immediate effect of the boost factor is the efficient formation of H$_2$ in gas with lower volume densities. Comparing the density PDFs (top panel) from a simulation with (Fig.~\ref{fig:sim:peq:boost10}) and without (Fig.~\ref{fig:simpressureP25}b) the boost factor confirms that the \ion{H}{I} to H$_2$ transition is shifted to $\approx 0.5\,\mathrm{dex}$ lower densities with $b_{\mathrm{dust,max}} =10$. 
The increased H$_2$ fraction only slightly decreases the pressure of the CNM by $\approx 0.1~\mathrm{dex}$ (right panel) and the transition from the WNM to the CNM (red line, big panel) remains unaffected. 

\begin{figure}
    \centering
    \includegraphics[width=\linewidth]{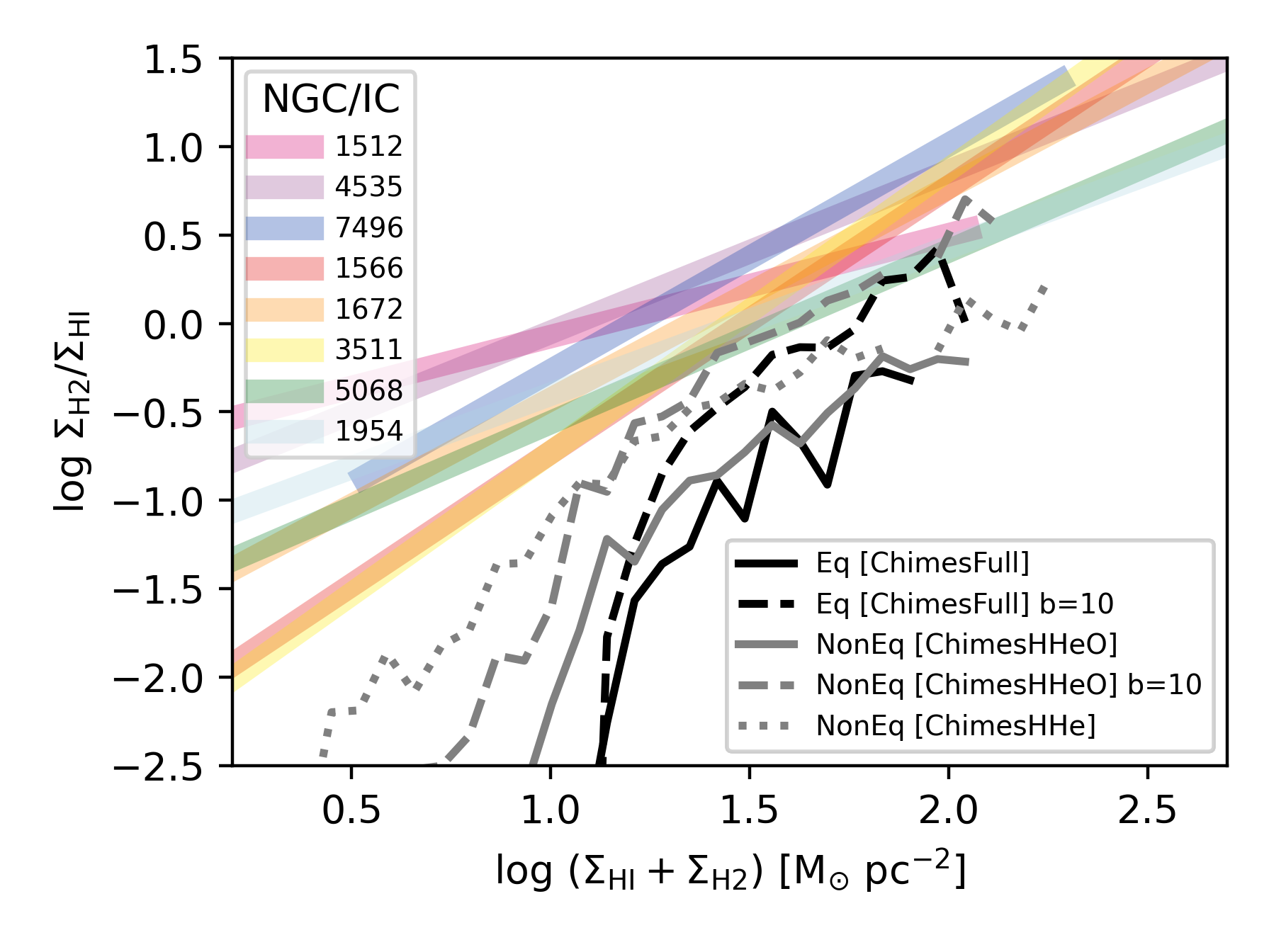}
    \caption{The molecular to atomic hydrogen transition, quantified as the ratio of the mass surface densities, $\log \Sigma_{\mathrm{H2}}/\Sigma_{\mathrm{HI}}$, vs. the total neutral hydrogen mass surface density, $\log (\Sigma_{\mathrm{H2}} +\Sigma_{\mathrm{HI}})$, in an isolated galaxy at $z=0$ with gas at constant solar metallicity for the fiducial parameters listed in Table~\ref{tab:fiducialparameter} (solid lines). The black (grey) lines represent simulations with (non-)equilibrium chemistry and the dashed lines are for simulations with boosted H$_2$ formation on dust grains with $b_{\mathrm{dust}}  = 10$. Each line represents the median of individual 1~kpc x 1~kpc gas patches for the four simulations shown in Fig.~\ref{fig:simpressureP25}. The grey dotted line shows the H$_2$ to \ion{H}{I} ratio for a non-equilibrium simulation with the ChimesHHe network. The thick bands are linear regression fits in $\log \Sigma_{\mathrm{H2}}/\Sigma_{\mathrm{HI}}$ and $\log (\Sigma_{\mathrm{H2}} +\Sigma_{\mathrm{HI}})$ including both measurement errors and non-detections, to observations of individual galaxies from \citet{Eibensteiner2024} (divided by 1.35 to correct for their included contribution from helium) and added for reference. }
    \label{fig:sim:H2fraction}
\end{figure}

The increased H$_2$ fraction per volume density propagates into the H$_2$ and \ion{H}{I} mass surface densities as illustrated in Fig.~\ref{fig:sim:H2fraction}. Here we show the ratio of H$_2$ and \ion{H}{I} mass surface densities, $\Sigma_{\mathrm{H2}}\,/\,\Sigma_{\mathrm{HI}}$, versus the total hydrogen mass surface density, $\Sigma_{\mathrm{HI}} + \Sigma_{\mathrm{H2}}$. The mass surface densities are calculated with the fast SPH interpolation in \textsc{SwiftSimIO} \citep{Borrow2020} on a $1\,\mathrm{kpc}\times1\,\mathrm{kpc}$ grid (comparable to the average beam size in \citealp{Eibensteiner2024}) based on the hydrogen species fraction of each gas particle.
Compared to the simulation with the fiducial parameter values and assuming chemical equilibrium (black solid line), the H$_2$ to \ion{H}{I} ratio is slightly increased if the species fractions of hydrogen, helium and oxygen are calculated in non-equilibrium (solid grey line). In simulations with a boost factor of $b_{\mathrm{dust,max}} =10$, the H$_2$ to \ion{H}{I} ratio is systematically increased for both equilibrium (black dashed line) and non-equilibrium (grey dashed line) species abundances. A similar increase is seen in non-equilibrium simulations that do not include oxygen in the chemical network (grey dotted line), as discussed in Sections~\ref{sec:equilibriumabundances:red} and \ref{sec:galaxy:eq}. 

Only the non-equilibrium variation with a boost factor of $b=10$ (grey dashed line) of this isolated galaxy has H$_2$ ratios as high as observations of Milky Way mass galaxies indicate (colored bands, \citealp{Eibensteiner2024}). Note that their H$_2$ ratio is sensitive to the assumed conversion factor, $\alpha_{\mathrm{CO}}$, used to convert the observed CO integrated intensity into an H$_2$ surface density. Furthermore, \citet{Trayford2025arXiv} shows that combining the hybrid cooling with a live dust model increases the H$_2$ fractions. The main reason for this is the higher proportion of small grains around the atomic-molecular transition densities, which boosts the dust cross sections and thus H$_2$ formation. Simulations of large cosmological volumes will reveal if the low H$_2$ fraction found in this simulated isolated galaxy is representative for the full galaxy population.

\begin{figure}
    \centering
    \includegraphics[trim={3.2cm 1cm 1.4cm 0.cm},clip,width=\linewidth]{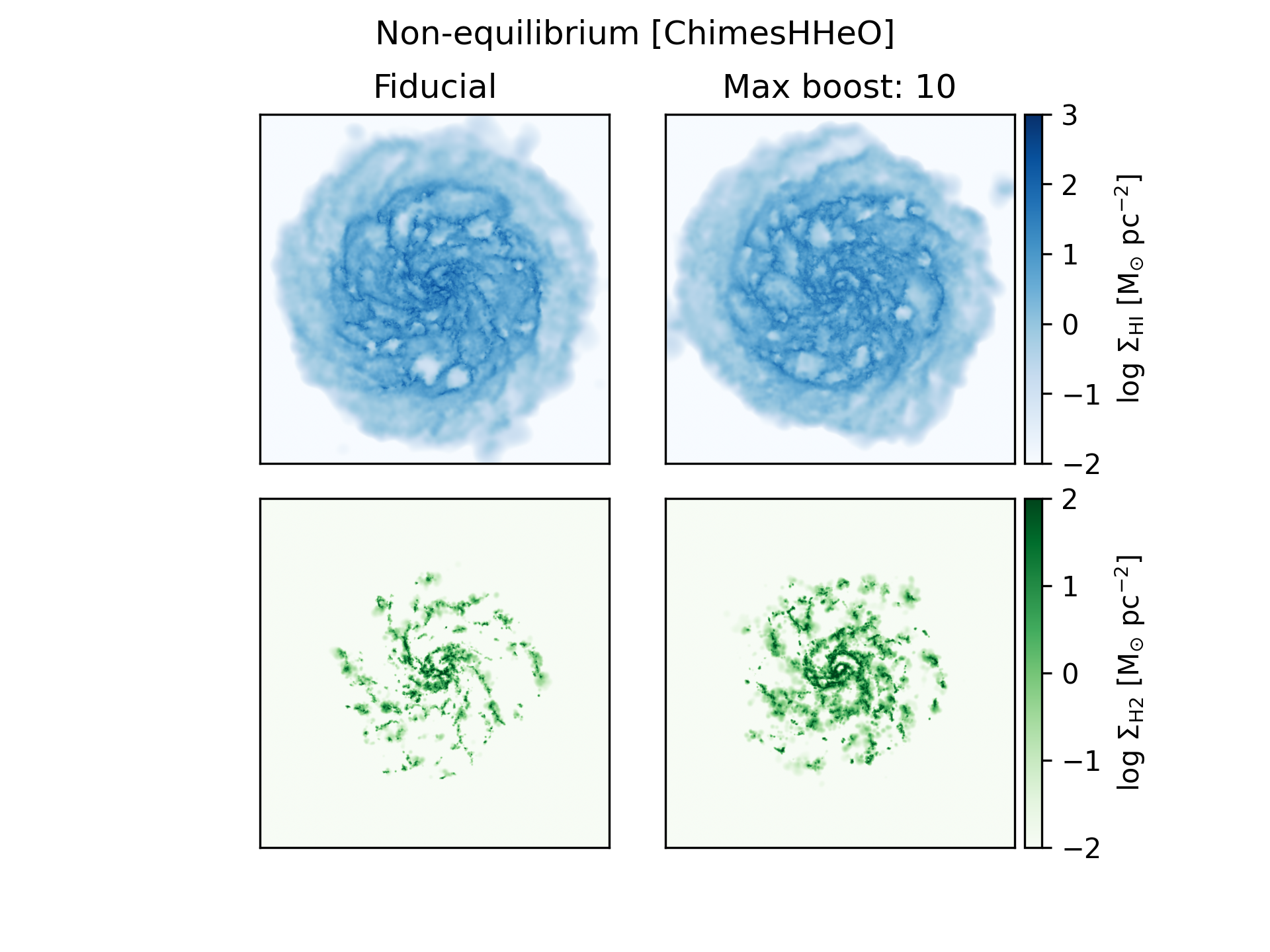}
    \caption{Mass surface density maps of \ion{H}{I} (top row) and H$_2$ (bottom row) for the fiducial parameter values (Table~\ref{tab:fiducialparameter}), using a chemical network including hydrogen, helium and oxygen to calculate the non-equilibrium species fractions (left column; grey solid line in Fig.~\ref{fig:sim:H2fraction}). In the simulation shown in the right column, the formation rate of H$_2$ on dust grains is boosted by a factor of $b_{\mathrm{dust,max}} = 10$ (grey dashed line in Fig.~\ref{fig:sim:H2fraction}; see text for details).}
    \label{fig:sim:massmaps}
\end{figure}

For a visual impression, the \ion{H}{I} (top row) and H$_2$ (bottom row) mass surface density maps ($50\times50\,\mathrm{kpc}$ image with $1024^2$ pixels) are shown in Fig.~\ref{fig:sim:massmaps} for the fiducial model in equilibrium with the full chemical network (left column) and the model with a dust boost parameter of $b_{\mathrm{dust}} = 10$ (right column). The left and right columns correspond to the grey solid and grey dashed lines in Fig.~\ref{fig:sim:H2fraction}), respectively. In nature (or in simulations with much higher mass resolution, see e.g.~\citealp{Thompson2024} for a simulation using \textsc{chimes} with a particle mass of $400\,\mathrm{M}_{\odot}$), the averaged surface densities will include individual molecular clouds with volume densities that exceed the volume densities resolved in the simulation by several orders of magnitude. Introducing the boost factor indeed compensates for this and the effects are apparent both at the highest surface densities, i.e. in the center of the galaxy, as well as in the lower surface density regions along the spiral arms.

\section{Discussion and conclusions}\label{sec:summary}

We present a hybrid (non-)equilibrium chemistry model, \textsc{hybrid-chimes}, to calculate the species fractions as well as the cooling and heating rates of gas in galaxy formation simulations. We parametrize the diffuse interstellar radiation field, the cosmic ray rate and the shielding column density through a density- and temperature-dependent reference column density, which is an approximation for the typical coherence length of self-gravitating structures. This treatment yields a powerful model that can be used in simulations that do not include explicit radiative transfer calculations. The work builds on \citetalias{PS20} and we discuss the updates relative to that work in Section~\ref{sec:method} and in Appendix~\ref{sec:app:PS20comp}. 
The fiducial parameter values of this work are listed in Table~\ref{tab:fiducialparameter} and the dependence of the thermal equilibrium relation between pressure and density on parameter variations is demonstrated in Fig.~\ref{fig:pressurequilibrium}. 

The species fractions and per element cooling and heating rates are tabulated for H, He, C, N, O, Ne, Si, Mg, S, Ca, and Fe, assuming chemical equilibrium, i.e. ionization equilibrium and steady state chemistry. In Section~\ref{sec:hybrid}, we present a hybrid model, \textsc{hybrid-chimes}, that combines a reduced non-equilibrium chemical network for a subset of elements with the pre-tabulated properties for the remaining elements calculated assuming chemical equilibrium. Cooling rates, such as from collisional excitation, recombination, free-free emission or from Compton processes, are proportional to the number density of free electrons. For hybrid cooling, the total electron density used in all cooling rates includes the electrons from the non-equilibrium ion species plus the electrons from the equilibrium ion species. The cooling rates of the equilibrium species therefore also include some of the important non-equilibrium effects. 

As mentioned in Section~\ref{sec:hybrid}, the chemistry and radiative cooling library \textsc{grackle} uses a similar hybrid approach to calculate cooling and heating rates. For a subset of either six, nine, or twelve primordial species, their abundances as well as cooling and heating rates are calculated in non-equilibrium. The metal cooling rates for species abundances in chemical equilibrium are calculated with \textsc{cloudy} for solar abundance ratios, assuming a constant or redshift-dependent radiation field (see \citealp{Smith2008} for details). 

Different from \textsc{grackle}, we tabulate the cooling and heating rates per element, which improves accuracy for non-solar abundance ratios. In addition, individual metal elements can easily be added to the non-equilibrium network in our hybrid model, albeit at a higher computational cost. In contrast to \textsc{grackle}, we use the same chemistry solver (\textsc{chimes}) for both the non-equilibrium and the equilibrium abundances (\textsc{grackle} uses \textsc{cloudy} for the metal species). This avoids inconsistencies in the detailed implementation of individual processes between non-equilibrium and equilibrium species (see Appendix~\ref{sec:app:PS20comp} for an example of the inconsistencies between \textsc{cloudy} and \textsc{chimes}). We also ensure that equilibrium and non-equilibrium calculations assume identical radiation fields, cosmic ray rates, dust content\footnote{An inconsistency in the dust properties is unavoidable when coupling to a dust evolution model \citep{Trayford2025arXiv}. In the simulation, gas with the same density, temperature, and metallicity may have different dust properties, which cannot be reproduced in the tables without adding multiple extra dimensions for dust content and dust composition. This inconsistency only affects the dust shielding of metal species that are assumed to be in chemical equilibrium and therefore pre-tabulated.}, and shielding columns. By scaling all cooling rates with the total (equilibrium plus non-equilibrium, see Fig.~\ref{fig:hybrid}) electron fraction, we increase the accuracy of non-equilibrium cooling rates by adding the electrons from metal species, and capture important non-equilibrium effects, even in the cooling rates of `equilibrium' elements. In \textsc{grackle}, the electron densities from primordial and metal species do not affect each other's cooling rates. This may lead to an overestimation of the metal cooling rate if optically thin equilibrium tables for metals, which include optically thin ion fractions and therefore free electrons from hydrogen and helium, are combined with the cooling rates from the chemistry network that uses the \citet{Rahmati2013} prescription for shielding (discussed in \citealp{Hu2017Grackle}). \citet{Emerick2019Grackle} produced metal cooling rate tables for \textsc{grackle} that include self-shielding of gas (but not shielding by dust) to alleviate this problem. The general issues related to assuming solar abundance ratios and the lack of connection between the equilibrium and non-equilibrium calculations through the electron densities remain unaddressed in that approach.

We caution that while using a reduced chemical network may be unavoidable due to the drastically increased computational costs for the full network, important reactions may be missing from small networks. We discuss in Section~\ref{sec:hybrid} that for solar metallicity the molecular hydrogen fraction is artificially increased at the \ion{H}{I}-H$_{2}$ transition if the chemical network only includes hydrogen and helium. Including oxygen alleviates this issue and restores the molecular hydrogen fraction from the full chemical network (Fig.~\ref{fig:Oreactions}).

Balancing the cooling and heating rates for constant metallicity and redshift defines thermal equilibrium temperatures and pressures as a function of gas density. The thermal equilibrium pressure, $P_{\mathrm{eq}}$, is commonly used to characterize the thermally stable phases of the ISM (e.g. \citealp{Wolfire2003, Kim2023}). We compare the median thermal pressures of gas particles in equilibrium chemistry simulations of isolated galaxies, $P_{\mathrm{med}}$, to $P_{\mathrm{eq}}$ and find large differences between the expected and actual density and pressure distributions of the warm and cold neutral phases. This difference is particularly large for the cooling and heating rates from \citetalias{PS20} (Fig.~\ref{fig:simpressurePS20}), which were calibrated using simulations. The thermal equilibrium temperature would not predict a multi-phase ISM, while simulations that incorporate these rates do produce both a warm and a cold phase in pressure equilibrium, albeit at lower pressures than the fiducial model in this work. We conclude that models assuming both chemical and thermal equilibrium should be used with caution because they may produce inaccurate predictions for the typical physical conditions of the ISM in simulations that assume chemical but not thermal equilibrium. 

We compare the pressures $P_{\mathrm{med}}$ and $P_{\mathrm{eq}}$ to the pressure equilibrium function from \citet{Wolfire2003} and found reasonable agreement for the fiducial model for solar metallicities. We emphasize that each \citet{Wolfire2003} curve is for specific values of the radiation field, the cosmic ray rate, the dust content, and the shielding column density. Furthermore, the  \citet{Wolfire2003} model describes the interface between warm and cold phases (also done e.g. in \citealp{BialySternberg2019}), while our model describes typical gas properties, which for high densities may be in the interior of a cold atomic or molecular gas cloud. Finally, our model is designed to be used in cosmological simulations of galaxy formation which include gas exposed to a large variety of conditions, ranging from the diffuse intergalactic medium to molecular clouds in high-redshift starbursting galaxies, which we accomplish by varying the parameters that are assumed to be constant for each individual \citet{Wolfire2003} model. 

The hybrid cooling model presented in this work, connected to the live multi-species and multi-sizes dust grain model (presented in \citealp{Trayford2025arXiv}), is central to the properties of the multi-phase interstellar medium in the upcoming simulation project \textsc{colibre} (Schaye et al. in preparation). 

\section*{Acknowledgements}

We thank the anonymous referee for a swift and constructive referee report.
The development for this work has greatly benefited from discussions with the \textsc{colibre} team. The analysis presented in Appendix~\ref{sec:app:highz} was inspired by discussions with Azadeh Fattahi and Shaun Brown, who also provided the cross section files that correspond to the Lyman Werner radiation from \citet{Incatasciato2023}. This research was funded in part by the Austrian Science Fund (FWF) grant number V~982-N. This paper made use of the following python packages: astropy \citep{astropy2022}, numpy \citep{numpy}, scipy \citep{scipy}, matplotlib \citep{matplotlib}, unyt \citep{unyt} and swiftsimio \citep{Borrow2020, Borrow2021}. This work used the DiRAC@Durham facility managed by the Institute for Computational Cosmology on behalf of the STFC DiRAC HPC Facility (www.dirac.ac.uk). The equipment was funded by BEIS capital funding via STFC capital grants ST/K00042X/1, ST/P002293/1, ST/R002371/1 and ST/S002502/1, Durham University and STFC operations grant ST/R000832/1. DiRAC is part of the National e-Infrastructure. The computational results presented have been achieved in part using the Vienna Scientific Cluster (VSC). JT acknowledges support of a STFC Early Stage Research and Development grant (ST/X004651/1). This project has received funding from the Netherlands Organization for Scientific Research (NWO) through research programme Athena 184.034.002.

%%%%%%%%%%%%%%%%%%%%%%%%%%%%%%%%%%%%%%%%%%%%%%%%%%
\section*{Data Availability}

The tables for selected models are available at both the main chimes webpage \href{https://richings.bitbucket.io/chimes/download.html}{richings.bitbucket.io/chimes/download.html} as well as \textsc{hybrid-chimes} project webpage \href{https://www.sylviaploeckinger.com/hybridchimes}{www.sylviaploeckinger.com/hybridchimes}. The tables follow the same format as those presented by \citet{PS20} to allow for a seamless update. However, special attention should be given to the sections titled `Caution when upgrading from PS20' within this work. Each table set contains one \texttt{hdf5} file per redshift that includes the cooling and heating rates as well as the species fractions for the full range in densities, temperatures and metallicities. Furthermore, we provide a collection of python scripts to produce new tables with updated parameter values at the same links.   

The chemical network \textsc{chimes} is publicly available from \href{https://richings.bitbucket.io/chimes/user_guide/index.html}{richings.bitbucket.io/chimes/user\_guide/index.html} and the public version of \textsc{swift} can be found on \href{https://www.swiftsim.com}{swiftsim.com}. The \textsc{swift} modules related to the \textsc{colibre} galaxy formation module will be integrated into the public version after the public release of \textsc{colibre}.



\bibliographystyle{mnras}
\bibliography{stellarfeedback} % if your bibtex file is called example.bib

%\begin{thebibliography}{99}
%\end{thebibliography}

%%%%%%%%%%%%%%%%%%%%%%%%%%%%%%%%%%%%%%%%%%%%%%%%%%

%%%%%%%%%%%%%%%%% APPENDICES %%%%%%%%%%%%%%%%%%%%%

\appendix

\section{Comparison to \citetalias{PS20}} \label{sec:app:PS20comp}

\begin{figure}
    \centering
        \includegraphics[width=\linewidth]{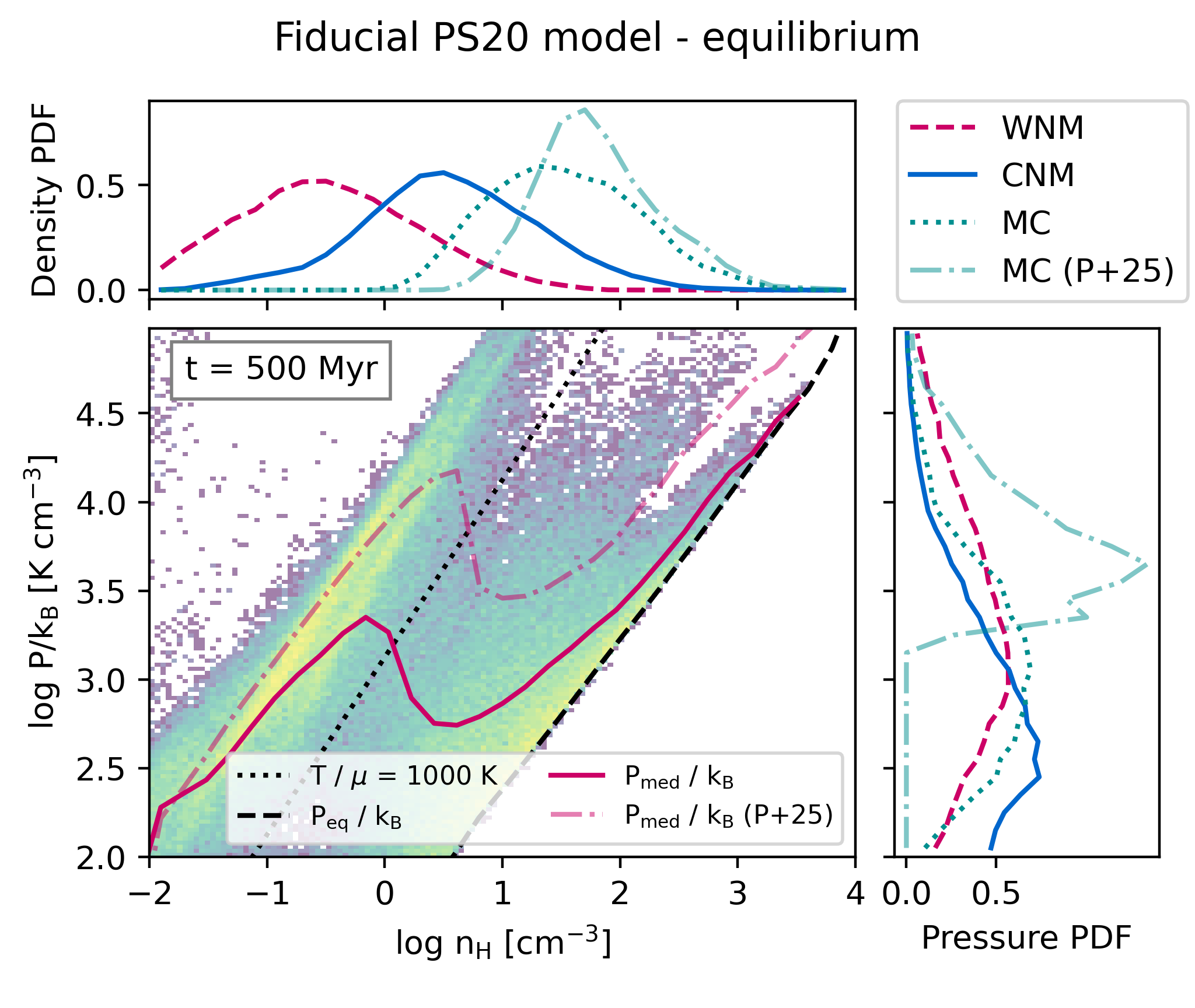}
    \caption{As Fig.~\ref{fig:simpressureP25}a but for the tabulated rates presented in \citetalias{PS20}. For reference, the median pressure line (red solid line, big panel) and the molecular PDFs (green dotted lines, small panels) from Fig.~\ref{fig:simpressureP25}a are repeated as dash-dotted lines.}
    \label{fig:simpressurePS20}
\end{figure}

This work is based on \citetalias{PS20} and here we discuss the differences between the fiducial model \citetalias{PS20} and this work. In Fig.~\ref{fig:simpressurePS20} we show the thermal pressure and density distributions of gas in an isolated galaxy with the fiducial \citetalias{PS20} cooling tables (UVB\_dust1\_CR1\_G1\_shield1). Applying the rates of \citetalias{PS20} in an isolated galaxy simulation (see Section~\ref{sec:galaxysimulations}) results in a multiphase ISM over a wide range of pressures, while the thermal equilibrium pressure (black dashed line) would be indicative of a single (cold) phase at pressures of $\log P/k_{\mathrm{B}}\,[\mathrm{K\,cm}^{-3}] > 2$. Using the thermal equilibrium temperatures or pressures of tabulated rates to predict the properties of the simulated multi-phase ISM would therefore lead to inaccurate expectations. Compared to the fiducial model in this work (Fig.~\ref{fig:simpressureP25}a), the CNM has shifted to lower densities (by $\approx 0.5\,\mathrm{dex}$) and therefore to lower pressures. 

Table~\ref{tab:fiducialparameter} lists the fiducial parameter values used in this work (3rd column) and in \citetalias{PS20} (4th column). The slightly lower normalization of the ISRF (by a factor of 0.54, Table~\ref{tab:fiducialparameter}) in \citetalias{PS20} is expected to reduce the photoelectric heating rate in the WNM and to reduce the CNM pressures (see also second column in Fig.~\ref{fig:pressurequilibrium}). This is indeed seen in Fig.~\ref{fig:simpressurePS20} when comparing the median pressure lines of simulations with the cooling rates from \citetalias{PS20} (solid line, large panel) and from this work (dash-dotted line, large panel), but the difference is larger than expected for a factor of 2 difference in the ISRF normalization (compare second panel of Fig.~\ref{fig:pressurequilibrium}). 

\begin{figure}
    \centering
    \includegraphics[width=\linewidth]{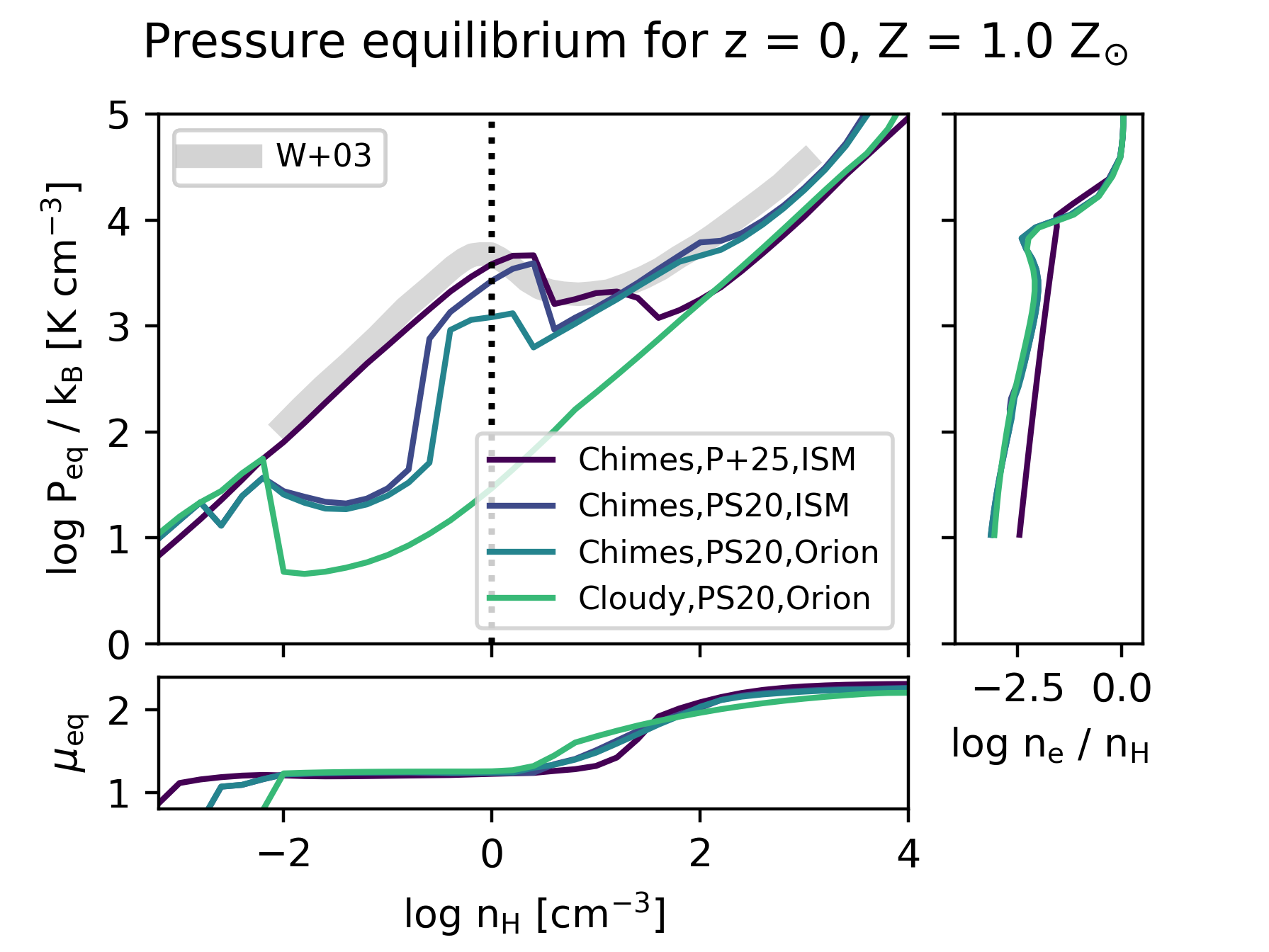}
    \caption{The thermal equilibrium pressure (top panel) and the mean particle mass at thermal equilibrium (bottom panel) calculated with \textsc{cloudy} or \textsc{chimes}, for the parameter values of this work (P+25) or from \citetalias{PS20}, and for the \textsc{orion} and \textsc{ism} grain sets (see labels).  }
    \label{fig:PS20comparison}
\end{figure}

In addition to the difference in the parameter values, the species fractions as well as the cooling and heating rates are calculated with \textsc{cloudy} version 17.03 \citep{Cloudy1998PASP} in \citetalias{PS20}, while in this work we use \textsc{chimes} \citep{chimes2014optthin, chimes2014shielded}. We recalculated the species abundances and cooling rates for the fiducial model from \citetalias{PS20} with \textsc{chimes} and compare them with the original tables. Fig.~\ref{fig:PS20comparison} shows the resulting equilibrium pressure, $P_{\mathrm{eq}}$, and the mean particle mass, $\mu_{\mathrm{eq}}$, at $P_{\mathrm{eq}}$. 

\begin{figure}
    \centering
    \includegraphics[width=\linewidth]{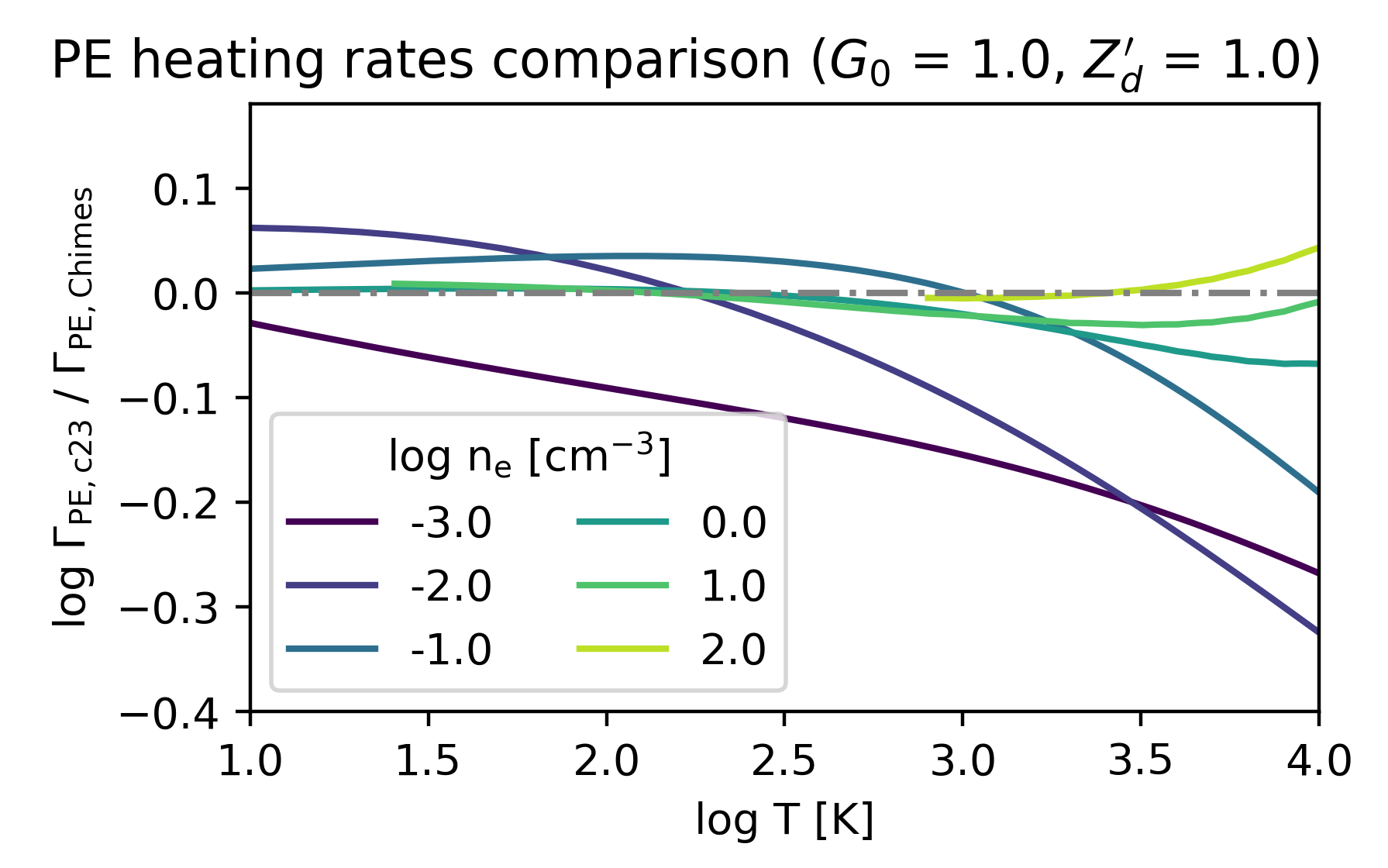}
    \caption{Ratio of the PE heating rates in \textsc{cloudy} version 23 \citep{Cloudy2023}, $\Gamma_{\mathrm{PE,c23}}$, and in \textsc{chimes}, $\Gamma_{\mathrm{PE,Chimes}}$, for the same radiation field ($G_0$, without shielding), the \textsc{ism} grain model, and for different constant electron densities, $n_{\mathrm{e}}$ (different line colors, see the legend). }
    \label{fig:PEcomparison}
\end{figure}

The large differences in $P_{\mathrm{eq}}$ at intermediate densities ($-1\lesssim \log n_{\mathrm{H}}\,[\mathrm{cm}^{-3}] \lesssim 1$) are related mainly to differences in the PE heating rate and the free electron fractions. As an example, we quantify the differences for $\log T\,[\mathrm{K}] = 2.5$ and $\log n_{\mathrm{H}}\,[\mathrm{cm}^{-3}] = 0$, a gas temperature and density for which the differences are largest. For the same \textsc{cloudy} grain model, the PE heating rate is a factor of three lower in \textsc{cloudy} (from \citealp{Weingartner2006}, both in version 17, as used in \citetalias{PS20}, and in the current version 23, \citealp{Cloudy2023}) compared to that in \textsc{chimes} (from \citealp{Wolfire2003}). Within \textsc{cloudy}, the \textsc{orion} grain model produces a factor of 2.8 lower PE heating than the \textsc{ism} grain model, while this difference shrinks to $\approx 20\,\mathrm{per\,cent}$ in \textsc{chimes}. Finally, the free electron fraction in \textsc{cloudy} is 20 (40) per cent lower than in \textsc{chimes} for the \textsc{orion} (\textsc{ism}) grain model at this example temperature and density. Because the PE heating rates of \textsc{chimes} and \textsc{cloudy} for the same radiation field and electron fractions agree within $0.15\,\mathrm{dex}$ at $\log T\,[\mathrm{K}] = 2.5$ (see Fig.~\ref{fig:PEcomparison}), the differences between the radiative transfer calculations in \textsc{cloudy} and the shielding approximations in \textsc{chimes} are likely the cause of the discrepancies. Given the sensitivity of the PE heating rate to details of the dust composition, an agreement within a factor of a few is reasonable.

Summarizing, while the \citetalias{PS20} parameter values have been updated in this work (Table~\ref{tab:fiducialparameter}), the thermal equilibrium temperature function also changes as a result of using \textsc{chimes} instead of \textsc{cloudy} and of using the \textsc{ism} instead of the \textsc{orion} grain model. 

\section{Background radiation variations before hydrogen reionization}\label{sec:app:highz}

\begin{figure}
    \centering
    \includegraphics[width=\linewidth]{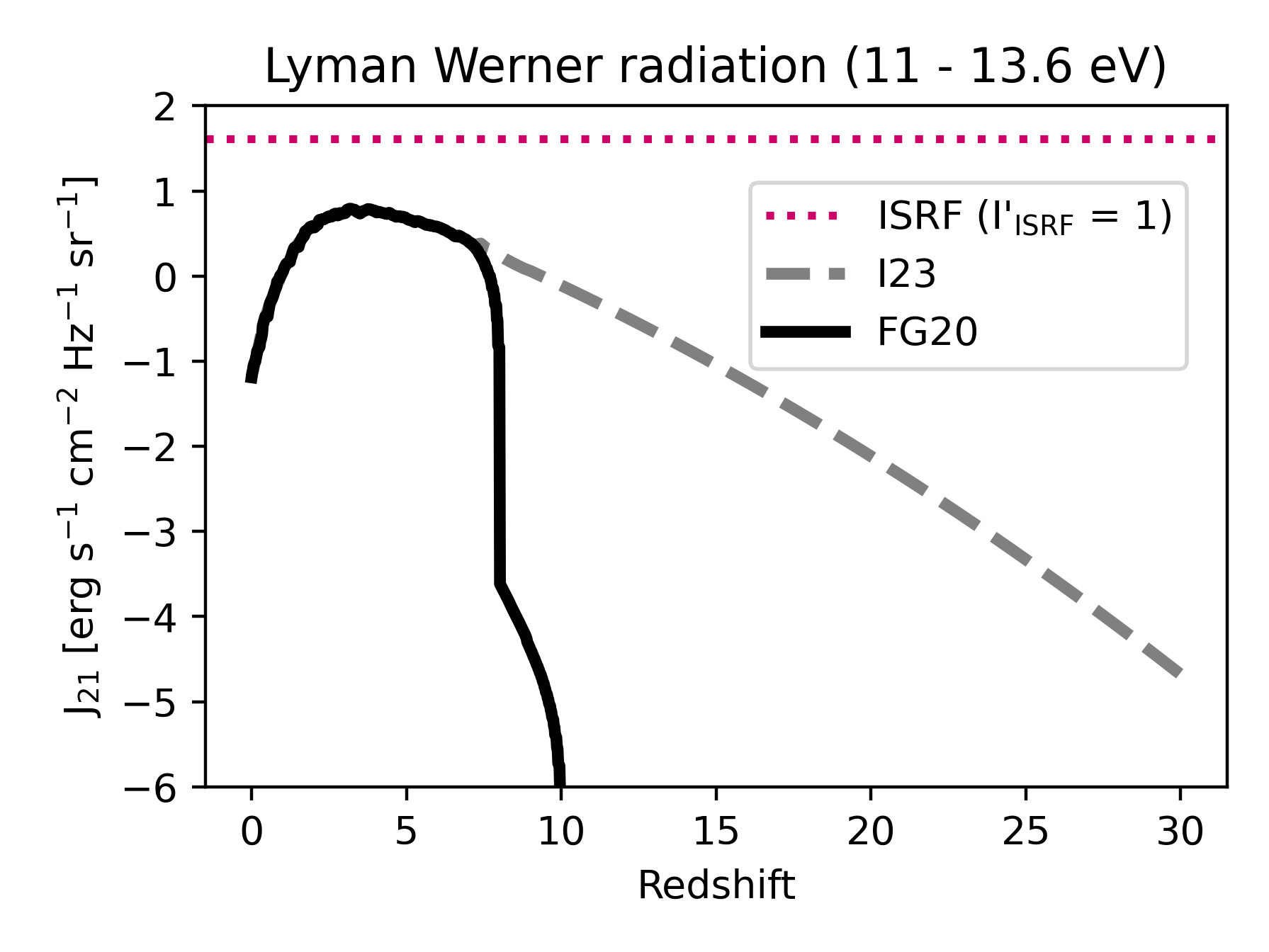}
    \caption{Average specific intensity of H$_2$ photodissociating radiation (LW radiation), $J_{\mathrm{21}} \,[\mathrm{erg\,s^{-1}\,cm^{-2}\, Hz^{-1}\,\mathrm{sr}^{-1}}]= 10^{-21} \, J_{\nu} $, for energies between 11 and 13.6~eV. The LW radiation for the UV background from \citet{FG20} is indicated by the black solid line while the specific intensity in the LW band from \citet{Incatasciato2023} is shown as the grey dashed line. The red dotted line is the specific intensity of the ISRF with a normalization of $I'_{\mathrm{ISRF}} = 1$. The specific LW radiation intensity is identical in both background radiation fields for $z\lesssim7$. For $z\gtrsim7$, the photodissociating radiation decreases drastically for the the UV background from \citet{FG20}. }
    \label{fig:LymanWerner}
\end{figure}

The background radiation field before hydrogen reionization is very poorly constrained for energies below the hydrogen ionization energy ($E<13.6\,\mathrm{eV}$). Lyman-Werner (LW) radiation, typically defined as the energy range between 11 and 13.6~eV, strongly affects the formation of the first galaxies, as it dissociates hydrogen molecules, which provide the main radiative cooling process in primordial gas for $T<10^4\,\mathrm{K}$. Photons in this energy range have a very long mean free path ($\approx 100\,\mathrm{comoving\,Mpc}$, \citealp{Ahn2009}), even at redshifts ($z\gtrsim7$) at which photons with higher energies are efficiently absorbed locally. 

The fiducial model presented in this work includes both a redshift-dependent homogeneous background radiation field from external sources (distant galaxies and quasars) and a density- and temperature-dependent radiation field that represents the diffuse interstellar radiation within a galaxy. For some applications, such as studying the formation of the first stars, the assumption of an established interstellar radiation field may not be valid. As in \citetalias{PS20}, we therefore also provide a set of tabulated rates and fractions that only include the homogeneous background radiation field. The fiducial background radiation used in this work is based on \citet{FG20} and is the same as that used in \citetalias{PS20}. For $E\leq 13.6\,\mathrm{eV}$, the spectra from \citetalias{PS20} (`modFG') are identical to the original \citet{FG20} spectra. For both, the flux of the LW radiation increases drastically by more than 5 orders of magnitudes between $z \approx9$ and $z \approx 7$, in contrast to estimates from the First Billion Year (FiBY) simulation project, which predicts a more gradual increase \citep{Incatasciato2023} of 2 orders of magnitudes over a much larger range in redshifts, between $z=20$ and $z=7$ (see Fig.~\ref{fig:LymanWerner}). In order to enable studies of the impact of these highly different LW fluxes on halo and galaxy formation, we provide an additional set of tables that use the LW fluxes from \citet{Incatasciato2023} instead of the LW fluxes from \citet{FG20} and that include a larger redshift range (up to $z=30$). In practice, the `I23' UVB is constructed by changing the normalization of the `modFG20' spectra for energies $\leq 13.6\,\mathrm{eV}$ so that the average Lyman-Werner specific intensity matches that from \citet{Incatasciato2023} at each redshift. The high-energy part of the spectrum ($E>13.6\,\mathrm{eV}$) is identical for $z\ge 9$ for both UVB radiation fields, because the \citet{FG20} UVB is only defined for $z\le 9$.   

\begin{figure}
    \centering
    \includegraphics[width=\linewidth]{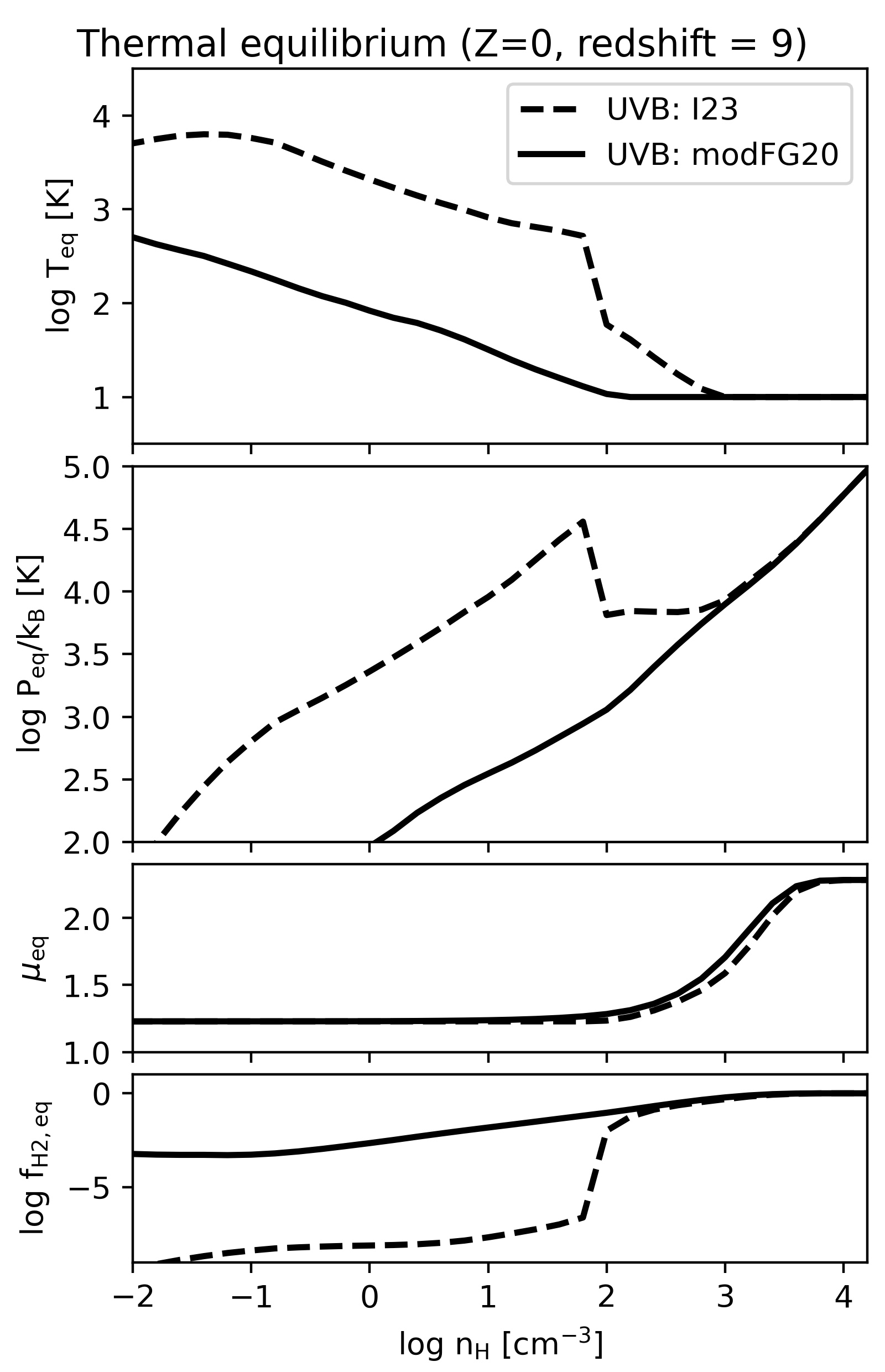}
    \caption{The individual panels show, from top to bottom, the temperature, $T_{\mathrm{eq}}$, pressure, $P_{\mathrm{eq}}$, mean particle mass, $\mu_{\mathrm{eq}}$, and H$_2$ fraction, $f_{\mathrm{H2,eq}}$ in thermal equilibrium for primordial gas ($Z=0$) at $z=9$. Each of the indicated models includes the homogeneous background radiation and self-shielding, but neither cosmic rays nor an interstellar radiation field. The black solid line represents the results for the fiducial background (`modFG20'), as used in \citetalias{PS20}, calculated with \textsc{chimes}. The black dashed line indicated the equilibrium properties of gas exposed to a radiation field that adds the `I23' \citep{Incatasciato2023} Lyman Werner radiation to the `modFG20' spectra.}
    \label{fig:Teqhighz}
\end{figure}

Fig.~\ref{fig:Teqhighz} compares the thermal equilibrium temperature (first panel), pressure (second panel), mean particle mass (third panel), and H$_2$ fraction (fourth panel) of primordial gas ($Z=0$) at $z=9$, for heating and cooling rates calculated assuming no ISRF and no cosmic rays and for the two different background radiation fields: `modFG20' (based on \citealp{FG20}, as in \citetalias{PS20}, black solid line) and `I23' (LW radiation field from \citealp{Incatasciato2023}, black dashed line).  The increased LW radiation substantially decreases the H$_2$ fraction of gas with intermediate densities ($\log n_{\mathrm{H}}\,[\mathrm{cm}^{-3}] \lesssim 2$, see bottom panel). The corresponding reduction of H$_2$ cooling results in equilibrium temperatures and pressures that are $\approx 1\,\mathrm{dex}$ higher than those for the fiducial background radiation field. At high densities ($\log n_{\mathrm{H}} \gtrsim 2$), the gas becomes self-shielded from LW radiation in both cases and their equilibrium properties agree.

%%%%%%%%%%%%%%%%%%%%%%%%%%%%%%%%%%%%%%%%%%%%%%%%%%

% Don't change these lines
\bsp	% typesetting comment
\label{lastpage}
\end{document}